\documentclass[letter,11pt]{article}

\usepackage{jheppub}
\usepackage{multirow}

\def\as{\alpha_S}
\def\t{{\bar t}}
\def\dy{{\Delta y}}
\def\dmody{{\vert\Delta y\vert}}
\def\mody{{\vert y\vert}}
\def\modyt{{\vert y_t\vert}}
\def\Mtt{m_{t\bar t}}
\def\PTtt{p_{T,t\bar t}}
\def\PTt{p_{T,t}}
\def\GeV{\, \rm GeV}
\def\AFB{\rm A_{FB}}
\def\AcumulFB{{\hat A}_{\rm FB}}

\title{\boldmath NNLO QCD predictions for fully-differential top-quark pair production at the Tevatron}

\author[a]{Micha\l{}  Czakon,}
\author[a]{Paul Fiedler,}
\author[b]{David Heymes}
\author[b]{and Alexander Mitov}

\affiliation[a]{Institut f\"ur Theoretische Teilchenphysik und Kosmologie,
RWTH Aachen University, D-52056 Aachen, Germany}
\affiliation[b]{Cavendish Laboratory, University of Cambridge, Cambridge CB3 0HE, UK}

\note{Preprint numbers: Cavendish-HEP-16/01, TTK-16-02}

\abstract{We present a comprehensive study of differential distributions for Tevatron top-pair events at the level of stable top quarks. All calculations are performed in NNLO QCD with the help of a fully differential partonic Monte-Carlo and are exact at this order in perturbation theory. We present predictions for all kinematic distributions for which data exists. Particular attention is paid on the top-quark forward-backward asymmetry which we study in detail. We compare the NNLO results with existing approximate NNLO predictions as well as differential distributions computed with different parton distribution sets. Theory errors are significantly smaller than current experimental ones with overall agreement between theory and data.}

\begin{document} 
\maketitle
\flushbottom

\section{Introduction}

Top-quark pair production is one of the cornerstones of the Tevatron physics program. Despite the relatively limited statistics for top events at Tevatron energies, both the CDF and D\O\ collaborations have presented a number of measurements of differential distributions \cite{Aaltonen:2009iz,CDF:2013gna,CDF:public,Abazov:2010js,Abazov:2014vga,D0:public} and differential top-quark forward-backward asymmetry ($\AFB$) \cite{Aaltonen:2012it,CDF:NoteAFBslope,Abazov:2014cca}.

Being a $p\bar p$ collider, the Tevatron produces top-quark pairs from initial states consisting predominantly of a light quark-antiquark pair. As a result, top-quark pair production at the Tevatron offers direct access to quark parton distribution functions (pdf) and is an order of magnitude more sensitive to charge asymmetries than the LHC. These two considerations are among the main motivation for the current work. 

The current paper extends our previous work \cite{Czakon:2014xsa} on top-quark $\AFB$ by presenting a detailed study of next-to-next-to leading order (NNLO) QCD corrections to differential $\AFB$ and related differential distributions in the following variables: $t\bar t$ rapidity difference $\dy \equiv y_t - y_\t$, $\dy$ and $\Mtt$ as well as $\dy$ and $t\bar t$ transverse momentum $\PTtt$. We also present NNLO QCD corrections to the slopes of $\AFB$ in the variables $\dy$ and $\Mtt$, as well as to the lowest few Legendre moments that have been measured by CDF \cite{CDF:2013gna,CDF:public} in the context of $\AFB$. We study the $\PTtt$ cumulative asymmetry which, as already indicated in Ref.~\cite{Czakon:2014xsa}, allows one to better understand the origin of higher-order QCD corrections to $\AFB$. Finally, we present the NNLO QCD prediction for the cumulative $\Mtt$ asymmetry and discuss it in the context of recent predictions \cite{Wang:2015lna} based on the Principle of Maximum Conformality  (PMC) \cite{Brodsky:2011ta}.

We further extend the scope of the current study by presenting NNLO QCD predictions for all major differential distributions for stable top-quark pairs. Specifically, we show predictions for the following one-dimensional differential distributions measured by the D\O\ Collaboration~\cite{Abazov:2014vga}: $t\bar t$ invariant mass $\Mtt$, transverse momentum $\PTt$ of the top quark (or antiquark) and absolute rapidity $\modyt$ of the top quark (or antiquark). We also present the top-quark differential distribution in $\cos\theta$ (defined in sec.~\ref{sec:costheta} below), together with the related Legendre moments, and compare the NNLO QCD predictions with measurements of the CDF Collaboration~\cite{CDF:2013gna}. Comparisons at the differential level will be helpful in better understanding Standard Model (SM) top-quark production at hadron colliders and will be useful in, for example, further improving top-quark mass extraction at the Tevatron. We compare the main NNLO kinematic distributions with approximate NNLO predictions that have been used in the past.

Although NNLO theoretical predictions for distributions of top-quark decay products are preferred, such a calculation is beyond the scope of the present work given the significant additional effort its implementation would require (despite the fact that differential NNLO top decay is known \cite{Gao:2012ja,Brucherseifer:2013iv}). We are planning to undertake such a calculation in the future.

Finally, we utilise a number of parton distribution sets to study the effect of different pdf's on the predicted differential cross-sections. 

The paper is organised as follows. In sec.~\ref{sec:calc} we discuss the calculation from technical perspective. In sec.~\ref{sec:dif-dist} we present and discuss the NNLO QCD corrections for the $\Mtt,\, \PTt$ and $\modyt$ differential distributions. Sec.~\ref{sec:AFB} is devoted to the top-quark forward-backward asymmetry. In sec.~\ref{sec:pdf} we compare differential distributions based on four pdf sets. A summary of our findings can be found in the last section. All predictions can be found in tables in the appendix.

\section{Details of the calculation}\label{sec:calc}

NLO corrections to top-quark pair production can nowadays be obtained in a multitude of complete Monte-Carlo frameworks (\textsc{Mcfm} \cite{Campbell:2010ff}, \textsc{Powheg} \cite{Alioli:2010xd}, \textsc{aMC@Nlo} \cite{Alwall:2014hca}, \textsc{Sherpa}
\cite{Gleisberg:2008ta}, \textsc{Helac-Nlo} \cite{Bevilacqua:2011xh}), including also the associated production with jets, vector bosons and Higgs. The most advanced calculations at this level of perturbation theory allow for a realistic modelling of the final state. In particular they involve the complete top-quark off-shell effects \cite{Denner:2010jp,Bevilacqua:2010qb,Denner:2012yc,Frederix:2013gra,Cascioli:2013wga,Heinrich:2013qaa}. We should also mention the most recent calculations of this type, where off-shell effects could even be included in associated production \cite{Denner:2015yca,Bevilacqua:2015qha}. As far as NNLO corrections are concerned, it should be possible to work in the Narrow Width Approximation as done at NLO in \cite{Bernreuther:2010ny,Melnikov:2009dn,Campbell:2012uf,Campbell:2014kua}. For now, however, our results are for stable top quarks.

Since our calculation is of NNLO precision, we point out that there has been tremendous progress in this field and many new results appeared \cite{GehrmannDeRidder:2008ug,GehrmannDeRidder:2007hr,Weinzierl:2008iv,Weinzierl:2009ms,DelDuca:2015zqa,Catani:2007vq,Grazzini:2008tf,Catani:2009sm,Ferrera:2011bk,Catani:2011qz,Grazzini:2013bna,Grazzini:2015nwa,Cascioli:2014yka,Gehrmann:2014fva,Grazzini:2015hta,Boughezal:2015dva,Boughezal:2015aha,Boughezal:2015ded,Campbell:2016jau,Ridder:2013mf,Currie:2013dwa,Chen:2014gva,Ridder:2015dxa,Boughezal:2011jf,Brucherseifer:2013iv,Brucherseifer:2013cu,Boughezal:2013uia,Boughezal:2015dra,Caola:2014daa,Brucherseifer:2014ama,Caola:2015wna}. This has been possible thanks to the development of subtraction schemes \cite{GehrmannDeRidder:2005cm,Somogyi:2006da,Somogyi:2006db,Somogyi:2008fc,Bolzoni:2010bt,Somogyi:2013yk,Czakon:2010td,Czakon:2011ve,Czakon:2014oma}, slicing methods \cite{Catani:2007vq,Gaunt:2015pea,Boughezal:2015dva} and the calculation of several two-to-two virtual amplitudes \cite{Anastasiou:2000kg,Anastasiou:2000ue,Anastasiou:2001sv,Glover:2001af,Bern:2002tk,Bern:2003ck,Czakon:2008zk,Baernreuther:2013caa,Bonciani:2008az,Bonciani:2009nb,Bonciani:2010mn,Bonciani:2013ywa,Gehrmann:2013cxs,Gehrmann:2014bfa,Henn:2014lfa,Caola:2014lpa,Papadopoulos:2014hla,Caola:2014iua,Gehrmann:2015ora,vonManteuffel:2015msa}. As far as top-quark pair production is concerned, besides our own calculation \cite{Czakon:2014xsa,Czakon:2015owf} only partial results are known at the differential level \cite{Abelof:2015lna,Abelof:2014fza}; there is also progress at the level of total cross-section \cite{Bonciani:2015sha} obtained with slicing methods. 

We next describe our tools and methods in more detail.
In principle, cross-section contributions in fixed-order perturbation theory can be classified according to the number of additional real emissions with respect to the Born configuration. This is equivalent to the number of virtual loops in the involved amplitudes. At NNLO we would have, according to this classification, three contributions: double-virtual, real-virtual and double-real. Due to the presence of initial state collinear singularities, we must add to this list also the collinear renormalisation contributions, which allow to obtain a finite partonic cross-section. These may be viewed as either convolutions of leading-order splitting functions with the NLO cross-section, or as convolutions of splitting functions (double for leading order, and single for NLO splitting functions) with the Born contribution. 

In order to efficiently deal with infrared singularities, however, this simple picture with a total of five contributions usually needs to be modified. In consequence, a calculation is ultimately organised according to a subtraction scheme, which modifies each one of the five contributions. Our calculation is performed within the framework of the sector-improved residue subtraction scheme \textsc{Stripper} \cite{Czakon:2010td,Czakon:2011ve,Czakon:2014oma}. The results of this work, as well as of our previous Tevatron $\AFB$ paper \cite{Czakon:2014xsa}, have been obtained with the original methods described in more detail in Refs.~\cite{Czakon:2010td,Czakon:2011ve}
\footnote{A subset of the results has been checked using the most recent complete implementation of the four-dimensional formulation of \textsc{Stripper} \cite{Czakon:2014oma}.}.
In the following we describe the original approach as it has been applied in the current work as well as in Ref.~\cite{Czakon:2014xsa}. In particular some of the results presented in Ref.~\cite{Czakon:2014xsa} (see Table I and related discussion) concern partial contributions and thus are dependent on the division into double-virtual, real-virtual and double-real parts.

A specific feature of the original formulation of \textsc{Stripper} was the uniform reliance on conventional dimensional regularisation (CDR). Thus, both real and virtual particles were in principle defined in $d=4-2\epsilon, \epsilon \neq 0$ dimensions. In practice, this implies that the momenta may involve higher dimensions, as is indeed the case in the double-real contribution, where we have to work in five dimensions. Furthermore, the cross-section contributions are not modified (with one exception described below), but rather a Laurent expansion in the regularisation parameter is obtained. In consequence, when we address the value of a particular contribution, we mean the finite part of the Laurent expansion, which depends on the integration measure chosen. For instance, our virtual integrals are defined with the minimal measure
\begin{equation}
e^{\epsilon \gamma_{\text{E}}} \mu^{2\epsilon}\int \frac{d^{4-2\epsilon} k}{i \pi^{2-\epsilon}} \; .
\end{equation}
The procedure outlined above  -- Laurent expansion plus choice of integration measure -- specifies our contributions, but with one exception: due to the divergent nature of phase-space integrals, one-loop amplitudes within the real-virtual contribution are, in principle, multiplied with inverse powers of $\epsilon$ which, in turn, results in the need to calculate the amplitude to order $\mathcal{O}(\epsilon^2)$, i.e. beyond its finite part. A similar problem occurs also in the double-virtual part, where we have to include the square of the one-loop amplitude. In the original calculation of the total cross-section \cite{Czakon:2013goa}, the contribution proportional to the $\mathcal{O}(\epsilon^2)$ part of the one-loop two-to-two amplitude was shifted from the real-virtual contribution to the double-virtual contribution. In the software used to obtain the results of the present publication we shifted there also the terms proportional to $\mathcal{O}(\epsilon)$, including those contained in the collinear renormalisation. This allowed us to check explicitly that they cancel from the calculation as first demonstrated in Ref.~\cite{Weinzierl:2011uz}.

Let us now specify the details of the setup, which is a straightforward extension of Refs.~\cite{Czakon:2013goa,Czakon:2012pz,Czakon:2012zr,Baernreuther:2012ws}. The two-loop virtual corrections are evaluated as in Refs.~\cite{Czakon:2008zk,Baernreuther:2013caa}, utilising the analytical form for the poles \cite{Ferroglia:2009ii}. We evaluate the one-loop squared amplitude afresh although it has been calculated previously \cite{Anastasiou:2008vd}. The finite part of the one-loop two-to-three amplitude is computed with a code used in the calculation of $pp \to t\t j$ at NLO \cite{Dittmaier:2007wz}. The main problem we face is the ``de-symmetrisation'' of the contributions, since flavour and parity symmetries were used for the calculation of the total cross-section, while they do not apply here. A second issue is the inclusion of collinear renormalisation contributions at the differential level, which were not needed previously \footnote{For total cross-sections, one could simply perform convolutions with analytically known total cross-sections at LO and NLO.}. Due to the use of CDR, finite collinear renormalisation contributions are present even in the case of equal renormalisation and factorisation scales, because both the phase space and the matrix elements have a non-trivial expansion in $\epsilon$. With the complete software we have verified explicitly the numerical cancellation of all poles at the level of distributions. Of course, we also observe complete agreement for the total cross-section computed with the program \textsc{Top++} \cite{Czakon:2011xx}. 

A final check on our setup comes from a comparison of the top-pair transverse momentum distribution with results obtained independently (see Ref.~\cite{Czakon:2014xsa} for details). Indeed, once the top-quark pair has non-vanishing transverse momentum, the cross-section does not exhibit NNLO infrared singularities anymore, but rather only NLO ones. Thus, for non-zero values of the pair transverse momentum it is possible to obtain the $\PTtt$ distribution from a NLO calculation of top-quark pair production in association with an additional jet.

\section{Differential distributions}\label{sec:dif-dist}

\subsection{General comments}\label{sec:diff-general}

In this section we present NNLO predictions for the $t\bar t$ invariant mass $\Mtt$, the transverse momentum $\PTt$ of the top quark and the absolute rapidity $\modyt$ of the top quark, and we compare with existing D\O\ measurements \cite{Abazov:2014vga}. We also present the top-quark differential distribution in $\cos\theta$ (defined in sec.~\ref{sec:costheta} below), together with the related Legendre moments, and compare the NNLO QCD predictions with measurements of the CDF Collaboration~\cite{CDF:2013gna}. 

Our calculation is performed with stable top quarks and, apart from explicit binning, no kinematic cuts are imposed. These parton-level results are then compared to experimental measurements that have been unfolded to the level of top quarks. 

The calculation is performed with fixed (i.e. non-running) scales $\mu_{F,R}=m_t$. Such a scale choice is likely sufficiently appropriate for the limited kinematic range considered by us in this work. The error due to missing higher order effects is estimated from independent variation of the factorisation and renormalisation scales $\mu_R\neq\mu_F\in (m_t/2,2m_t)$, subject to the restriction $0.5\leq \mu_R/\mu_F\leq 2$ \cite{Cacciari:2008zb}, a procedure that has been validated with the NNLO inclusive $t\t$ cross-section \cite{Czakon:2013goa,Czakon:2012pz,Czakon:2012zr,Baernreuther:2012ws}.

Where applicable we present both absolute and normalised differential distributions. The normalised distributions are defined in such a way that their integral is unity for any value of the $\mu_{F,R}$ and for any pdf. Scale variation for all differential distributions, irrespective of their normalisation, is performed separately in each bin. As expected, once normalised, differential distributions exhibit much smaller scale variation. It is worth noting that normalised differential distributions have different sensitivity to the value of $m_t$ compared to the ones with absolute normalisation. This different $m_t$-dependence would be relevant, for example, for extracting $m_t$ from differential distributions.

Throughout the paper we use $m_t=173.3\GeV$ and, unless explicitly noted, we use the MSTW2008 (68\% cl) \cite{Martin:2009iq} pdf set. We always convolute partonic cross-sections with pdf's of matching accuracy (i.e. LO with LO, NLO with NLO, etc). Unless explicitly indicated, no electroweak (EW) corrections are included. 
  
At NLO the pdf error is derived as usual, i.e. using the prescription for computing pdf uncertainty specific to each of the four pdf sets we use in this paper (specified below). Due to the large computational cost at NNLO, however, we do not compute the NNLO pdf error directly but follow a different strategy for its estimation. 

As a first handle on the pdf dependence in NNLO QCD we compare predictions derived with the central members of four different pdf sets: {\tt MSTW2008nnlo68cl}, {\tt CT10nnlo} \cite{Gao:2013xoa}, {\tt NNPDF23\_nnlo\_FFN\_NF5\_as\_0118} \cite{Ball:2012cx} and {\tt HERAPDF15NNLO\_EIG} \cite{CooperSarkar:2011aa}. The results of this comparison can be found in sec.~\ref{sec:pdf}. Second, for the MSTW2008 pdf set only, we derive an approximate pdf error with the help of the following procedure:
\footnote{We are grateful to Juan Rojo for bringing this procedure to our attention.}
denoting by $d\sigma_p$ any differential partonic cross-section at order $p={\rm LO,NLO,NNLO}$ and by $f\!\! f_{p}^{(i)}$ the order-$p$ partonic fluxes constructed from a pdf member $i,~i\geq 0$, we {\it assume} that the ratio:
\begin{equation}
{d\sigma_{\rm NNLO}\otimes f\!\! f_{\rm NNLO}^{(i)}\over d\sigma_{\rm NLO}\otimes f\!\! f_{\rm NNLO}^{(i)}} \approx {\rm independent}~{\rm of}~i, ~ {\rm for}~ {\rm all} ~ i\geq 0\,.
\end{equation}
We only calculate $d\sigma_{\rm NNLO}\otimes f\!\! f_{\rm NNLO}^{(0)}$ and $d\sigma_{\rm NLO}\otimes f\!\! f_{\rm NNLO}^{(i)}$ (the latter is simply an NLO cross-section convoluted with NNLO pdf, whose calculation is inexpensive). Thus, we arrive at the following approximation for an NNLO differential distribution with a pdf member $i\geq 1$:
\begin{equation}
d\sigma_{\rm NNLO}\otimes f\!\! f_{\rm NNLO}^{(i)} = d\sigma_{\rm NLO}\otimes f\!\! f_{\rm NNLO}^{(i)} ~\times~ {d\sigma_{\rm NNLO}\otimes f\!\! f_{\rm NNLO}^{(0)}\over d\sigma_{\rm NLO}\otimes f\!\! f_{\rm NNLO}^{(0)}} \, .
\label{eq:pdfapproximation}
\end{equation}
Eq.~(\ref{eq:pdfapproximation}) above allows us to compute an approximate NNLO prediction for all pdf members and, from there, to derive an approximate pdf error at NNLO following the usual pdf error estimation procedure appropriate for the MST2008 set. {\it A posteriori} such approximate pdf--error--estimating procedure is also justified by the observation that for the Tevatron kinematic ranges considered in this work the scale error is always dominant over the pdf one and thus the precise value of the pdf error is not very important.

The Monte-Carlo (MC) integration error of our results is generally small even at the differential level. For inclusive quantities like the total cross-section and the inclusive asymmetry $\AFB$, the MC error is typically at the permil-level. In all bins for which data is available, the MC error is around 1\% or less, i.e. it is negligible. In some bins where the cross-sections are very small, the MC errors could become sizeable. Clearly, to reduce the MC error in such bins special effort has to be made but this is not really necessary for the goals of the present work. In the following we specify the MC error for each individual distribution.

\subsection{$\Mtt$ distribution}\label{sec:Mtt}

\begin{figure}[t]
\hskip -4mm
\includegraphics[width=0.52\textwidth]{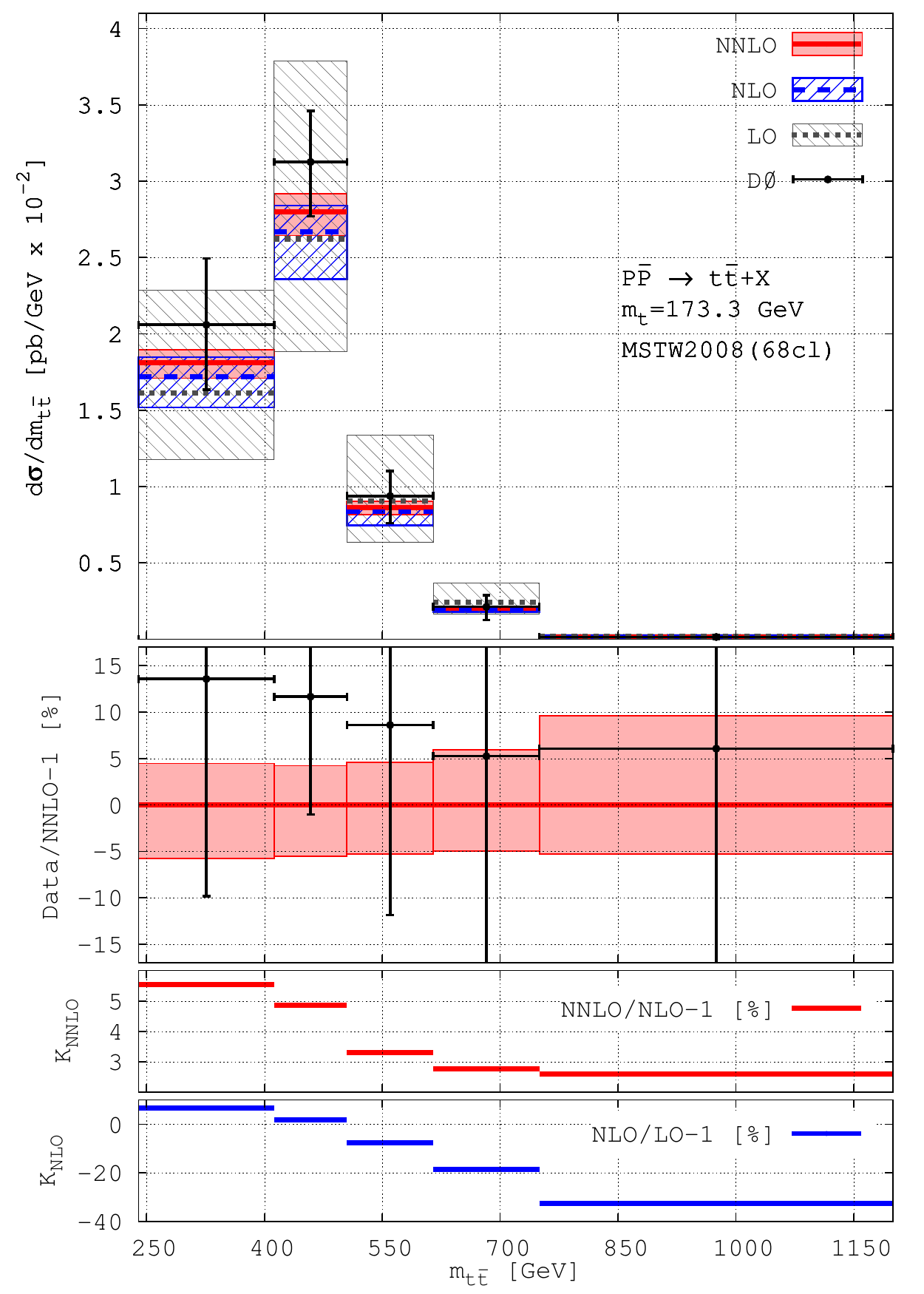}
\includegraphics[width=0.52\textwidth]{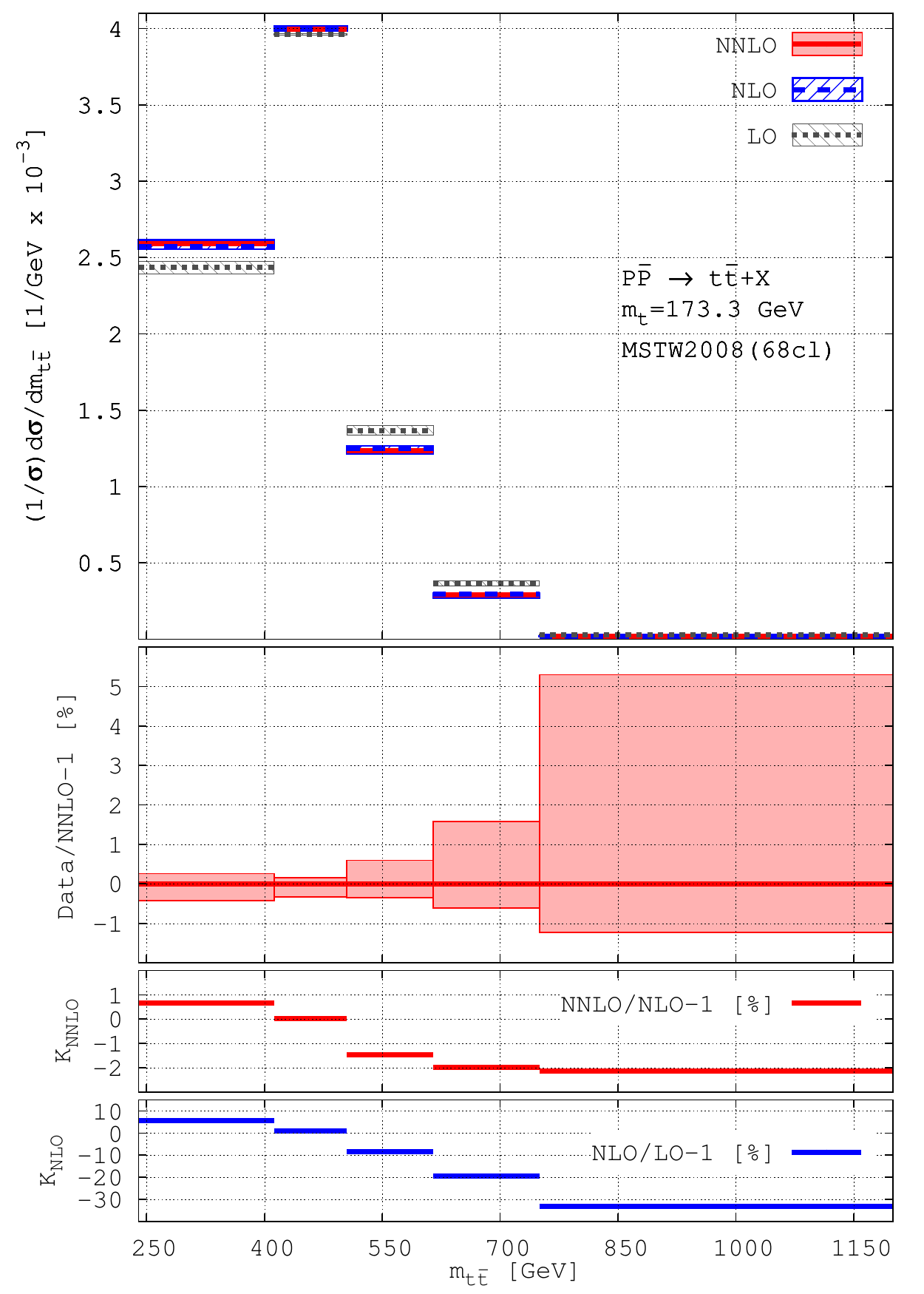}
\caption{\label{fig:Mtt} The $\Mtt$ distribution computed through NNLO in QCD and compared to data from the D\O\ Collaboration~\cite{Abazov:2014vga}. The plot on the left shows the absolute normalisation; the one on the right shows the same distribution but normalised to unity. The plots show the ratio of data to NNLO QCD as well as the NNLO/NLO and NLO/LO K-factors $K_{\rm NNLO}$ and $K_{\rm NLO}$. The error of the theory predictions at NLO and NNLO are from adding scales and pdf in quadrature.}
\end{figure}
In fig.~\ref{fig:Mtt} we show the single-differential distribution $d\sigma/d\Mtt$, where $\Mtt$ is the invariant mass of the $t\t$ pair. The bins correspond to the ones used in the D\O\ analysis~\cite{Abazov:2014vga}: the data is split into five bins of unequal width spanning the interval $240\GeV \leq \Mtt \leq 1200\GeV$. Events with $\Mtt>1200\GeV$ have been collected in a separate overflow bin; these events are not shown in fig.~\ref{fig:Mtt} but their contribution can be found in appendix~\ref{sec:appendix} table~\ref{tab:Mtt}.

In fig.~\ref{fig:Mtt}(left) we present the differential distribution (in absolute normalisation) at LO, NLO and NNLO QCD and compare it with available D\O\ data. Data and NNLO QCD agree in all five bins. The experimental errors are significantly larger than the theory ones. To facilitate possible future more precise measurements, as well as studies of the sensitivity of differential distributions with respect to $m_t$, we present in fig.~\ref{fig:Mtt}(right) also the corresponding normalised theoretical prediction in LO, NLO and NNLO QCD. 

To better clarify the size of higher order radiative corrections we also show the K-factors $K_{\rm NNLO}$ and $K_{\rm NLO}$ defined, respectively, as the ratios NNLO/NLO and NLO/LO. From the plot of the normalised $\Mtt$ distribution we conclude that both the NLO and NNLO K-factors have similar behaviours as functions of $\Mtt$: higher-order effects tend to increase the spectrum close to absolute threshold $\Mtt\sim2m_t$ and decrease it past the peak of the distribution. The addition of the NNLO correction has an important stabilising effect on the predicted spectrum: not only the scale error decreases significantly but the size of the K-factor decreases by a factor of ten. This is a very welcoming feature of the NNLO result and it suggests much improved theoretical control over the shape of this distribution at large $\Mtt$. Similar observation has been made for the LHC in Ref.~\cite{Czakon:2015owf}.

The relative MC integration error is estimated to be below 1\% for all bins shown in fig.~\ref{fig:Mtt}. The relative MC error for the overflow bin $\Mtt\geq 1200\GeV$ (shown in appendix~\ref{sec:appendix} table~\ref{tab:Mtt}) is estimated to be about 3-4\%.

\subsection{$p_T$ distribution of the top quark}\label{sec:PTt}

\begin{figure}[t]
\hskip -4mm
\includegraphics[width=0.52\textwidth]{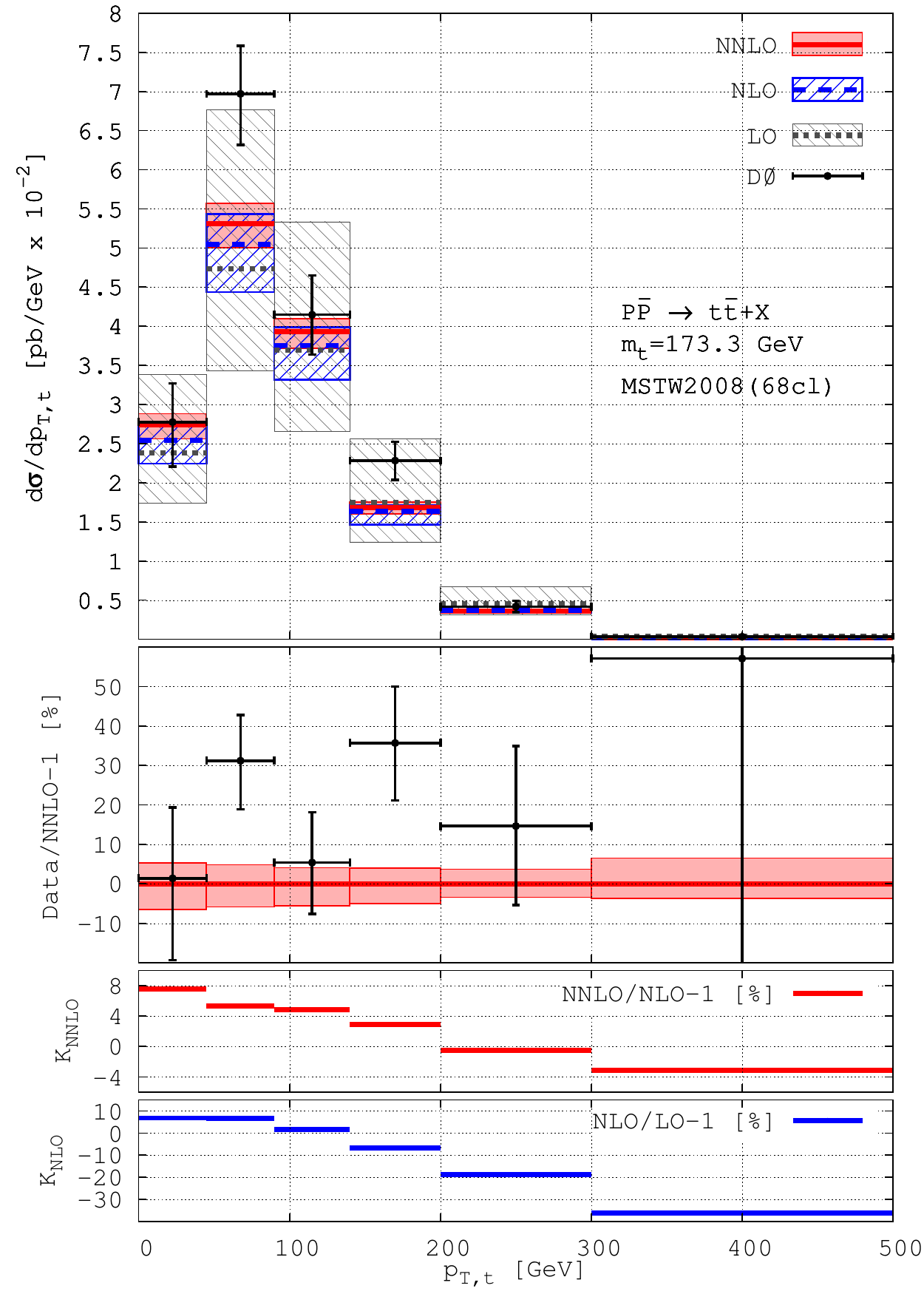}
\includegraphics[width=0.52\textwidth]{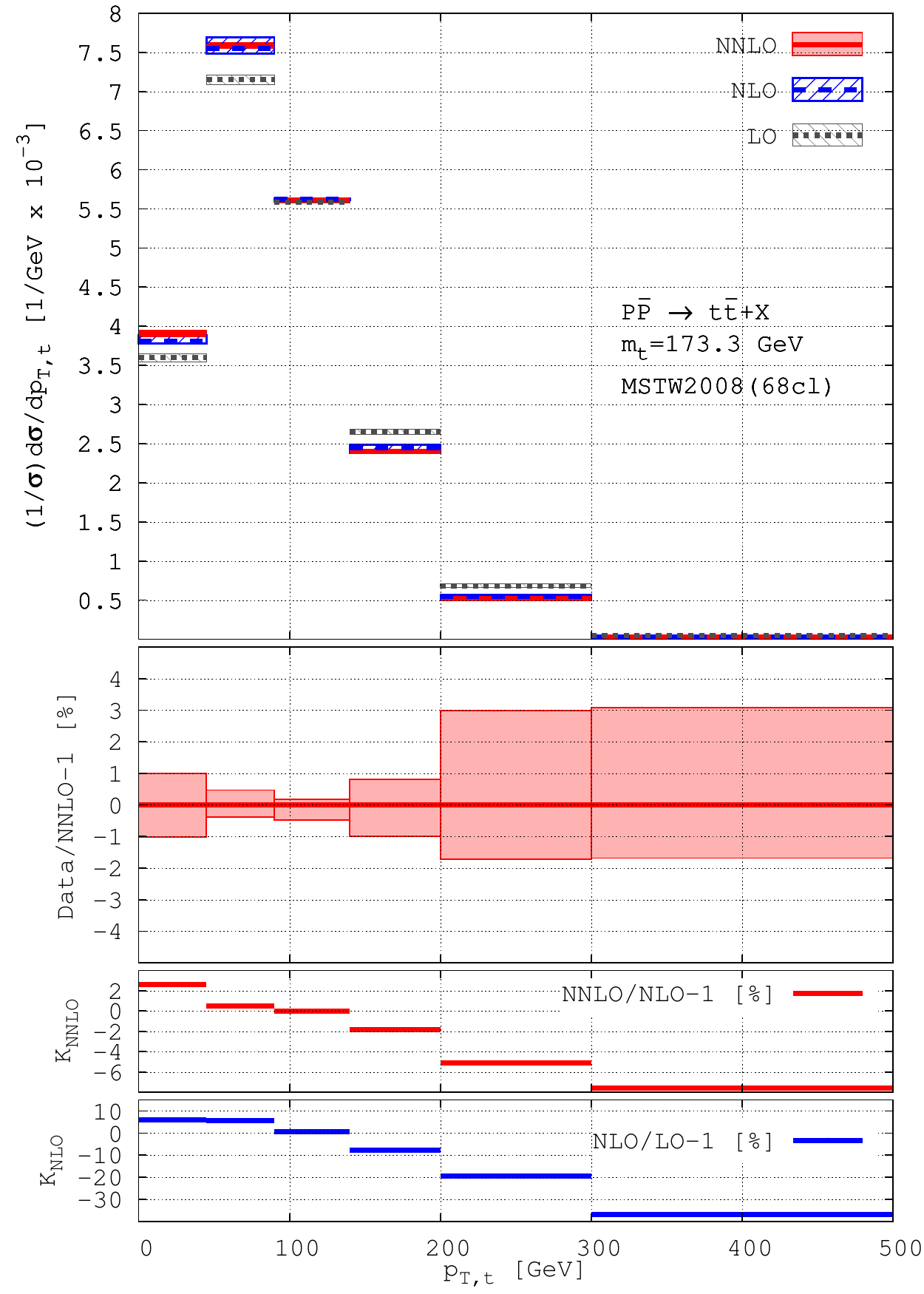}
\caption{\label{fig:PTt} The $\PTt$ spectrum computed through NNLO in QCD and compared to data from the D\O\ Collaboration~\cite{Abazov:2014vga}. The plot on the left shows its absolute normalisation, while the one on the right the same distribution but normalised to unity. The plots  also show the ratio of data to NNLO QCD as well as the NNLO/NLO and NLO/LO K-factors $K_{\rm NNLO}$ and $K_{\rm NLO}$. The error of the theory predictions at NLO and NNLO is derived by adding in quadrature errors from scales and pdf.}
\end{figure}
In fig.~\ref{fig:PTt} we show the single inclusive $p_T$ spectrum of the top quark in absolute normalisation (left) and normalised to unity (right). The bins correspond to the ones used in the D\O\ analysis~\cite{Abazov:2014vga}: the data is split in six unequal-size bins spanning the interval $(0,500)\GeV$. Computed events with $p_T>500\GeV$ have been collected in a separate overflow bin; they are not shown in fig.~\ref{fig:PTt}; their contribution can be found in appendix~\ref{sec:appendix} table~\ref{tab:PTt}.

The D\O\ data is for the $p_T$ of average top/antitop while our calculations are for the $p_{T,t}$ (or $p_{T,\bar t}$). We have checked that the $p_{T,t}$ and $p_{T, \bar t}$ spectra agree within the MC errors.

The relative MC integration error is estimated to be below 1\% for all bins with $\PTt\leq 300\GeV$. The highest bin in fig.~\ref{fig:PTt}, $300\leq\PTt\leq 500\GeV$, has MC error that approaches 2\%, while the MC error for the overflow bin $\PTt\geq 500\GeV$  (shown in appendix~\ref{sec:appendix} table~\ref{tab:PTt}) is around 5\%. 

In fig.~\ref{fig:PTt}(left) we present the differential distribution, in absolute normalisation, at LO, NLO and NNLO QCD and compare it with available D\O\ data. Data and NNLO QCD agree in four of the six bins, while in two of the bins data exceeds theory by, roughly, $2\sigma$. As for the $\Mtt$ distribution, the experimental errors are significantly larger than the theory ones. A dedicated comparison of the normalised $\PTt$ distribution with possible future measurements might be helpful in revealing the interplay between differential distributions and $m_t$.

To better clarify the importance of higher-order radiative corrections we also show the K-factors $K_{\rm NNLO}$ and $K_{\rm NLO}$. From the plot of the normalised $\PTt$ distribution we conclude that, as for the $\Mtt$ distribution, the NLO and NNLO K-factors have similar behaviour as functions of $\PTt$: higher order effects tend to increase the spectrum for small $\PTt$ and decrease it past the peak of the distribution. The NNLO correction again has sizeable stabilising effect on the predicted spectrum: not only the scale error decreases significantly but the size of the K-factor decreases by a factor of five, or more.

\subsection{Rapidity distribution of the top quark}\label{sec:mody}

\begin{figure}[h]
\hskip -4mm
\includegraphics[width=0.52\textwidth]{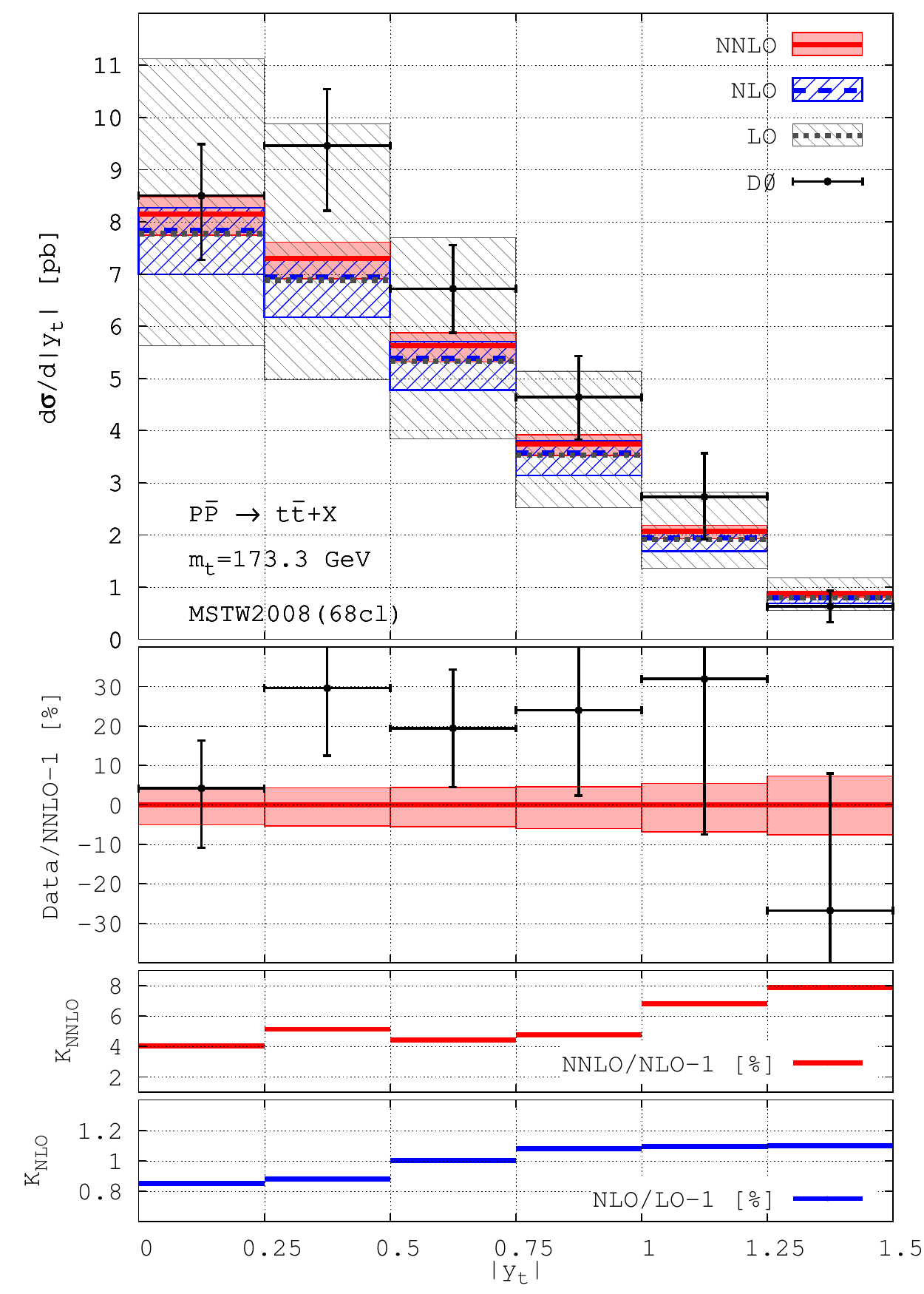}
\includegraphics[width=0.52\textwidth]{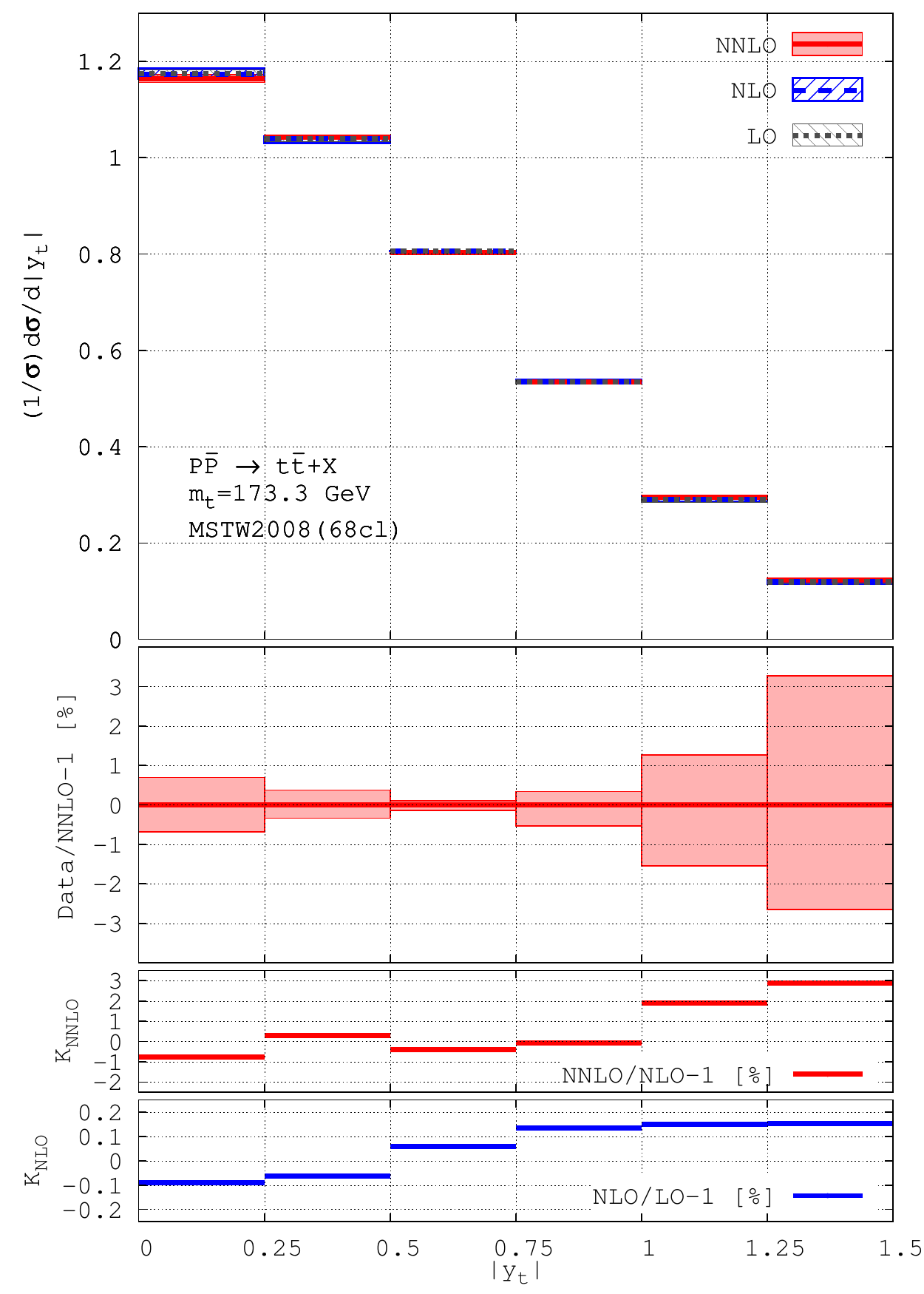}
\caption{\label{fig:mody} The $\modyt$ distribution computed through NNLO in QCD and compared to data from the D\O\ Collaboration~\cite{Abazov:2014vga}. The plot on the left shows its absolute normalisation, while the one on the right shows the same distribution but normalised to unity. The plots  also show the ratio of data to NNLO QCD as well as the NNLO/NLO and NLO/LO K-factors $K_{\rm NNLO}$ and $K_{\rm NLO}$. The error of the theory predictions at NLO and NNLO is derived by adding in quadrature errors from scales and pdf.}
\end{figure}
In fig.~\ref{fig:mody} we show the absolute rapidity $\modyt$ distribution of the top quark. The bins correspond to the ones used in the D\O\ analysis~\cite{Abazov:2014vga}: the data is split in six equal-width bins spanning the interval $0\leq\modyt\leq1.5$. Computed events with $\modyt>1.5$ have been collected in a separate overflow bin; they are not shown in fig.~\ref{fig:mody}; their contribution can be found in appendix~\ref{sec:appendix} table~\ref{tab:mody}.

The D\O\ data is for the average top/antitop $\mody$ while our calculations are for the top quark's $\modyt$. We have checked that the top and antitop $\mody$ distributions agree within the MC error. The relative MC integration error is estimated to be within 1\% for all bins on fig.~\ref{fig:mody} as well as for the overflow bin $\modyt>1.5$.

In fig.~\ref{fig:mody}(left) we present the differential distribution (in absolute normalisation) at LO, NLO and NNLO QCD and compare it with available D\O\ data. Data and NNLO QCD marginally agree in five of the six bins, while in one of the bins data exceeds theory by less than $2\sigma$. As for the $\Mtt$ and $\PTt$ distributions, the experimental error of the $\modyt$ distribution is significantly larger than the theory one. 

To better clarify the size of higher order radiative corrections we also show the K-factors $K_{\rm NNLO}$ and $K_{\rm NLO}$. From the plot of the normalised $\modyt$ distribution we observe that both the NLO and NNLO K-factors tend to increase, almost linearly, with $\modyt$. Unlike the $\Mtt$ and $\PTt$ distributions, however, the NNLO K-factor increases in size with respect to $K_{\rm NLO}$ by a factor of, roughly, five. This observation demonstrates the particular significance of the NNLO corrections in this observable. One should also note that the error of the NNLO correction is smaller than the NLO one by about a factor of two and the NNLO result is consistent with the NLO error band in all bins, which confirms that perturbative convergence is firmly present in this observable.

\subsection{Top-quark $\cos\theta$ distribution and Legendre moments}\label{sec:costheta}

\begin{figure}[h]
\hskip -4mm
\includegraphics[width=0.52\textwidth]{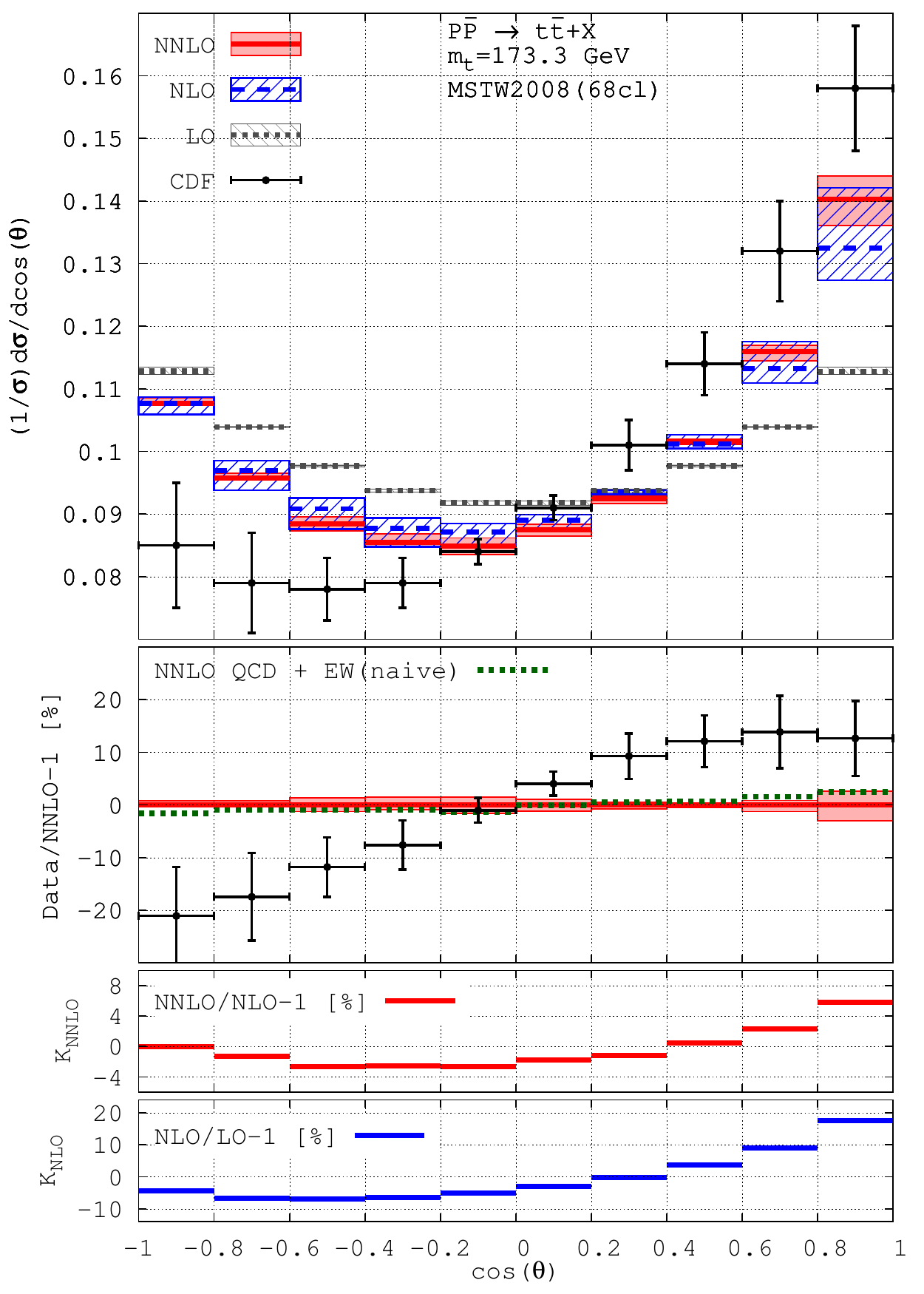}
\includegraphics[width=0.52\textwidth]{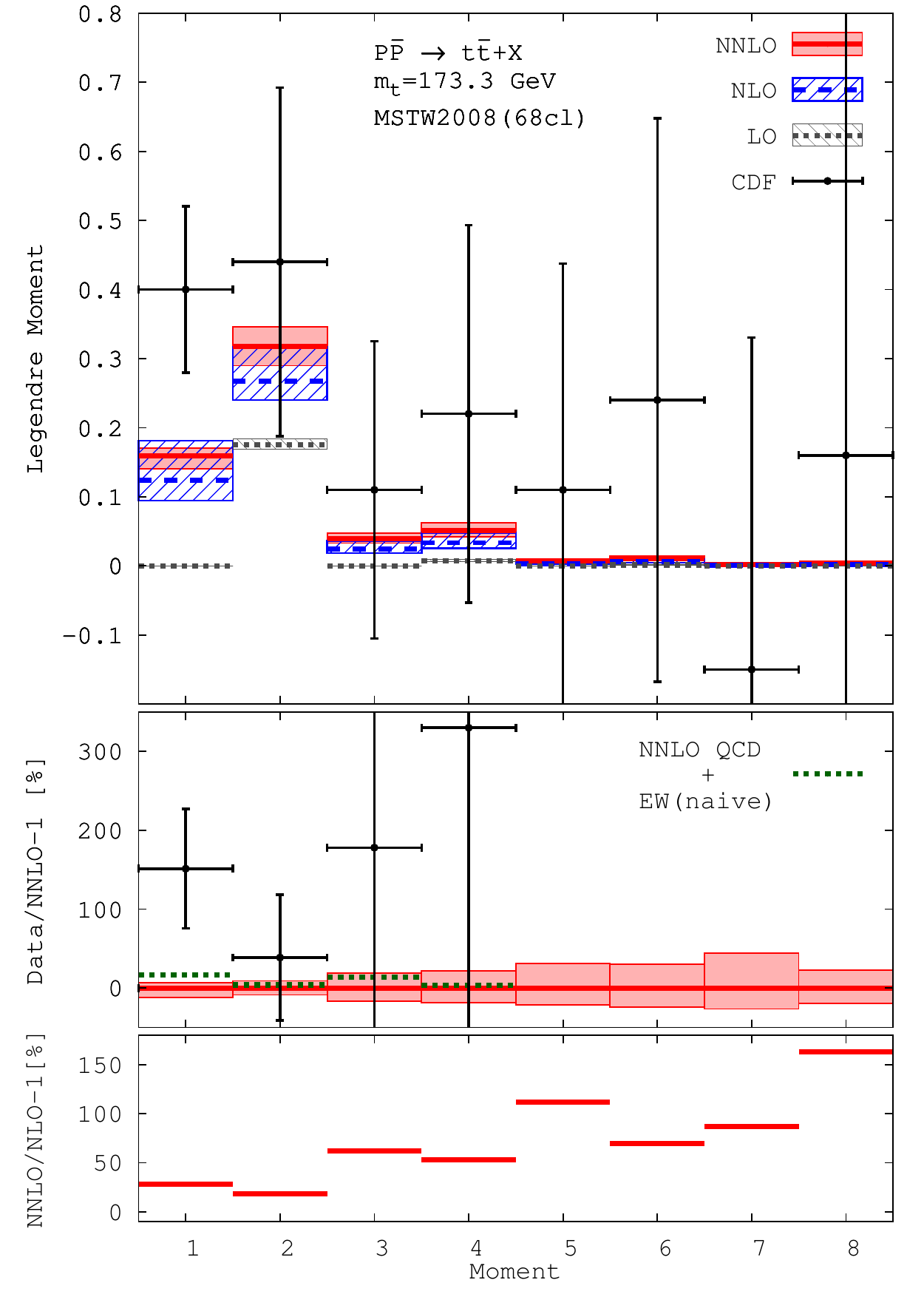}
\caption{\label{fig:costheta} The normalised top-quark $\cos\theta$ distribution (left) and related Legendre moments (right) through NNLO QCD compared to data from Ref.~\cite{CDF:2013gna}. Also shown is a naive estimate of the EW corrections (see text) as well as the K-factors $K_{\rm NNLO}$ and $K_{\rm NLO}$. The error of the theory predictions is based on scale variation only.}
\end{figure}

Next we present the NNLO QCD corrections to the differential distribution $(1/\sigma)d\sigma/d\cos\theta$, where $\theta$ is the angle between the top quark and the incoming proton in the $t\t$ rest frame. This angular distribution was measured by the CDF Collaboration~\cite{CDF:2013gna}. The data for the normalised distribution $(1/\sigma)d\sigma/d\cos\theta$ is available from \cite{CDF:public}; it is split in ten equal-width bins that span the full interval $-1\leq\cos\theta\leq 1$. The normalisation chosen in Ref.~\cite{CDF:2013gna} is such that the sum of the values of all bins equals unity, i.e. in effect the values in each bin correspond to the integral of the cross-section over that bin. The theory prediction, through NNLO QCD, is compared with data in fig.~\ref{fig:costheta}(left), see also appendix~\ref{sec:appendix} table~\ref{tab:costheta}. The effect of the NNLO correction is generally towards decreasing both the discrepancy with data and the scale dependence of the NLO prediction.

In fig.~\ref{fig:costheta}(left) we show the ratio Data/NNLO which is helpful in visualising the significance of the difference between the NNLO QCD prediction and data. We observe that the deviation between the two is never more than approximately $2\sigma$, while for most of the bins it is less than that or there is agreement between the two. Given the role this distribution plays in the analysis of $\AFB$, and the important role of the EW corrections to $\AFB$, it is interesting to estimate the effect of the EW corrections when added to the NNLO QCD ones. We do not have EW corrections computed in a form that is readily combinable with our QCD calculation. Therefore, as a rough estimate of the EW corrections, we take the difference between the known NLO+EW result \cite{Bernreuther:2012sx} and our NLO calculation. We attribute the difference to pure non-QCD corrections and add them to the NNLO QCD ones:
{\center NNLO QCD + EW(naive) = NNLO QCD + NLO(QCD+EW\cite{Bernreuther:2012sx}) - NLO QCD (this work).}
\newline

The NLO (QCD+EW) result has been taken from Ref.~\cite{CDF:public} which, in turn, has been provided by the authors of Ref.~\cite{Bernreuther:2012sx}. While the setups for the calculation of the $\cos\theta$ distribution (as well as the related Legendre moments, see below) in this work and in Ref.~\cite{Bernreuther:2012sx} differ in several aspects, we have cross-checked the pure NLO QCD results with Ref.~\cite{Bernreuther:2012sx}
\footnote{We wish to thank Werner Bernreuther for his help with this comparison.}
and found that they are in reasonable agreement (for the Legendre moments the agreement is only good for the first four moments due to differences in the way the moments are computed). 

The corresponding ratio (NNLO QCD+EW(naive))/NNLO is shown in fig.~\ref{fig:costheta}(left). While our estimate of the EW corrections is imperfect it should be sufficient to get an idea of the size of the EW corrections. From that figure we conclude that the EW corrections are comparable to the size of the error of the NNLO QCD corrections and are thus not negligible. Furthermore, they tend to decrease the difference between NNLO QCD and data. Still, the EW corrections are not very large and thus their inclusion, or not, is not significantly affecting the comparison of SM theory with data (especially given the sizeable error of the available data).

In fig.~\ref{fig:costheta}(left) we also show the NLO and NNLO K-factors. Similarly to the $\Mtt$ and $\PTt$ distributions discussed above, we find that the K-factors $K_{\rm NNLO}$ and $K_{\rm NLO}$ have similar shapes, while the NNLO K-factor has significantly smaller size compared to the NLO one. 

The CDF collaboration has also presented \cite{CDF:2013gna} the results for the first eight Legendre moments of the $\cos\theta$ distribution. The relation between moments and the distribution reads
\begin{equation}
{d\sigma\over d\cos\theta} = \sum_{\ell = 0}^\infty a_\ell P_\ell(\cos\theta)\, ,~~{\rm where}~~~ a_\ell = {2\ell+1\over 2}\int_{-1}^1P_\ell(\cos\theta) {d\sigma\over d\cos\theta} d\cos\theta\, ,
\label{eq:Leg-a}
\end{equation}
and $P_\ell(z)$ are the usual Legendre polynomials. 

The Legendre moment analysis of the angular distribution is well-suited for discussing the top-quark $\AFB$, see also sec.~\ref{sec:afb-legmom} below. The normalisation (both for data and our calculation) is such that the zeroth moment is $a_0=1$ (i.e. the moments correspond to the normalised distribution $(1/\sigma)d\sigma/d\cos\theta$ and are obtained from the ones defined in eq.~(\ref{eq:Leg-a}) by dividing by $\sigma/2$).
\footnote{We wish to thank Jon Wilson for helpful clarifications regarding Ref.~\cite{CDF:2013gna}.}

The corresponding moments are shown in fig.~\ref{fig:costheta}(right), see also appendix~\ref{sec:appendix} table~\ref{tab:legmom}. Similarly to the $\cos\theta$ distribution we also present a naive estimate of the EW corrections for the first four moments only. Given the rapidly increasing size of the errors in the higher moments we feel that restricting the comparison to the first four moments $a_1,\dots,a_4$ is justified. We observe that the NNLO QCD correction is sizeable, and becomes especially large for the higher moments, as can be seen from the NNLO K-factor. Still it is within the error band of the NLO correction and thus consistent with error estimates based on scale variation. We observe that the EW correction is particularly relevant for the first moment $a_1$ where it exceeds the size of the error estimate of the NNLO QCD. This finding is in line with the well-recognized importance of EW corrections for the inclusive $\AFB$.

The MC error in each bin of the $\cos\theta$ distribution (with absolute normalisation) is around few permil in each bin. The MC error of the Legendre moments grows rapidly for higher moments: for $a_{1,2,3}$ it is below 5 permil; for $a_4$ it is around 2\%, while for $a_8$ it exceeds 30\%. 

Finally, we would like to mention that our calculation of the Legendre moments is not based on summing over the bins in fig.~\ref{fig:costheta}(left) but, in order to avoid bin-size effects, the moments are computed by summing the contribution from each partonic event, similarly to the way all distributions are computed.

\subsection{Comparison with approximate NNLO/resummed NLO QCD results}\label{sec:approxNNLO}

\begin{figure}[h]
\hskip -4mm
\includegraphics[width=1.05\textwidth]{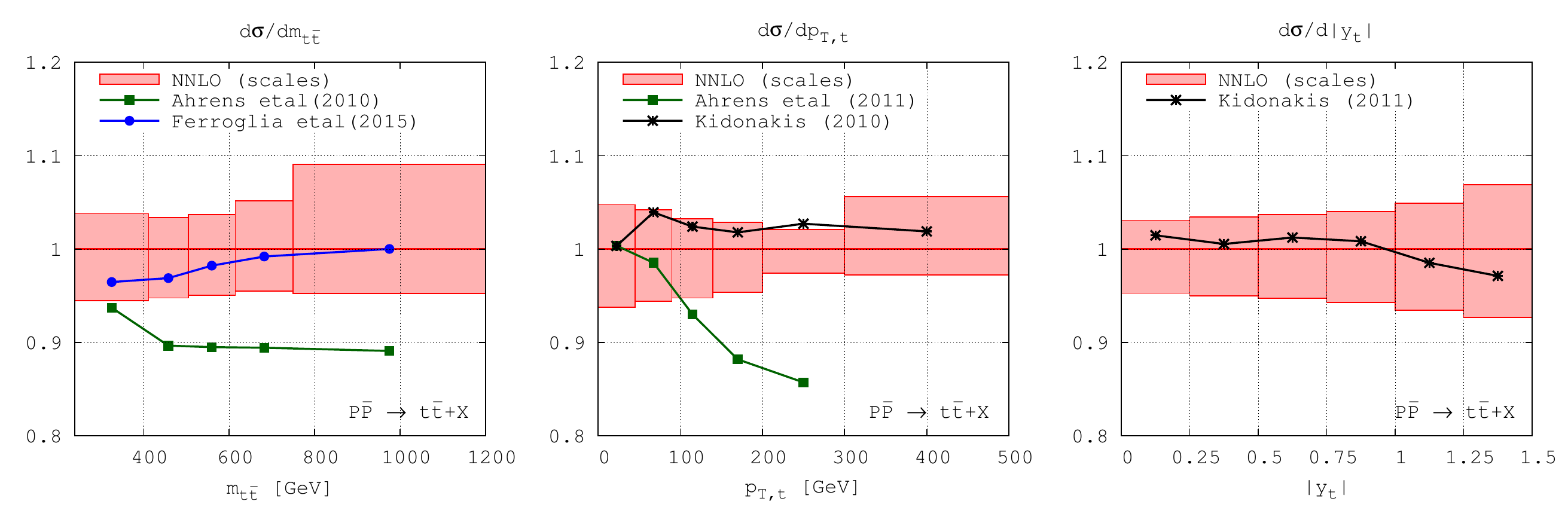}
\caption{\label{fig:approxNNLO} Ratios of various approximate NNLO/resummed NLO predictions and the exact NNLO QCD results for three differential distributions.}
\end{figure}

Until now, differential Tevatron top-quark measurements have been compared to theoretical predictions derived in either approximate NNLO or soft-gluon-resummed NLO. Such predictions are fixed-order NLO accurate and include partial NNLO contributions originating from the expansion of soft-gluon-resummed predictions (or possibly all-order towers if the results are resummed). It will be instructive to compare the presently-derived fully differential exact NNLO results with such ``NLO+" predictions. To that end in fig.~\ref{fig:approxNNLO} we show the ratio of various approximate NNLO/resummed NLO predictions to the exact NNLO QCD result.

For the $\Mtt$ distribution we compare with Ref.~\cite{Ahrens:2010zv} as well as the more recent work \cite{Ferroglia:2013zwa,Ferroglia:2015ivv}. For the purpose of this comparison the calculation of Refs.~\cite{Ferroglia:2013zwa,Ferroglia:2015ivv} has been performed with the same parameters as the NNLO calculation ($m_t$, pdf set and binning)
\footnote{We are grateful to Li Lin Yang for sending us the numbers from Refs.~\cite{Ferroglia:2013zwa,Ferroglia:2015ivv}}.
No corrections beyond NNLO are included in the prediction of Ref.~\cite{Ferroglia:2013zwa,Ferroglia:2015ivv}. The $\Mtt$ prediction from Ref.~\cite{Ahrens:2010zv} shown in fig.~\ref{fig:approxNNLO} is adapted from fig.~9 in Ref.~\cite{Abazov:2014vga}: the results from \cite{Ahrens:2010zv} are NLO+NNLL (next-to-next-to-leading log); resummation is performed in momentum space with a default scale choice $\mu = \Mtt$\;; however, unlike fig.~9 in \cite{Abazov:2014vga}, they are shown here with their normalisation not rescaled to exact NNLO. In fig.~\ref{fig:approxNNLO} we notice that the approximate NNLO result of Ref.~\cite{Ferroglia:2013zwa,Ferroglia:2015ivv} agrees with the exact NNLO one within the scale error of the NNLO result.

For the top-quark $\PTt$ distribution we compare with predictions from Refs.~\cite{Ahrens:2011mw,Kidonakis:2010dk}. The values shown here differ from fig.~11 of Ref.~\cite{Abazov:2014vga}; they correspond to the original calculations \cite{Ahrens:2011mw,Kidonakis:2010dk} and have been provided to us by the authors of Refs.~\cite{Ahrens:2011mw,Kidonakis:2010dk} for the purpose of this comparison. The result of Ref.~\cite{Ahrens:2011mw} is NLO+NNLL in momentum space, uses default scale $\mu = 2m_t$ and is shown here with its normalisation not rescaled to exact NNLO. The result of Ref.~\cite{Kidonakis:2010dk} is for $m_t=173\GeV$. In fig.~\ref{fig:approxNNLO} we notice that the approximate NNLO result of Ref.~\cite{Kidonakis:2010dk} agrees with the exact NNLO one within the scale error of the NNLO result.

Finally, we compare the $\modyt$ distribution with prediction from Ref.~\cite{Kidonakis:2011zn} (computed with $m_t=173\GeV$; the values shown are adapted from fig.~10 of Ref.~\cite{Abazov:2014vga}). As can be concluded from fig.~\ref{fig:approxNNLO}, the approximate result is consistent with the exact one within the scale error of the NNLO result. 

In conclusion, in order to fully document our presentation and avoid possible miscommunications
\footnote{Unfortunately, such a step is required since these numbers are not explicitly available in any publication. We thank Andreas Jung as well as the authors of references \cite{Ahrens:2010zv,Ferroglia:2013zwa,Ferroglia:2015ivv,Ahrens:2011mw,Kidonakis:2010dk,Kidonakis:2011zn} for providing us with their numbers and for cross-checking them.},
we next specify the bin values for all approximate NNLO/resummed NLO differential distributions shown in fig.~\ref{fig:approxNNLO}:
\begin{itemize}
\item $\Mtt$ (in units of $[10^{-2}{\rm pb}/\GeV]$):\\ 
Ref~\cite{Ahrens:2010zv}:~$(1.7,2.51,0.773,0.181,0.0126)$;\\ 
Refs.~\cite{Ferroglia:2013zwa,Ferroglia:2015ivv}: ~$(1.75,2.712,0.8483,0.2008,0.01414)$.
\item $\PTt$ (in units of $[10^{-2}{\rm pb}/\GeV]$):\\ 
Ref~\cite{Ahrens:2011mw}:~$(2.748,5.235,3.66,1.485,0.3125)$;\\ 
Ref.~\cite{Kidonakis:2010dk}: ~$(2.747,5.522,4.029,1.714,0.3745,0.02075)$.
\item $\modyt$ (in units of $[{\rm pb}]$):\\
Ref.~\cite{Kidonakis:2011zn}: ~$(8.276,7.34,5.697,3.773,2.039,0.8356)$.
\end{itemize}

\section{Top-quark $\AFB$ and related differential distributions}\label{sec:AFB}

\subsection{General Comments}\label{sec:AFB-general}

In this section we extend the study of Ref.~\cite{Czakon:2014xsa} and present detailed results for both the differential asymmetry and corresponding (doubly-) differential distributions in the $t\bar t$ rapidity difference $\dy \equiv y_t - y_\t$, in $\dy$ and $\Mtt$ and in $\dy$ and $\PTtt$. As in Ref.~\cite{Czakon:2014xsa} we define the differential asymmetry as the ratio of a numerator and denominator each computed through the NNLO QCD corrections of order ${\cal O}(\as^4)$. Following \cite{Aaltonen:2012it}, we define the differential asymmetry as
\begin{equation}
{\AFB} = {\sigma^{+}_{\rm bin} - \sigma^{-}_{\rm bin} \over \sigma^{+}_{\rm bin} + \sigma^{-}_{\rm bin} }
~~~,~~~
\sigma^\pm_{\mathrm{bin}} = \int \theta(\pm \Delta y) \theta_{\mathrm{bin}} \mathrm{d}\sigma\, .
\label{eq:AFB-diff}
\end{equation}
The binning function $\theta_{\mathrm{bin}}$ takes values zero or unity such that it restricts the kinematics of the $t\bar t$ pair to the corresponding bins (defined in the following). Setting $\theta_{\mathrm{bin}}=1$ in eq.~(\ref{eq:AFB-diff}) yields the inclusive asymmetry $\AFB$. 

An alternative definition for the inclusive $\AFB$ was considered in Ref.~\cite{Czakon:2014xsa} (such that the numerator/denominator ratio is expanded in powers of $\as$). Since in this work we do not show any new result for the inclusive $\AFB$ we do not need to introduce this definition here.

Unlike Ref.~\cite{Czakon:2014xsa}, in this work we include the pdf error (derived as described in sec.~\ref{sec:dif-dist}). As anticipated in Ref.~\cite{Czakon:2014xsa}, the $\AFB$ pdf error is negligible when compared to the scale error.

\subsection{$\dy$ differential distribution and asymmetry}\label{sec:afb-y}

\begin{figure}[h]
\hskip -4mm
\includegraphics[width=0.52\textwidth]{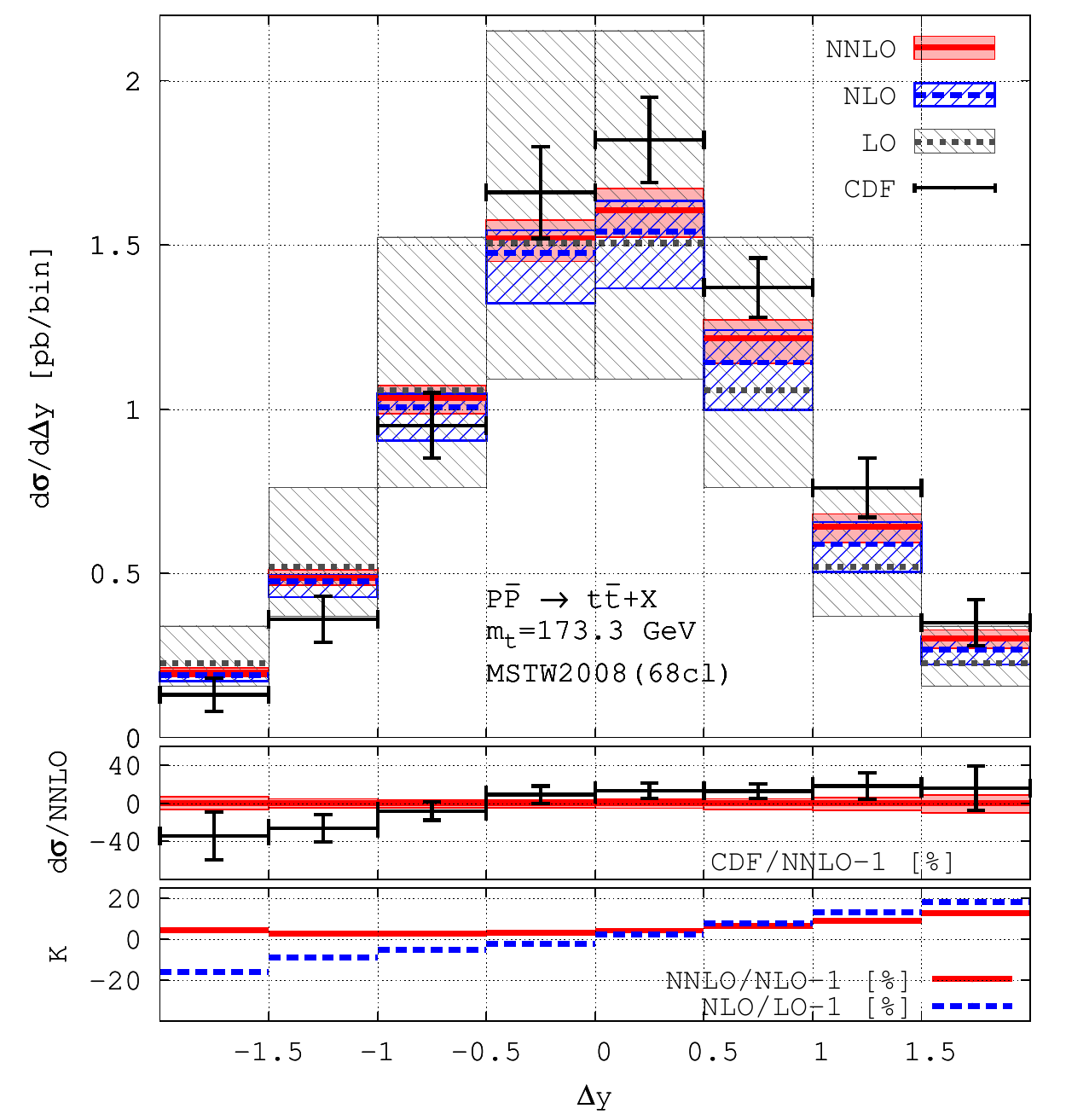}
\includegraphics[width=0.52\textwidth]{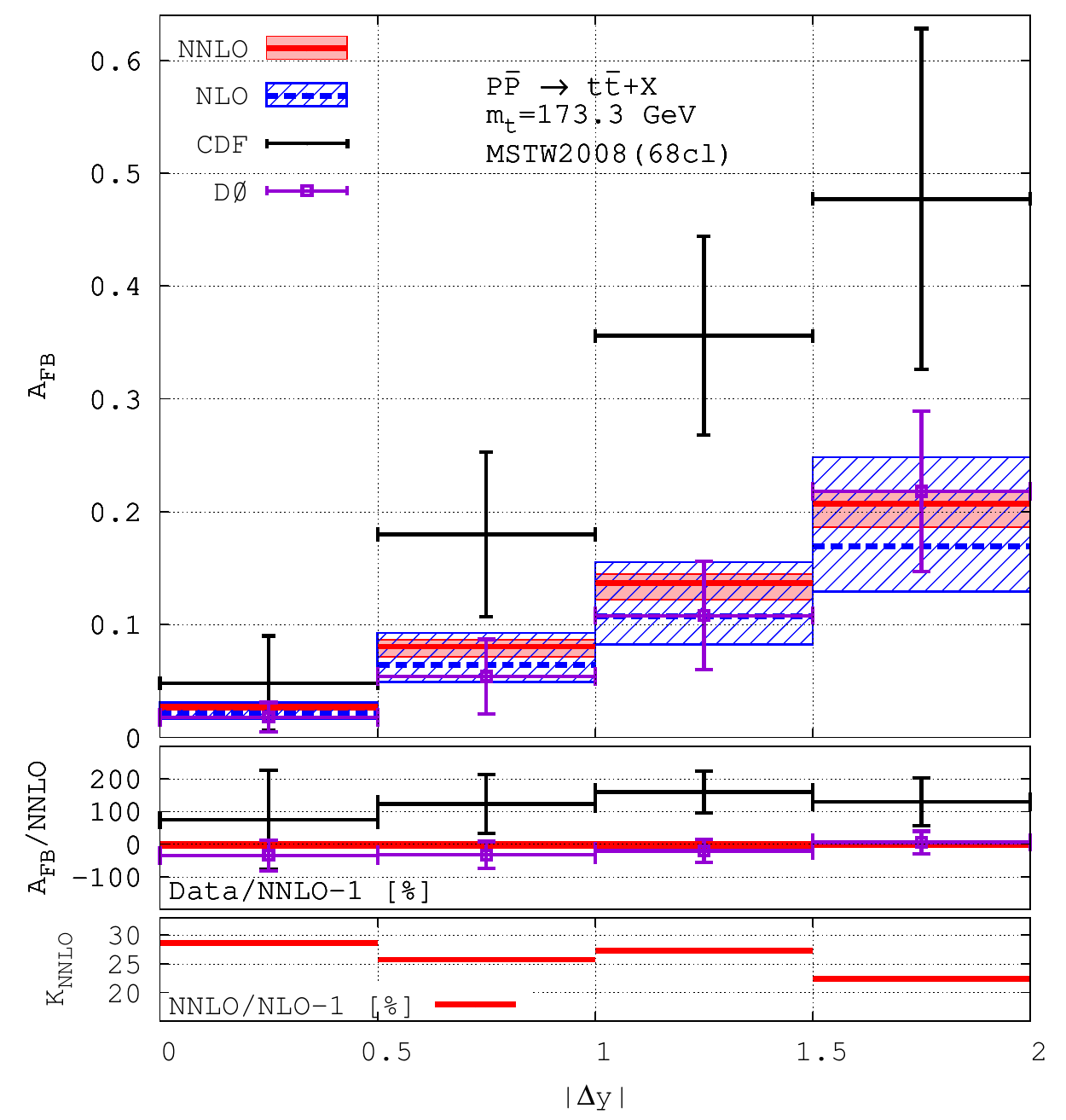}
\caption{\label{fig:afb-y} The differential distribution $d\sigma/d\dy$ (left) and related differential asymmetry $\AFB(\dmody)$ (right). Comparison includes SM theory through NNLO QCD and CDF and D\O\ data. The end-bins contain overflow events. The error of the theory prediction is derived from scale and pdf variation.}
\end{figure}

In fig.~\ref{fig:afb-y} we show the $\dmody$ dependence of $\AFB$ (right; see also appendix~\ref{sec:appendix} table~\ref{tab:afb-y}) and the corresponding differential distribution $d\sigma/d\dy$ (left; see also appendix~\ref{sec:appendix} table~\ref{tab:diff-y}). We use the same bins as Ref.~\cite{Czakon:2014xsa} which, in turn, match the CDF bins in Ref.~\cite{Aaltonen:2012it}. 

The differential asymmetry is divided into four equal-width bins. The bin with highest $\dmody$ contains overflow events. The theoretical prediction through NNLO QCD is shown in fig.~\ref{fig:afb-y}(right), see also appendix~\ref{sec:appendix} table~\ref{tab:afb-y}, and compared with data from CDF \cite{Aaltonen:2012it} and D\O\ \cite{Abazov:2014cca,D0:public} collaborations. We also plot the data normalised to the central NNLO QCD prediction as well as the NNLO K-factor (the NLO K-factor is not defined for $\AFB$ since the LO result is zero).

We notice that the K-factor is nearly constant with $\dmody$ and, at around 25\%, is rather sizeable. Looking at the estimated errors, we notice that the NNLO result has significantly smaller errors than the NLO one (by about a factor of three) and moreover the NNLO error band is fully contained within the NLO one. This feature demonstrates that this observable possesses good perturbative convergence.

The comparison between NNLO theory predictions for $\AFB$ and data has already been discussed in Ref.~\cite{Czakon:2014xsa}. Its main feature is the agreement of NNLO QCD with the D\O\ data while the CDF measurement is higher than theory. The significance of this discrepancy is between $1 \sigma$ and $2 \sigma$. With the exception of the bin with highest $\dmody$, the significance of the discrepancy seems to be growing with $\dmody$.

A compact way of presenting the $\dmody$ dependence of $\AFB$ is through its slope \cite{Aaltonen:2012it}. The least-squares linear fit to the NNLO QCD prediction, assuming zero intercept, reads:
\begin{equation}
\AFB(\dmody) = \alpha_y\dmody, ~~{\rm where} ~~ \alpha_y = 0.114^{+0.006}_{-0.012}\,.
\label{eq:beta_y}
\end{equation}
The error in eq.~(\ref{eq:beta_y}) includes both scale and pdf variation, although the contribution from the pdf error is marginal (its omission would only change the ``+"-error in (\ref{eq:beta_y}) from 0.006 to 0.005). The slope in NLO QCD reads $\alpha_y^{\rm NLO} = 0.092^{+0.042}_{-0.022}$.

\begin{figure}[h]
\hskip -4mm
\includegraphics[width=0.52\textwidth]{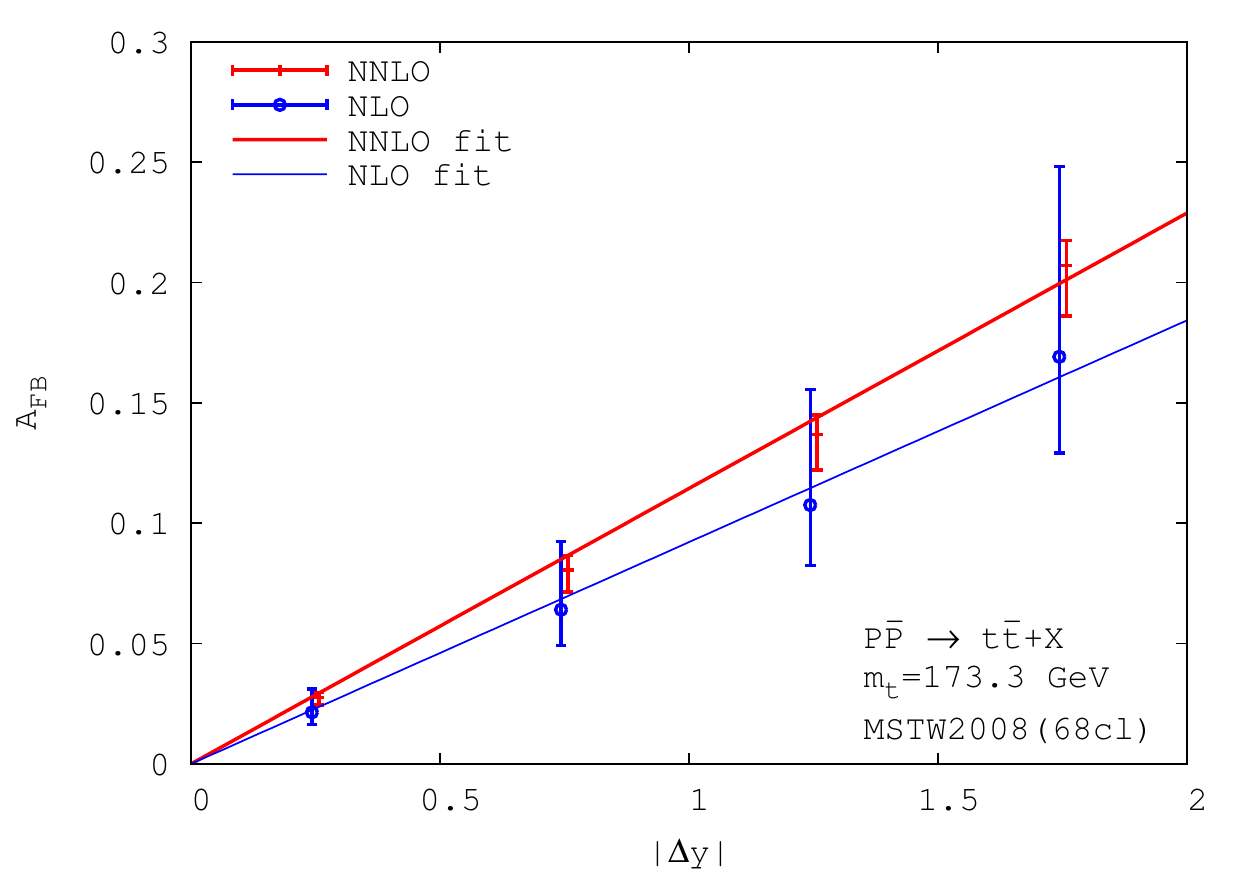}
\includegraphics[width=0.52\textwidth]{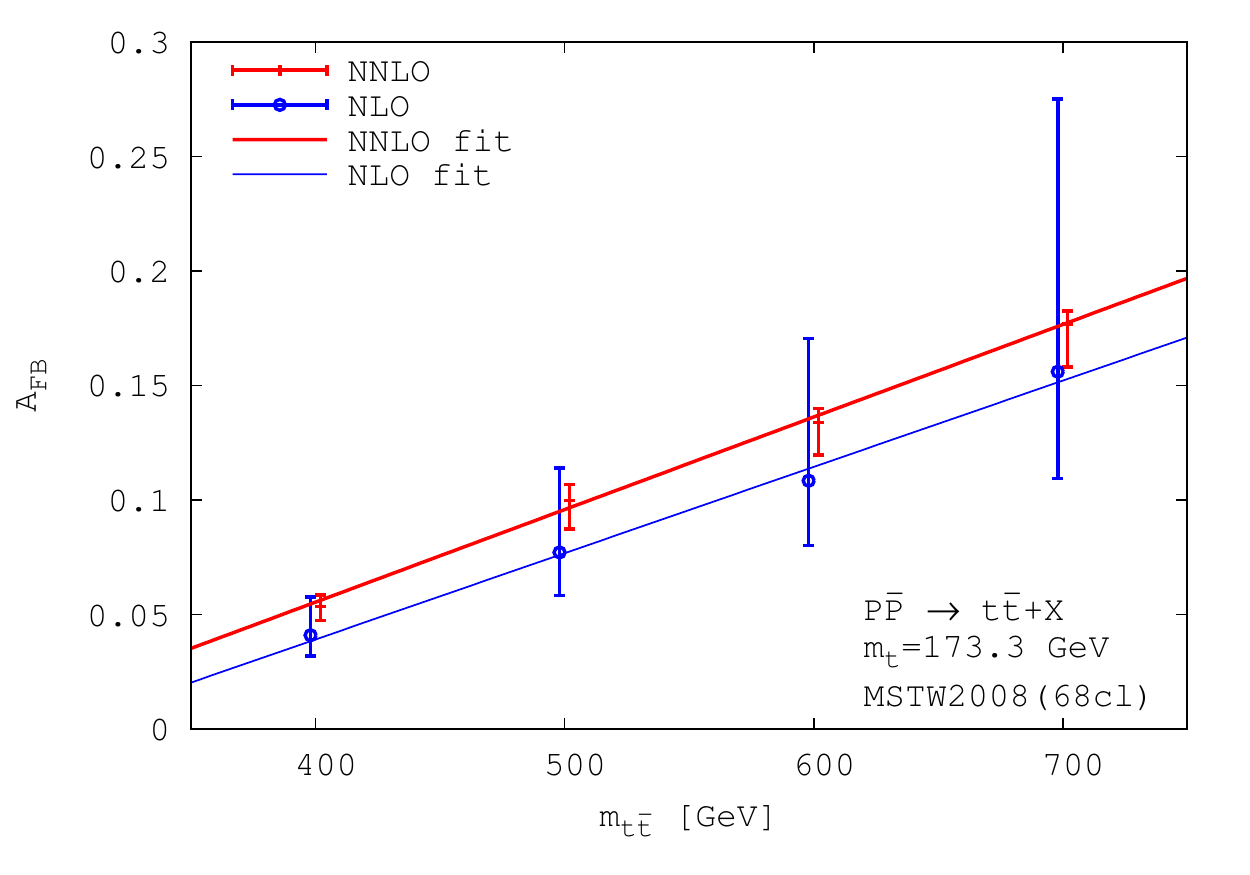}
\caption{\label{fig:slope} A least-squares linear fit (see text) to the central values of the NLO and NNLO QCD $\dmody$ (left) and $\Mtt$ (right) asymmetries versus the exact calculation in all available bins. To better gauge the quality of the fits, the full theory error (scales and pdf added in quadrature) in each bin is also shown.}
\end{figure}
In fig.~\ref{fig:slope}(left) we compare the linear fits eq.~(\ref{eq:beta_y}) to the central value of $\AFB$, at NLO and NNLO, with the actual calculated $\AFB$ bin values. To give a better perspective for the quality of the linear fit we also show the theoretical error in each bin. From this figure we notice that the predicted asymmetry has similar functional behaviour in NLO and NNLO QCD. The $\AFB(\dmody)$ functional dependence is likely not linear and the departure from linearity appears to be slowly growing with $\dmody$. Due to the relatively large size of the NLO error the deviation from linearity could be ignored at NLO. At NNLO, however, this deviation is more significant as it appears comparable to the size of the theory error.

The CDF collaboration has recently measured the slope of $\AFB(\dmody)$ in dilepton final states \cite{CDF:NoteAFBslope}. The measured slope is $\alpha_y^{\rm CDF(\ell\ell)}=0.14 \pm 0.15$ which agrees with the NNLO QCD prediction eq.~(\ref{eq:beta_y}). The latest CDF combination \cite{CDF:NoteAFBslope} yields $\alpha_y^{\rm CDF({\rm comb})}=0.227 \pm 0.057$ and is $2\sigma$ above NNLO QCD (\ref{eq:beta_y}). The latest D\O\ measurement \cite{Abazov:2014cca} in lepton-plus-jets final state $\alpha_y^{\rm D\O }= 0.154 \pm 0.043$ is consistent with the NNLO QCD prediction eq.~(\ref{eq:beta_y}).

The differential distribution $d\sigma/d\dy$ is divided into eight equal-width bins. The two end-bins with largest $\dmody$ contain overflow events. The theoretical prediction through NNLO QCD is shown in fig.~\ref{fig:afb-y}(left), see also appendix~\ref{sec:appendix} table~\ref{tab:diff-y}, and compared with available data from the CDF \cite{Aaltonen:2012it} collaboration. We also plot the data normalised to the central NNLO QCD prediction as well as the NLO and NNLO K-factors.

Similarly to the $\dmody$ dependent $\AFB$, we notice that perturbative convergence is present in this distribution: the NNLO K-factor is much flatter than the NLO one and its size is smaller. Moreover, the NNLO result has significantly smaller errors than the NLO one (by about a factor of two) and the NNLO error band is consistent with the NLO one.

The level of agreement between NNLO QCD and CDF data appears to be much better than that for the related differential asymmetry. Indeed, in most of the bins data and theory are consistent within errors, and in the bins where discrepancy is present it is below $1.5 \sigma$. We hope this result may prove useful in future analyses of the asymmetry in this observable. 

The MC error on the differential asymmetry $\AFB(\dmody)$ is below 1\% in each bin. The MC error on the differential distribution $d\sigma/d\dy$ is around couple of permil in each bin. Such high-precision in the calculation of the differential asymmetry could not be achieved by simply subtracting the corresponding bins of the differential distribution. To that end we have performed an independent, high-precision calculation of the asymmetric contributions in each bin of the differential distribution. Only then it is possible to extract an asymmetry with small statistical error. In practise, we do not need to compute only the asymmetric contribution to each bin; we still allow some symmetric contributions as long as they are not numerically dominant over the asymmetric ones. Excluding the main symmetric contributions to the differential distribution like the LO one and the $gg$-initiated partonic channels turns out to be sufficient for this purpose.

\subsection{$\Mtt$ distribution and asymmetry}

\begin{figure}[h]
\hskip -4mm
\includegraphics[width=0.52\textwidth]{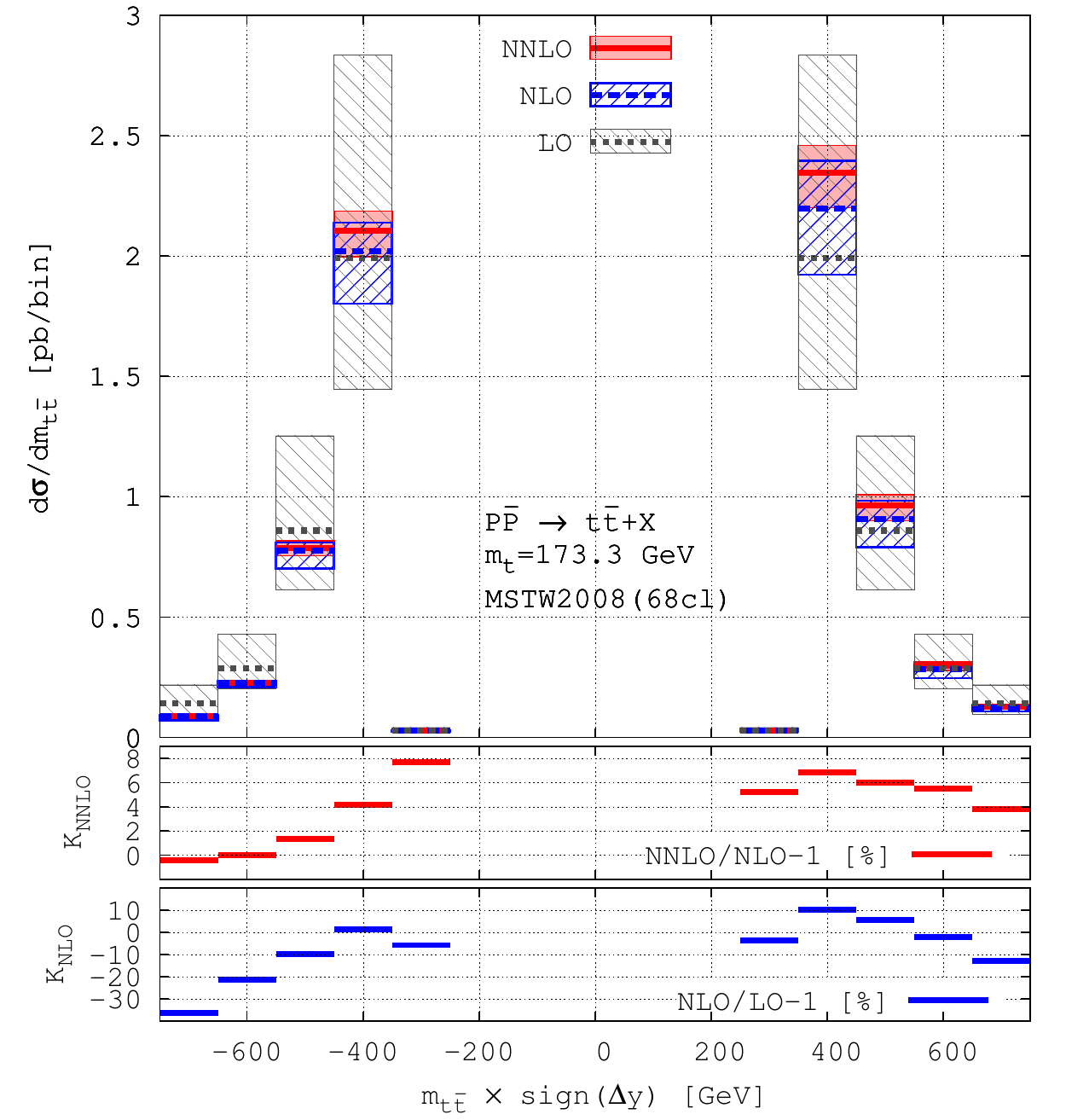}
\includegraphics[width=0.52\textwidth]{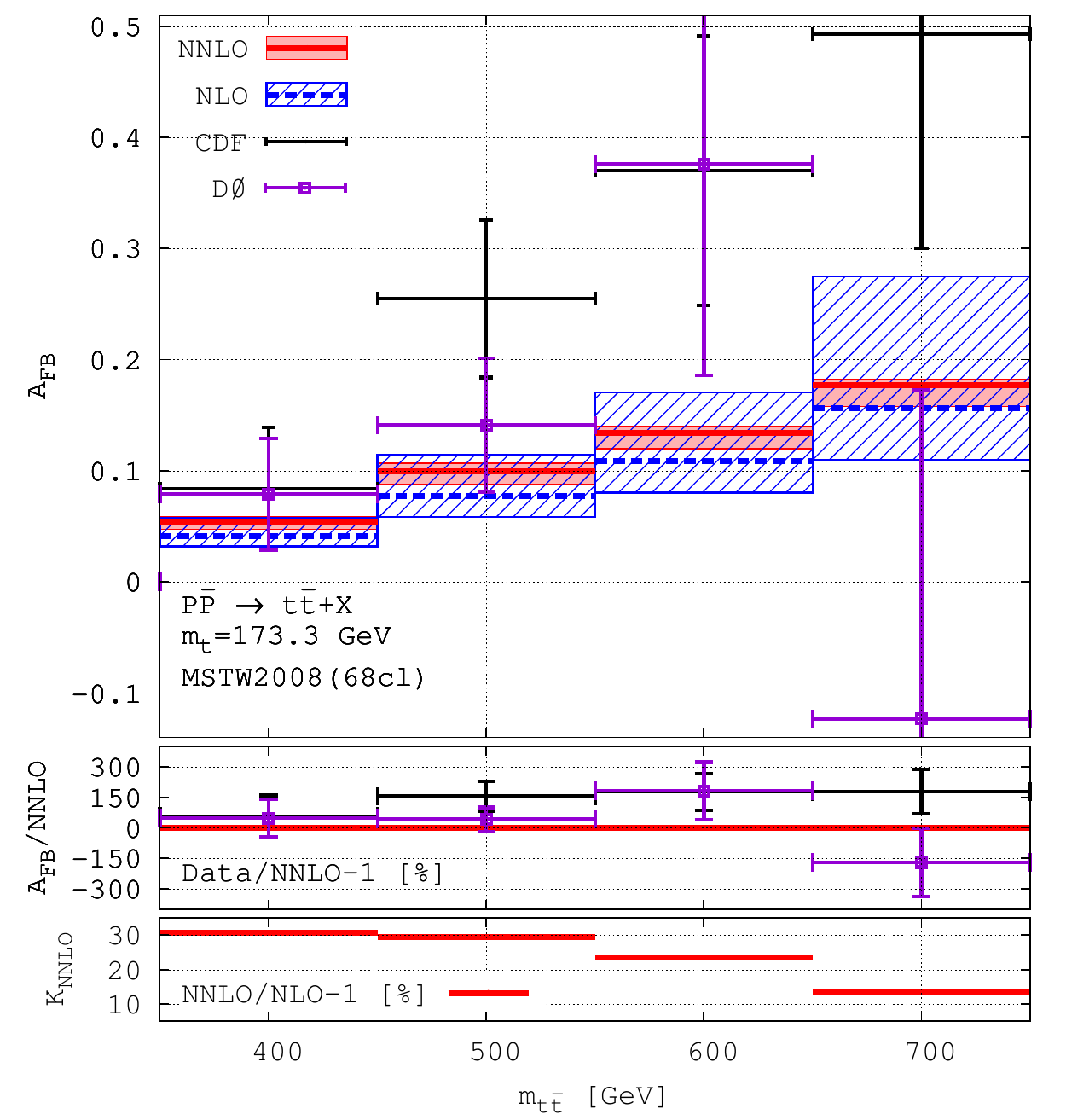}
\caption{\label{fig:afb-mtt} The differential distribution $d^2\sigma/d\dy d\Mtt$ (left) and related differential asymmetry $\AFB(\Mtt)$ (right). Comparison includes SM theory through NNLO QCD and CDF and D\O\ data. The end-bins contain overflow events. The error of the theory prediction is derived from scale and pdf variation.}
\end{figure}
In fig.~\ref{fig:afb-mtt}(right) we show the $\Mtt$ dependence of $\AFB$ (see also appendix~\ref{sec:appendix} table~\ref{tab:afb-mtt}). We use the same bins as in Ref.~\cite{Czakon:2014xsa} which, in turn, match the bins of the CDF analysis \cite{Aaltonen:2012it}. 

The differential asymmetry is divided into four equal-width bins. The bins with lowest/largest $\Mtt$ contain overflow events. The theoretical prediction through NNLO QCD is compared in fig.~\ref{fig:afb-mtt}(right) with data from CDF \cite{Aaltonen:2012it} and D\O\ \cite{Abazov:2014cca,D0:public} collaborations. We also present the data normalised to the central NNLO QCD prediction as well as the NNLO K-factor (the NLO K-factor is not defined for $\AFB$ since the LO result is zero).

The K-factor $K_{\rm NNLO}$ is decreasing with $\Mtt$: close to threshold it is as large as 30\% and decreases to around 10\% in the highest $\Mtt$ bin. The estimated error of the NNLO result is significantly smaller than the NLO one (by about a factor of three or even more at high $\Mtt$). We also notice that the NNLO error band is fully contained within the NLO one. We conclude that this observable possesses good perturbative convergence.

The comparison between the NNLO theory prediction for $\AFB$ and data has already been discussed in Ref.~\cite{Czakon:2014xsa}. Here we will only note the near-perfect agreement of NNLO QCD with the D\O\ data (only one of the four bins shows a deviation, which is slightly above $1 \sigma$) and that the CDF measurement tends to be higher than NNLO QCD: the two agree in the bin with smallest $\Mtt$ while in the other three bins CDF data is above theory by up to about $2 \sigma$.

As was the case for $\AFB(\dmody)$, a compact way for presenting the $\Mtt$-differential asymmetry is through its slope. The least-squares linear fit to the QCD prediction, without any assumption on its behaviour at absolute threshold $\Mtt=2m_t$, reads:
\begin{equation}
\AFB(\Mtt) = \alpha_M\Mtt + \beta_M \,,
\label{eq:beta_M}
\end{equation}
and the values of the pair of coefficients $\alpha_M$ and $\beta_M$ for the central, lowest and highest predicted values are given in table~\ref{tab:slope}.
\begin{table}[h]
\begin{center}
\begin{tabular}{| c | c | c | c | c | c | c |}
\hline
 & ${\rm NLO_{cent}}$ & ${\rm NLO_{min}}$ & ${\rm NLO_{max}}$  & ${\rm NNLO_{cent}}$ & ${\rm NNLO_{min}}$ & ${\rm NNLO_{max}}$ \\ \hline
$\alpha_M \times 10^{3}\GeV$ & 0.377 & 0.255 & 0.709 & 0.404 & 0.364 & 0.405 \\ \hline
$\beta_M$ & -0.111 & -0.070 & -0.235 & -0.106 & -0.097 & -0.101 \\ \hline
\end{tabular}
\caption{\label{tab:slope} Values of the pairs of coefficients $(\alpha_M, \beta_M)$ from eq.~(\ref{eq:beta_M}) for the central/lowest/maximal computed bin values in NLO and NNLO QCD.}
\end{center}
\end{table}

In fig.~\ref{fig:slope}(right) we compare the linear fits eq.~(\ref{eq:beta_M}) to the central value of $\AFB$, at NLO and NNLO, with the calculated central $\AFB$ bin values. To give a better perspective for the quality of the linear fit we also show the theoretical error in each bin. We conclude that the $\Mtt$ functional dependence of $\AFB$ is consistent with being linear in this $\Mtt$ range.

The CDF collaboration has measured \cite{Aaltonen:2012it} the slope of $\AFB(\Mtt)$ and found the value $\alpha_M^{\rm CDF}=1.55\pm 0.48 ~[10^{-3}\GeV^{-1}]$ which is higher than the NNLO QCD prediction eq.~(\ref{eq:beta_M}) (a direct comparison between the two should be done with caution, however, because the intercept $\beta_M$ was not specified in Ref.~\cite{Aaltonen:2012it}). The corresponding D\O\ slope \cite{Abazov:2014cca} reads $\alpha_M^{\rm D\O }= 0.39~[10^{-3}\GeV^{-1}]$ (with $\beta_M=-0.055$). We do not quote the errors of the measurement; for those we refer the reader to Ref.~\cite{Abazov:2014cca}. The slope $\alpha_M^{\rm D\O }$ is consistent with the NNLO QCD prediction in eq.~(\ref{eq:beta_M}).

The differential distribution $d^2\sigma/d\dy d\Mtt$ is divided into ten bins as shown in fig.~\ref{fig:afb-mtt}(left), see also appendix~\ref{sec:appendix} table~\ref{tab:diff-mtt}. The two bins with largest $\Mtt$ contain overflow events. We note the slight difference in the binning between the differential distribution in fig.~\ref{fig:afb-mtt}(left) and the differential asymmetry fig.~\ref{fig:afb-mtt}(right): we make the contribution from the bin $250\GeV\leq\Mtt \leq 350\GeV$ explicit in fig.~\ref{fig:afb-mtt}(left) while, in order to match the binning of the CDF $\AFB$ analysis, have absorbed it into the $350\GeV\leq\Mtt \leq 450\GeV$ bin in fig.~\ref{fig:afb-mtt}(right).

We observe that the NLO and NNLO K-factors of the differential distribution have reasonably similar shapes and $K_{\rm NNLO}$ is smaller than $K_{\rm NLO}$. Similarly to the other differential distributions considered above, the NNLO error band is smaller than the NLO one (by about a factor of two) and the NNLO result is consistent with the error estimate of the NLO QCD prediction. These features indicate good perturbative convergence in this observable.

The MC error on the differential asymmetry $\AFB(\Mtt)$ is below 1\% in each bin. The MC error on the differential distribution $d^2\sigma/d\dy d\Mtt$ is around 1\% in the two central bins with $250\GeV\leq\Mtt \leq 350\GeV$ and below 4 permil in the remaining bins. Such high-precision in the calculation of the differential asymmetry is achieved following the strategy for the calculation of $\AFB(\dmody)$ described in sec.~\ref{sec:afb-y}.

\subsection{The first Legendre moment of the $\cos\theta$ distribution}\label{sec:afb-legmom}

As an alternative way at looking at the top-quark $\AFB$, the CDF collaboration measured \cite{CDF:2013gna} the Legendre Moments of the differential distribution $d\sigma/d\cos\theta$. The main idea is based on the realisation that the forward-backward asymmetry is almost exclusively confined to the first Legendre moment, thus offering an alternative assessment of this asymmetry. In sec.~\ref{sec:costheta} we described the calculation of the NNLO QCD correction to these moments. The results can  be found in fig.~\ref{fig:costheta}(right) as well as appendix~\ref{sec:appendix} table~\ref{tab:legmom}. We observe that the CDF measurement of $a_1$ is about $1.7 \sigma$ above the theory prediction after {\it naively} accounting for EW corrections. As can be anticipated from the inclusive $\AFB$, NNLO QCD and EW corrections each decrease this discrepancy.

\subsection{$\PTtt$ distribution and differential asymmetry}

\begin{figure}[h]
\hskip -4mm
\includegraphics[width=0.52\textwidth]{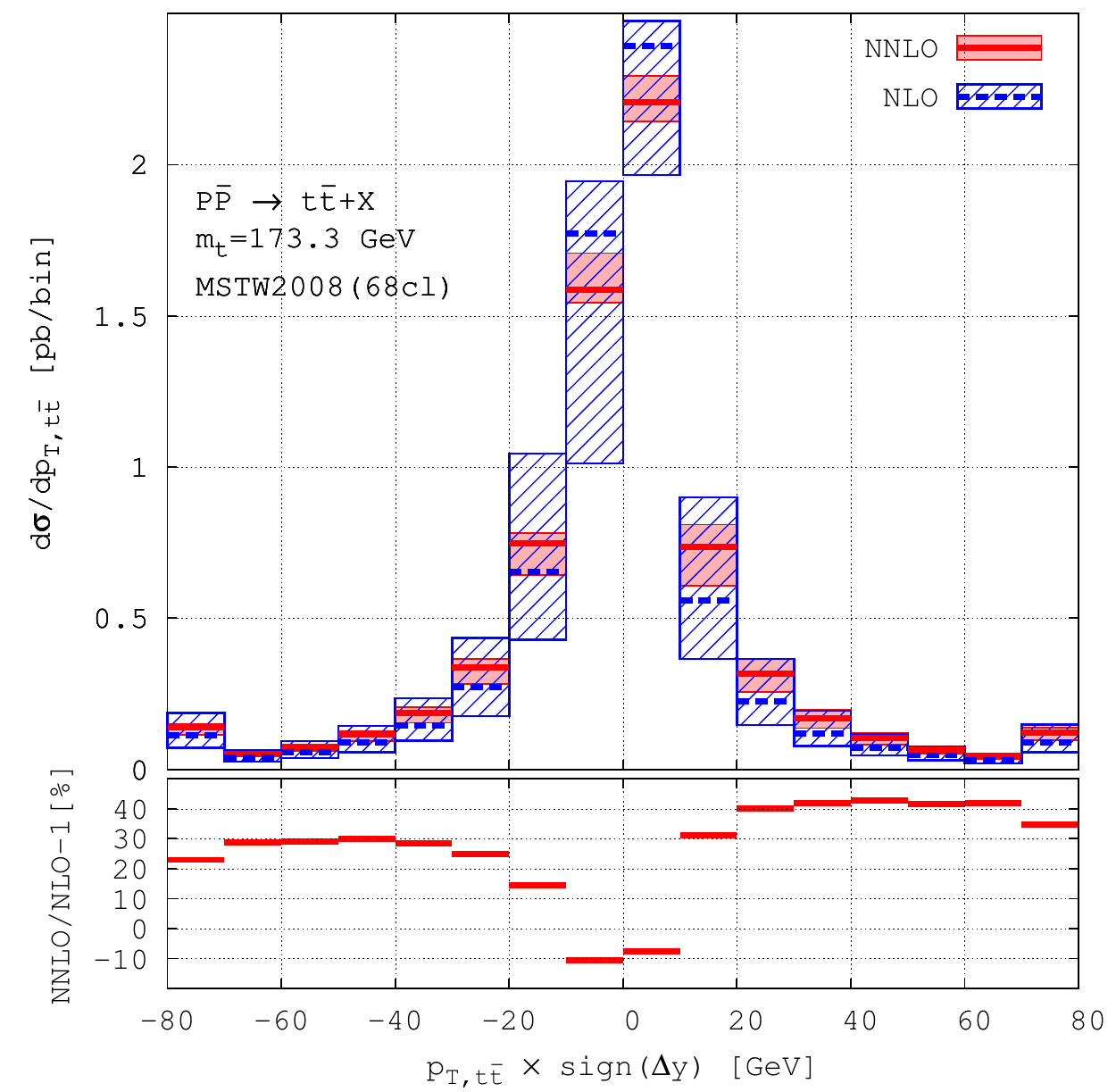}
\includegraphics[width=0.52\textwidth]{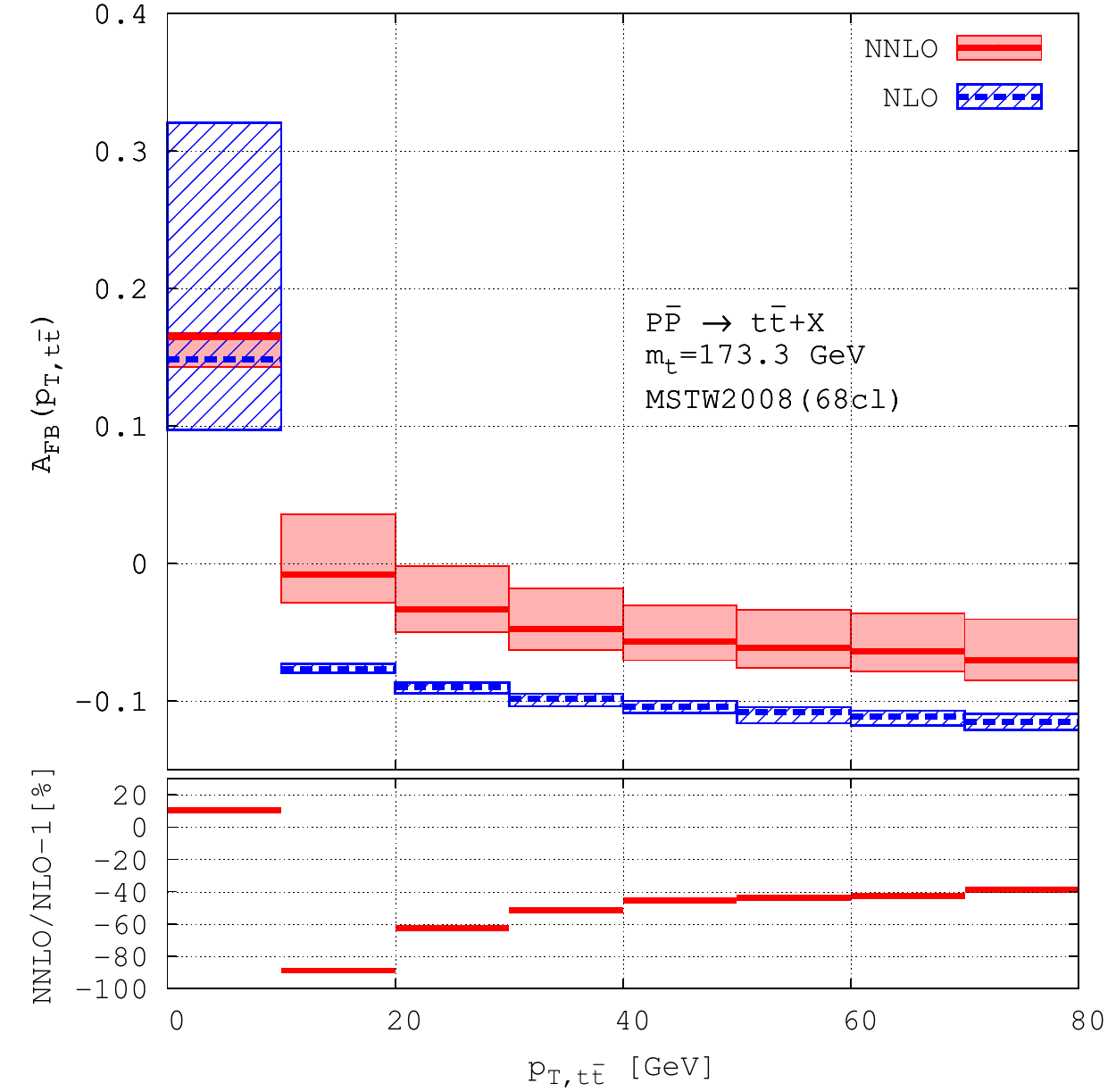}
\caption{\label{fig:afb-pt} The differential distribution $d^2\sigma/d\dy d\PTtt$ (left) and related differential asymmetry $\AFB(\PTtt)$ (right). Included is SM theory through NNLO QCD. The end-bins contain overflow events. The error of the theory prediction is derived from scale and pdf variation.}
\end{figure}
In the following we study the $\PTtt$-dependent forward-backward asymmetry which is of special theoretical interest. The NLO and NNLO QCD prediction for the asymmetry is shown in fig.~\ref{fig:afb-pt}(right) and in appendix~\ref{sec:appendix} table~\ref{tab:afb-ptt}. We use the same bins as in Ref.~\cite{Czakon:2014xsa} that, in turn, match the bins of the CDF analysis \cite{Aaltonen:2012it} (however no parton-level results for $\AFB(\PTtt)$ that we could compare to have been published). The differential asymmetry is divided into eight equal-width bins. The bin with largest $\PTtt$ contains overflow events. 

The shape of the $\PTtt$-asymmetry for $\PTtt>0$ can be derived with purely NLO calculation in the process ($t\bar t j$) and that part of the asymmetry has been understood for quite some time \cite{Dittmaier:2007wz}. We have verified in Ref.~\cite{Czakon:2014xsa} the consistency of our inclusive $t\bar t$ NNLO calculation with NLO $t\bar t j$ predictions from Refs.~\cite{Melnikov:2010iu,Melnikov:2011qx,Hoeche:2013mua} and found perfect agreement with an independent evaluation performed with the package \textsc{Helac-Nlo} \cite{Bevilacqua:2011xh}. The difference between NNLO and NLO corrections to the $\PTtt$ asymmetry for $\PTtt\geq 10 \GeV$ follows the pattern noticed in CDF data \cite{Aaltonen:2012it} and is, moreover, consistent with the analysis of Ref.~\cite{Gripaios:2013rda}.

For reference, the differential distribution $d^2\sigma/d\dy d\PTtt$ is shown in fig.~\ref{fig:afb-pt}(left) as well as in appendix~\ref{sec:appendix} table~\ref{tab:diff-ptt}. It is divided into sixteen bins of equal width and the two bins with largest $\PTtt$ contain overflow events. As for the differential asymmetry, the behaviour of this distribution away from the point $\PTtt=0$ is well understood and has been extensively studied in the context of $t\bar t j$ production in NLO QCD \cite{Dittmaier:2007wz,Melnikov:2010iu,Melnikov:2011qx,Hoeche:2013mua}. In particular, we do not show the LO QCD contribution since it enters only the two central bins containing the point $\PTtt=0$. The corresponding prediction can be found in appendix~\ref{sec:appendix} table~\ref{tab:diff-ptt}.

The relative MC error on the central value of the differential asymmetry $\AFB(\PTtt)$ is below 1\% in each of the eight bins. In some of the bins,  and for some scale choices, the predicted bin asymmetry becomes very close to zero (see fig.~\ref{fig:afb-pt}(right)) and, as can be anticipated, in such cases the relative MC error becomes much larger. Such large relative MC errors, however, are harmless and do not adversely impact the error estimate in the corresponding bins. The rather asymmetric error in the first bin $0\leq\PTtt\leq 10\GeV$ is not due to statistical effects since in this bin the relative MC error is at the sub-permil level for all $\mu_{F,R}$ values. The relative MC error on the differential distribution $d^2\sigma/d\dy d\PTtt$ is below 4 permil in all bins. The calculation of the differential asymmetry follows the strategy for minimising the MC error described in sec.~\ref{sec:afb-y}.

\subsection{$\PTtt$ cumulative asymmetry}

One of the unexpected findings of Ref.~\cite{Czakon:2014xsa} was that the NNLO QCD corrections to the inclusive $\AFB$ were significant, much larger than what had been anticipated from arguments based on soft-gluon resummation \cite{Almeida:2008ug,Ahrens:2011uf}. These soft-gluon-based predictions are compatible with the parton shower based analysis of the top-quark $\AFB$ performed in Ref.~\cite{Skands:2012mm} but not with the soft-gluon resummation prediction of Ref.~\cite{Kidonakis:2011zn}, which are larger (and recently updated in Ref.~\cite{Kidonakis:2015ona}). We presume the difference in the predictions between Ref.~\cite{Kidonakis:2011zn} and Refs.~\cite{Almeida:2008ug,Ahrens:2011uf} is due to different subleading terms. This is an often present ambiguity in resummed calculations matched to fixed order results of lower accuracy (NLO in this case). Our viewpoint on such subtleties has been explained at length in Ref.~\cite{Cacciari:2011hy}; further discussion of this problem goes beyond the scope of this paper.

A detailed comparison between the NNLO QCD corrections to $\AFB$ and the soft-gluon resummation based predictions was performed in Ref.~\cite{Czakon:2014xsa} and we do not repeat it here. Our goal in the following is to elaborate on an observation made in Ref.~\cite{Czakon:2014xsa}, namely, that the difference between the NNLO fixed order predictions and the ones based on soft-gluon resummation matched to NLO could potentially be understood by considering the $\PTtt$ differential asymmetry. The physics behind this idea is the following: soft-gluon resummation in $t\bar t$ production applies to kinematic configurations which are of almost 2-to-2 type, i.e. configurations where the final state consists of a top pair which takes almost all the energy available to the partonic reaction and is, possibly, accompanied by very soft radiation that carries very little energy. Since the initial state has zero transverse momentum, one necessarily arrives at kinematic configurations consisting of $t\bar t$ pairs with small $\PTtt$. It is hard to quantify on purely theoretical grounds how small that $\PTtt$ would be, but as a guidance one can use the fact that the top pair $p_T$ is peaked below $10\GeV$ \cite{Mrenna:1996cz,Li:2013mia}. Thus we expect that the bulk of the contributions from soft gluon resummation would be at small $\PTtt$ (presumably in the first bin $0\leq\PTtt\leq 10\GeV$) and will be decreasing fast with $\PTtt$.
\footnote{One should keep in mind that in practise, implementations of soft gluon resummation typically generate noticeable contributions even in kinematical regions that are far from the relevant partonic threshold. We are not concerned with such effects here.}

To that end we define the {\it cumulative} forward-backward asymmetry $\AcumulFB(\PTtt^{\rm cut})$. It has a single bin of variable width $\PTtt^{\rm cut}$, where $0\leq\PTtt\leq \PTtt^{\rm cut}\,\,$, and
\begin{equation}
{\AcumulFB}(\PTtt^{\rm cut}) = {\hat N\over \hat D} \equiv {\sigma^{+}_{\rm c} - \sigma^{-}_{\rm c} \over \sigma^{+}_{\rm c} + \sigma^{-}_{\rm c} }
~~~,~~~
\sigma^\pm_{\mathrm{c}} = \int \theta(\pm \Delta y) \theta(\PTtt^{\rm cut}-\PTtt) \mathrm{d}\sigma\, ,
\label{eq:AcumulFB-diff}
\end{equation}
Eq.~(\ref{eq:AcumulFB-diff}) implicitly defines a cumulative numerator $\hat N$ and denominator $\hat D$ (in units of $pb$). For example, $\AcumulFB(10\GeV)$ corresponds to the first (leftmost) bin of $\AFB$ in fig.~\ref{fig:afb-pt}(right), while $\AcumulFB(80\GeV)$ corresponds to the inclusive asymmetry (recall that the last bin contains also the overflow events with $\PTtt\geq 80\GeV$). The cumulative asymmetry may be better suited for studying the $\PTtt$ dependence since it is not as singular as the usual $\PTtt$ differential asymmetry (because in any bin corrections from all relevant perturbative orders contribute).

\begin{figure}[h]
\hskip -4mm
\includegraphics[width=1.0\textwidth]{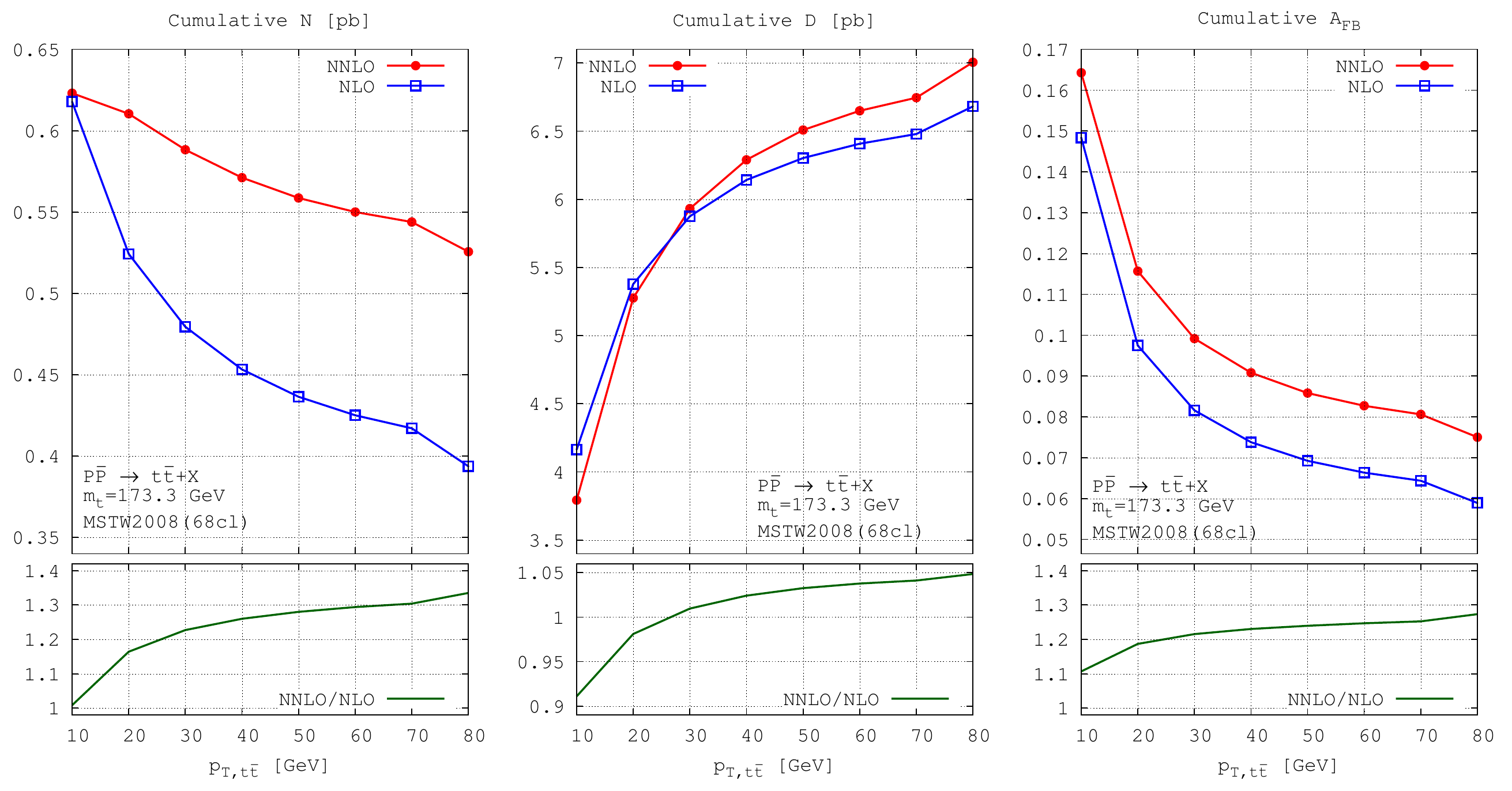}
\caption{\label{fig:afb-cumul} NLO and NNLO QCD corrections to the cumulative numerator, denominator and asymmetry defined in eq.~(\ref{eq:AcumulFB-diff}). The visible non-smoothness of the lines reflects the $10\GeV$ binning of the underlying calculation. The presence of overflow events in the highest bin is clearly noticeable. The values at $\PTtt=80\GeV$ correspond to the inclusive numerator, denominator and asymmetry.}
\end{figure}
\begin{table}[h]
\begin{center}
\begin{tabular}{| c || c | c | c | c | c | c | c | c |}
\hline
$\PTtt~ [\GeV]$ & 10 & 20 & 30 & 40 & 50 & 60 & 70 & $\geq$ 80 \\ \hline\hline
${\hat N}^{\rm NLO}$ [pb] & 0.618 &  0.524 &  0.479 &  0.453 &  0.436 &  0.425 &  0.417 &  0.394 \\ \hline
${\hat N}^{\rm NNLO}$ [pb] & 0.623 &  0.610 &  0.588 &  0.571 &  0.559 &  0.550 &  0.544 &  0.526 \\ \hline
${\hat D}^{\rm NLO}$ [pb] & 4.164 &  5.378 &  5.876 &  6.142 &  6.303 &  6.407 &  6.479 &  6.682 \\ \hline
${\hat D}^{\rm NNLO}$ [pb] & 3.793 &  5.276 &  5.932 &  6.290 &  6.508 &  6.649 &  6.745 &  7.005 \\ \hline
$\AcumulFB^{\rm NLO}$ & 0.148 &  0.097 &  0.082 &  0.074 &  0.069 &  0.066 &  0.064 &  0.059 \\ \hline
$\AcumulFB^{\rm NNLO}$ & 0.164 &  0.116 &  0.099 &  0.091 &  0.086 &  0.083 &  0.081 &  0.075 \\ \hline
\end{tabular}
\caption{\label{tab:afb-cumul} Values for the cumulative numerator, denominator and asymmetry appearing in fig.~\ref{fig:afb-cumul}.}
\end{center}
\end{table}
The results for the cumulative numerator, denominator and asymmetry are shown in fig.~\ref{fig:afb-cumul} and table~\ref{tab:afb-cumul}. Fig.~\ref{fig:afb-cumul} clearly demonstrates the observation made in Ref.~\cite{Czakon:2014xsa}: the numerator $\hat N$ receives tiny NNLO correction for small $\PTtt\,$ (in particular in the first bin $\PTtt\leq 10\GeV$), and the difference in $\AFB$ is solely due to the change in the denominator $\hat D$. Therefore, since the denominator is itself symmetric in $\dy$ (see eq.~(\ref{eq:AcumulFB-diff})), the intrinsic asymmetry in this bin is the same in NLO and NNLO QCD. Once one goes to higher $\PTtt\,$ the NNLO correction to $\hat N$ grows fast while the rate of change in the denominator $\hat D$ is much slower. From this we conclude that the difference between the inclusive $\AFB$ computed in NNLO and NLO QCD originates from events that are accompanied by hard radiation, or at least radiation that is harder than what is required for being in the soft-gluon resummation regime. It seems to us that a measurement of the cumulative $\AFB$ might be very beneficial also because the difference between NNLO and NLO corrections is very weakly dependent on $\PTtt\,$ which might allow for more conclusive separation of higher order effects in this observable.

\subsection{$\Mtt$ cumulative asymmetry}\label{sec:Mtt-cumulative}

\begin{figure}[h]
\hskip -4mm
\includegraphics[width=1.0\textwidth]{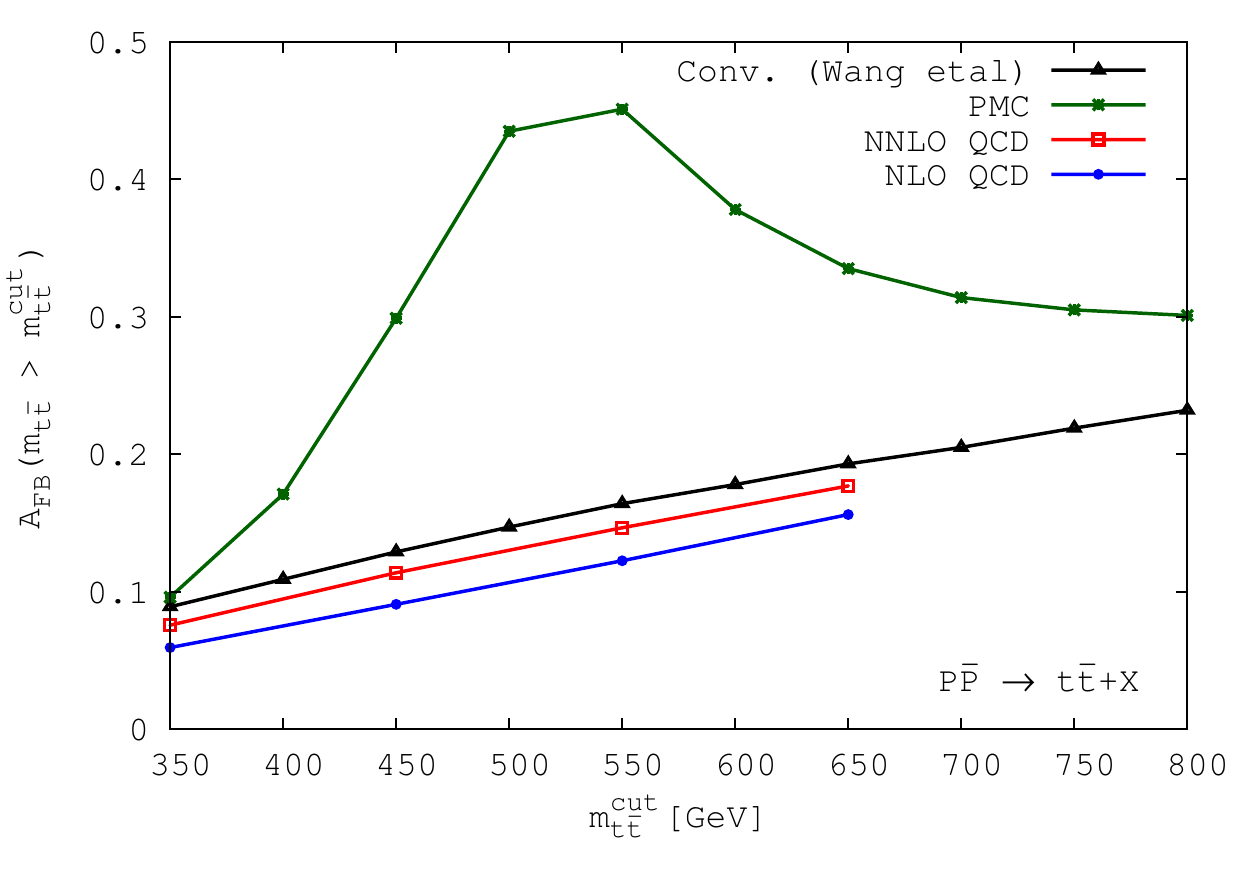}
\caption{\label{fig:Mtt-cumulative} Predictions for the $\Mtt$ cumulative asymmetry: pure QCD at NLO and NNLO (as derived in this work),  NLO prediction of Ref.~\cite{Wang:2015lna} including EW corrections, as well as the PMC scale-setting prediction of Ref.~\cite{Wang:2015lna}.}
\end{figure}

The $\Mtt$ cumulative asymmetry $\AcumulFB(\Mtt > \Mtt^{\rm cut})$ has recently been discussed in Ref.~\cite{Wang:2015lna}. In fig.~\ref{fig:Mtt-cumulative} we present the predictions for this asymmetry in NLO and NNLO QCD (with unexpanded numerator and denominator) as well as two predictions from Ref.~\cite{Wang:2015lna}. 

The first prediction of Ref.~\cite{Wang:2015lna} is based on conventional scale-setting; it differs from our NLO calculation in the inclusion of EW corrections, the use of expanded definition for the asymmetry as well as minor differences due to value of $m_t$ and different pdf set. Although our predictions cannot be compared directly, it is clear from fig.~\ref{fig:Mtt-cumulative} that the predictions are rather similar. A detailed comparison between inclusive $\AFB$ predictions based on expanded and unexpanded $\AFB$ definition, and with/without EW corrections, can be found in Ref.~\cite{Czakon:2014xsa}.

The second prediction of Ref.~\cite{Wang:2015lna} is based on the PMC/BLM scale-setting procedure. As already pointed out in Ref.~\cite{Wang:2015lna} the conventional and PMC predictions are substantially different from each other. This difference in behaviours is mainly due to the qualitatively different scale at which the renormalised coupling is evaluated in the two approaches. In the conventional scale-setting approach used by us, $\mu_R$ is set to $m_t$ while in the PMC approach $\mu_R$ depends strongly on $\Mtt$ (see table {\rm III} in Ref.~\cite{Wang:2015lna}): as $\Mtt$ increases from threshold to around 800 $\GeV$, the renormalisation scale at first strongly decreases and then starts to grow fast again. Its minimum is reached around $\Mtt\sim 500\GeV$ where it is smaller than the value at threshold by a factor of almost four. The maximal value for $\mu_R$ is reached at maximal $\Mtt$ where the scale is larger than its threshold value. Such a behaviour is easily contrasted with the conventional scale-setting approach where, even for dynamic scales, one typically expect a monotonic increase of $\mu_R$ with increasing $\Mtt$. Moreover, one expects that in the limited range of $\Mtt$ used for the calculation of the NNLO result, fixed and dynamic scales would lead to consistent predictions within scale errors (see also recent discussion for the LHC \cite{Czakon:2015owf}).

We conclude that the two scale-setting approaches produce very different predictions for the $\Mtt$ cumulative $\AcumulFB$ and it should be easy to distinguish between the two with data, especially in the region around $\Mtt\sim 500\GeV$.  We would also like to point out that the NNLO prediction based on conventional scale-setting with $\mu_R=m_t$ exhibits the ``increasing-decreasing" behaviour pointed out in Ref.~\cite{Wang:2015lna}, albeit much less pronounced than in the PMC scale-setting approach.

\section{Comparisons between different pdf sets}\label{sec:pdf}

An alternative way of assessing the pdf dependence in theory predictions is to compare calculations with different pdf sets. In this section we compare NNLO QCD predictions based on four state-of-the-art pdf sets: CT10, HERA 1.5, MSTW2008 and NNPDF 2.3. We compare the central pdf members for central scale choice $\mu_F=\mu_R=m_t$.

\begin{figure}[h]
\hskip -4mm
\includegraphics[width=1.0\textwidth]{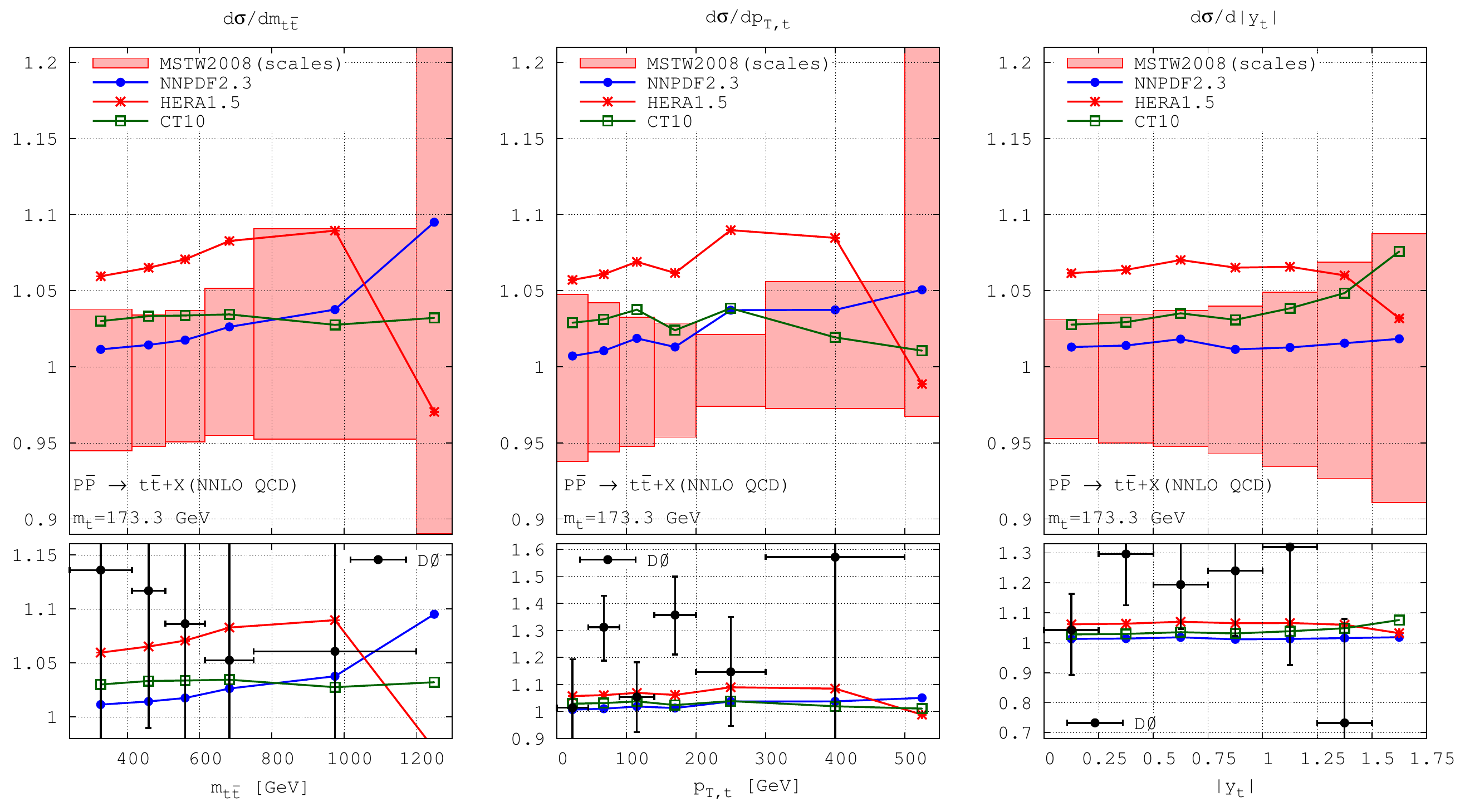}
\caption{\label{fig:pdf-compare} NNLO QCD prediction for three differential distributions (in $\Mtt,\,\PTt$ and $\modyt$) with four pdf sets. Given are the ratios of the CT10, HERA 1.5 and NNPDF 2.3 based predictions with respect to MSTW2008. For reference also the scale dependence of the MSTW2008 prediction is shown (red band). For improved visibility, in the lower plots we compare the same predictions with the available data from the D\O\ Collaboration~\cite{Abazov:2014vga}.}
\end{figure}
\begin{figure}[h]
\hskip -4mm
\includegraphics[width=1.0\textwidth]{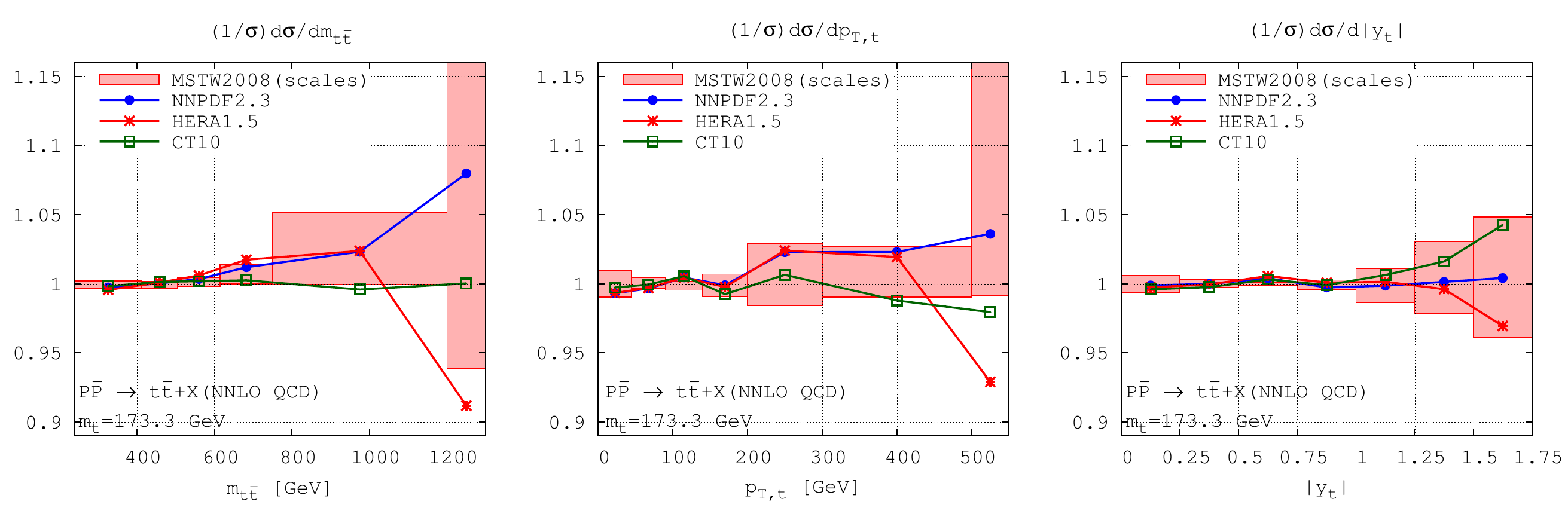}
\caption{\label{fig:pdf-compare-norm} As in fig.~\ref{fig:pdf-compare} but for the normalised to unity distributions.}
\end{figure}
In fig.~\ref{fig:pdf-compare} we present the ratio of CT10, HERA 1.5 and NNPDF 2.3 with respect to MSTW2008 (the predictions for the latter pdf set could be found in the previous sections). We study the following three differential distributions (with absolute normalisation): $\Mtt,\,\PTt$ and $\modyt$. Additionally, in the upper plots we present the scale error of the MSTW2008 result, while in the lower plots we compare with available data from the D\O\ collaboration~\cite{Abazov:2014vga}.

We observe that the spread among the pdf sets is comparable to the size of the NNLO scale variation and only the HERA 1.5 prediction lies outside the scale error band. Since in the kinematic range considered in this work pdf error is (much) smaller than the one due to scale variation, it seems that the spread in predictions based on different pdf sets may be not fully compatible with the pdf error estimates of the individual pdf sets. We also observe that all pdf sets agree with the available data. Although the spread of theory predictions is much smaller than the size of the experimental error it could nevertheless be interesting to speculate if the currently available data has the power to constrain pdf.

It was pointed out in Ref.~\cite{Czakon:2015pga} that a separate fit to normalisation and shapes is desired in pdf studies. Indeed, the absolute normalisation could be affected by systematic effects which are harder to control (one of those, as we pointed out in the previous discussions, is the top mass which affects normalisation and shapes rather differently). In fig.~\ref{fig:pdf-compare-norm} we compare the predictions for the normalised to unity $\Mtt,\,\PTt$ and $\modyt$ distributions. Unlike the case of absolute normalisations, we now observe a remarkable agreement between all pdf sets in the full kinematic ranges. Moreover, the agreement is within the estimated (from scales) theory error of the MSTW2008 prediction. The latter fact is quite remarkable since the scale error of the normalised predictions is much smaller than the error of the predictions with absolute normalisation and, for most of the kinematical range, is in the 1\% range. The various pdf sets start to diverge from each other only towards the end-bins. As for the distributions with absolute normalisation, the results for the overflow bins in the $\Mtt$ and $\PTt$ distributions should be interpreted with care given the MC error (not shown) is around 4-5\% for all pdf sets. On the other hand, the estimated MC error in the last bin of the $\modyt$ distribution is only around 1\% and therefore the spread observed between the various pdf's in that bin is, likely, a significant effect. The above observations are very interesting in the context of the expectation set in Ref.~\cite{Czakon:2015pga} that a separate fit to normalisation and shapes is needed in pdf studies as well as the well-appreciated fact that the large-$x$ pdf region can effectively be constrained with top-quark data \cite{Czakon:2013tha}.

Comparisons between various pdf sets, in NLO QCD and for LHC 7 TeV, have recently been performed in Ref.~\cite{Aad:2014zka}. The results we present in this work represent the first comparison between pdf sets  in full NNLO QCD at the differential level (albeit for a different collider). Our findings are in rough agreement with the ones in Ref.~\cite{Aad:2014zka} but with one exception: in the normalised $\Mtt$ comparison of Ref.~\cite{Aad:2014zka} one can clearly notice that the HERA prediction is distinct (on the scale of the theory error) from the other pdf sets, while in our Tevatron calculation we do not observe such trend. It will be very interesting to clarify the origin of this difference (different perturbative orders versus different colliders) by directly comparing LHC predictions based on different pdf sets.

\section{Conclusions}

In this work we present a complete set of NNLO QCD predictions for stable top-quark production at the Tevatron. The predictions are for the $y_t,~\Mtt,~\PTt,~\PTtt$ and $\cos\theta$ differential distributions measured by the CDF and D\O\ collaborations. We present LO, NLO and NNLO predictions, study the convergence of perturbation theory in each distribution and present the relevant $K$-factors. All results are given in tables for convenience and future use. All distributions are computed with the MSTW2008 pdf set. Additionally we compare predictions for three differential distributions, with both absolute and unit normalisation, derived with four different pdf sets. For distributions with absolute normalisation we observe spread among the different pdf sets which is comparable with the size of the scale error. Normalised distributions, however, show remarkable independence of the choice of pdf set. Such stability may be useful in future analyses, for example, in order to disentangle the dependence on $m_t$. 

We have also presented detailed predictions for many $\AFB$-related differential observables. In particular we present predictions for the slopes of the $\dy$- and $\Mtt$-dependent asymmetry and Legendre moments. We also present predictions for the cumulative $\PTtt$ asymmetry which we have used to analyse in depth the origin of NNLO QCD correction to $\AFB$. We point out that the $\PTtt$ cumulative asymmetry is much better behaving than the usual $\PTtt$-asymmetry and conclude that a future measurement of this cumulative asymmetry would be valuable. We also present a prediction for the $\Mtt$ cumulative asymmetry which we compare with a prediction based on the PMC scale-setting approach. The predictions in the conventional and PMC scale-setting approach differ significantly, making it possible for a future measurement to easily distinguish between the two. 

We have made significant effort to derive results with very high quality. Typically, the Monte-Carlo integration error in each bin of the differential distributions is at the few-permil level and is thus totally negligible. For the differential asymmetry the relative MC error is up to around one percent per bin. 

Throughout the present work we use fixed scales $\mu_R=\mu_F=m_t$ despite that, arguably, running scales are better suited in describing differential distributions. We have several reasons for doing so. The first reason is of technical nature. Secondly, and arguably most importantly, in the limited kinematic ranges considered in the present work, the use of dynamic scales is not strictly required. We expect that the use of dynamic scales would not take the predictions outside of the estimated theory error range. Given also that experimental errors are significantly larger than the theory ones, it does not appear that the question of including dynamic scales will be of relevance to top-physics at the Tevatron.

In conclusion, our hope is that this work offers a complete set of state-of-the-art theory predictions for top-quark production at the Tevatron which should be up-to-date until, at least, theory predictions for NNLO top-quark production with top-quark decay become available.

\begin{acknowledgments}
We thank Stefan Dittmaier for kindly providing us with his code for the evaluation of the one-loop virtual corrections. M.C. thanks the CERN Theoretical Physics Department and Emmanuel College Cambridge for hospitality during the completion of this work. A.M. thanks the IPPP and Trevelyan College at Durham University for their hospitality through a COFUND Senior Research Fellowship during the completion of this work. The work of M.C. was supported in part by grants of the DFG and BMBF. The work of D.H. and A.M. is supported by the UK Science and Technology Facilities Council [grants ST/L002760/1 and ST/K004883/1]. 
\end{acknowledgments}

\appendix

\section{Tables with theory predictions}\label{sec:appendix}

\begin{table}[h]
\begin{center}
\begin{tabular}{| c | c | c | c |}
\hline
$\Mtt$ & \multicolumn{3}{|c|}{$d\sigma/d\Mtt$ [pb/bin]} \\
\hline 
 & {\rm LO} & {\rm NLO} & {\rm NNLO} \\ \hline
	[240	 ; 	412.5]	&	$	2.78	^{+	1.16	}_{-	0.75	} ~ \times ~ 10^{	0	}$	&	$	2.96	^{+	0.21	}_{-	0.34	}$$^{+	0.07	}_{-	0.06	}$$^{+	0.22	}_{-	0.35	} ~ \times ~ 10^{	0	}$	&	$	3.13	^{+	0.12	}_{-	0.17	}$$^{+	0.07	}_{-	0.05	}$$^{+	0.14	}_{-	0.18	} ~ \times ~ 10^{	0	}$	\\	\hline
	[412.5	 ; 	505]	&	$	2.43	^{+	1.08	}_{-	0.68	} ~ \times ~ 10^{	0	}$	&	$	2.47	^{+	0.14	}_{-	0.28	}$$^{+	0.07	}_{-	0.05	}$$^{+	0.16	}_{-	0.29	} ~ \times ~ 10^{	0	}$	&	$	2.59	^{+	0.09	}_{-	0.14	}$$^{+	0.07	}_{-	0.05	}$$^{+	0.11	}_{-	0.14	} ~ \times ~ 10^{	0	}$	\\	\hline
	[	505	 ; 	615	]	&	$	9.95	^{+	4.74	}_{-	2.94	} ~ \times ~ 10^{	-1	}$	&	$	9.20	^{+	0.31	}_{-	0.99	}$$^{+	0.28	}_{-	0.18	}$$^{+	0.42	}_{-	1.00	} ~ \times ~ 10^{	-1	}$	&	$	9.50	^{+	0.35	}_{-	0.47	}$$^{+	0.26	}_{-	0.18	}$$^{+	0.44	}_{-	0.50	} ~ \times ~ 10^{	-1	}$	\\	\hline
	[	615	 ; 	750	]	&	$	3.27	^{+	1.68	}_{-	1.02	} ~ \times ~ 10^{	-1	}$	&	$	2.66	^{+	0.06	}_{-	0.25	}$$^{+	0.09	}_{-	0.05	}$$^{+	0.11	}_{-	0.25	} ~ \times ~ 10^{	-1	}$	&	$	2.73	^{+	0.14	}_{-	0.12	}$$^{+	0.08	}_{-	0.06	}$$^{+	0.16	}_{-	0.14	} ~ \times ~ 10^{	-1	}$	\\	\hline
	[750	 ; 	1200]	&	$	9.21	^{+	5.26	}_{-	3.08	} ~ \times ~ 10^{	-2	}$	&	$	6.20	^{+	0.00	}_{-	0.88	}$$^{+	0.23	}_{-	0.14	}$$^{+	0.23	}_{-	0.89	} ~ \times ~ 10^{	-2	}$	&	$	6.36	^{+	0.58	}_{-	0.30	}$$^{+	0.20	}_{-	0.15	}$$^{+	0.61	}_{-	0.34	} ~ \times ~ 10^{	-2	}$	\\	\hline
	[	1200	 ; 	$\infty$	]	&	$	2.82	^{+	2.12	}_{-	1.12	} ~ \times ~ 10^{	-4	}$	&	$	1.07	^{+	0.10	}_{-	0.78	} ~ \times ~ 10^{	-4	}$	&	$	1.27	^{+	0.47	}_{-	0.14	} ~ \times ~ 10^{	-4	}$	\\	\hline
%
%
\hline
$\Mtt$ & \multicolumn{3}{|c|}{$(1/\sigma)d\sigma/d\Mtt$ [1/bin]} \\
\hline 
 & {\rm LO} & {\rm NLO} & {\rm NNLO} \\ \hline
	[240 ; 412.5]	&	$	4.20	^{+	0.07	}_{-	0.07	} ~ \times ~ 10^{	-1	}$	&	$	4.44	^{+	0.08	}_{-	0.03	}$$^{+	0.00	}_{-	0.02	}$$^{+	0.08	}_{-	0.03	} ~ \times ~ 10^{	-1	}$	&	$	4.47	^{+	0.01	}_{-	0.02	}$$^{+	0.01	}_{-	0.01	}$$^{+	0.01	}_{-	0.02	} ~ \times ~ 10^{	-1	}$	\\	\hline
	[412.5	 ; 	505]	&	$	3.66	^{+	0.00	}_{-	0.00	} ~ \times ~ 10^{	-1	}$	&	$	3.70	^{+	0.02	}_{-	0.01	}$$^{+	0.01	}_{-	0.00	}$$^{+	0.02	}_{-	0.01	} ~ \times ~ 10^{	-1	}$	&	$	3.70	^{+	0.01	}_{-	0.01	}$$^{+	0.00	}_{-	0.00	}$$^{+	0.01	}_{-	0.01	} ~ \times ~ 10^{	-1	}$	\\	\hline
	[	505	 ; 	615	]	&	$	1.50	^{+	0.03	}_{-	0.03	} ~ \times ~ 10^{	-1	}$	&	$	1.38	^{+	0.01	}_{-	0.04	}$$^{+	0.01	}_{-	0.00	}$$^{+	0.02	}_{-	0.04	} ~ \times ~ 10^{	-1	}$	&	$	1.36	^{+	0.01	}_{-	0.00	}$$^{+	0.01	}_{-	0.00	}$$^{+	0.01	}_{-	0.00	} ~ \times ~ 10^{	-1	}$	\\	\hline
	[	615	 ; 	750	]	&	$	4.94	^{+	0.24	}_{-	0.21	} ~ \times ~ 10^{	-2	}$	&	$	3.98	^{+	0.11	}_{-	0.33	}$$^{+	0.04	}_{-	0.01	}$$^{+	0.12	}_{-	0.33	} ~ \times ~ 10^{	-2	}$	&	$	3.90	^{+	0.05	}_{-	0.00	}$$^{+	0.03	}_{-	0.02	}$$^{+	0.06	}_{-	0.02	} ~ \times ~ 10^{	-2	}$	\\	\hline
	[750	 ; 	1200]	&	$	1.39	^{+	0.12	}_{-	0.10	} ~ \times ~ 10^{	-2	}$	&	$	9.28	^{+	0.49	}_{-	1.71	}$$^{+	0.14	}_{-	0.10	}$$^{+	0.51	}_{-	1.72	} ~ \times ~ 10^{	-3	}$	&	$	9.08	^{+	0.47	}_{-	0.01	}$$^{+	0.12	}_{-	0.11	}$$^{+	0.48	}_{-	0.11	} ~ \times ~ 10^{	-3	}$	\\	\hline
	[	1200	 ; 	$\infty$	]	&	$	4.25	^{+	0.91	}_{-	0.68	} ~ \times ~ 10^{	-5	}$	&	$	1.61	^{+	0.31	}_{-	1.19	} ~ \times ~ 10^{	-5	}$	&	$	1.82	^{+	0.58	}_{-	0.11	} ~ \times ~ 10^{	-5	}$	\\	\hline
\end{tabular}
\caption{\label{tab:Mtt} The $\Mtt$ differential distribution in LO, NLO and NNLO QCD. The format is $central\pm scales\pm pdf\pm total$. At LO, as well as for the last bin, only the scale error is given. We stress that the normalisation is {\it per bin} and thus differs from the one shown in fig.~\ref{fig:Mtt} (to convert between the two one needs to divide by the bin width). The MC error (not shown) is estimated in sec.~\ref{sec:Mtt}.}
\end{center}
\end{table}
%

\begin{table}[h]
\begin{center}
\begin{tabular}{| c | c | c | c |}
\hline
$\PTt$ & \multicolumn{3}{|c|}{$d\sigma/d\PTt$ [pb/bin]} \\
\hline 
 & {\rm LO} & {\rm NLO} & {\rm NNLO} \\ \hline
	[	0	 ; 	45	]	&	$	1.07	^{+	0.45	}_{-	0.29	} ~ \times ~ 10^{	0	}$	&	$	1.15	^{+	0.09	}_{-	0.13	}$$^{+	0.03	}_{-	0.02	}$$^{+	0.09	}_{-	0.14	} ~ \times ~ 10^{	0	}$	&	$	1.23	^{+	0.06	}_{-	0.08	}$$^{+	0.03	}_{-	0.02	}$$^{+	0.07	}_{-	0.08	} ~ \times ~ 10^{	0	}$	\\	\hline
	[	45	 ; 	90	]	&	$	2.13	^{+	0.91	}_{-	0.59	} ~ \times ~ 10^{	0	}$	&	$	2.27	^{+	0.17	}_{-	0.27	}$$^{+	0.05	}_{-	0.05	}$$^{+	0.18	}_{-	0.27	} ~ \times ~ 10^{	0	}$	&	$	2.39	^{+	0.10	}_{-	0.13	}$$^{+	0.06	}_{-	0.04	}$$^{+	0.12	}_{-	0.14	} ~ \times ~ 10^{	0	}$	\\	\hline
	[	90	 ; 	140]	&	$	1.85	^{+	0.82	}_{-	0.52	} ~ \times ~ 10^{	0	}$	&	$	1.88	^{+	0.11	}_{-	0.21	}$$^{+	0.05	}_{-	0.04	}$$^{+	0.12	}_{-	0.22	} ~ \times ~ 10^{	0	}$	&	$	1.97	^{+	0.06	}_{-	0.10	}$$^{+	0.05	}_{-	0.04	}$$^{+	0.08	}_{-	0.11	} ~ \times ~ 10^{	0	}$	\\	\hline
	[140	 ; 	200]	&	$	1.05	^{+	0.49	}_{-	0.31	} ~ \times ~ 10^{	0	}$	&	$	9.81	^{+	0.32	}_{-	1.02	}$$^{+	0.35	}_{-	0.14	}$$^{+	0.47	}_{-	1.03	} ~ \times ~ 10^{	-1	}$	&	$	1.01	^{+	0.03	}_{-	0.05	}$$^{+	0.03	}_{-	0.02	}$$^{+	0.04	}_{-	0.05	} ~ \times ~ 10^{	0	}$	\\	\hline
	[200	 ; 	300]	&	$	4.51	^{+	2.23	}_{-	1.37	} ~ \times ~ 10^{	-1	}$	&	$	3.67	^{+	0.07	}_{-	0.31	}$$^{+	0.13	}_{-	0.07	}$$^{+	0.14	}_{-	0.32	} ~ \times ~ 10^{	-1	}$	&	$	3.65	^{+	0.08	}_{-	0.09	}$$^{+	0.11	}_{-	0.08	}$$^{+	0.14	}_{-	0.12	} ~ \times ~ 10^{	-1	}$	\\	\hline
	[300	 ; 	500]	&	$	6.59	^{+	3.68	}_{-	2.18	} ~ \times ~ 10^{	-2	}$	&	$	4.20	^{+	0.00	}_{-	0.81	}$$^{+	0.12	}_{-	0.15	}$$^{+	0.12	}_{-	0.82	} ~ \times ~ 10^{	-2	}$	&	$	4.07	^{+	0.23	}_{-	0.11	}$$^{+	0.14	}_{-	0.10	}$$^{+	0.27	}_{-	0.15	} ~ \times ~ 10^{	-2	}$	\\	\hline
	[500	 ; 	$\infty$ ]	&	$	5.54	^{+	3.88	}_{-	2.11	} ~ \times ~ 10^{	-4	}$	&	$	2.21	^{+	0.18	}_{-	1.51	} ~ \times ~ 10^{	-4	}$	&	$	2.25	^{+	0.54	}_{-	0.07	} ~ \times ~ 10^{	-4	}$	\\	\hline 
%
%
\hline
$\PTt$ & \multicolumn{3}{|c|}{$(1/\sigma)d\sigma/d\PTt$ [1/bin]} \\
\hline 
 & {\rm LO} & {\rm NLO} & {\rm NNLO} \\ \hline
	[	0	 ; 	45	]	&	$	1.62	^{+	0.02	}_{-	0.02	} ~ \times ~ 10^{	-1	}$	&	$	1.71	^{+	0.03	}_{-	0.01	}$$^{+	0.00	}_{-	0.01	}$$^{+	0.03	}_{-	0.01	} ~ \times ~ 10^{	-1	}$	&	$	1.76	^{+	0.02	}_{-	0.02	}$$^{+	0.00	}_{-	0.01	}$$^{+	0.02	}_{-	0.02	} ~ \times ~ 10^{	-1	}$	\\	\hline
	[	45	 ; 	90	]	&	$	3.22	^{+	0.03	}_{-	0.03	} ~ \times ~ 10^{	-1	}$	&	$	3.40	^{+	0.07	}_{-	0.02	}$$^{+	0.00	}_{-	0.02	}$$^{+	0.07	}_{-	0.03	} ~ \times ~ 10^{	-1	}$	&	$	3.41	^{+	0.02	}_{-	0.01	}$$^{+	0.01	}_{-	0.01	}$$^{+	0.02	}_{-	0.01	} ~ \times ~ 10^{	-1	}$	\\	\hline
	[	90	 ; 	140]	&	$	2.79	^{+	0.00	}_{-	0.00	} ~ \times ~ 10^{	-1	}$	&	$	2.81	^{+	0.01	}_{-	0.00	}$$^{+	0.00	}_{-	0.00	}$$^{+	0.01	}_{-	0.01	} ~ \times ~ 10^{	-1	}$	&	$	2.81	^{+	0.00	}_{-	0.01	}$$^{+	0.00	}_{-	0.00	}$$^{+	0.01	}_{-	0.01	} ~ \times ~ 10^{	-1	}$	\\	\hline
	[140	 ; 	200]	&	$	1.59	^{+	0.02	}_{-	0.02	} ~ \times ~ 10^{	-1	}$	&	$	1.47	^{+	0.02	}_{-	0.04	}$$^{+	0.02	}_{-	0.00	}$$^{+	0.02	}_{-	0.04	} ~ \times ~ 10^{	-1	}$	&	$	1.44	^{+	0.01	}_{-	0.01	}$$^{+	0.01	}_{-	0.00	}$$^{+	0.01	}_{-	0.01	} ~ \times ~ 10^{	-1	}$	\\	\hline
	[200	 ; 	300]	&	$	6.81	^{+	0.25	}_{-	0.22	} ~ \times ~ 10^{	-2	}$	&	$	5.49	^{+	0.17	}_{-	0.50	}$$^{+	0.07	}_{-	0.03	}$$^{+	0.18	}_{-	0.50	} ~ \times ~ 10^{	-2	}$	&	$	5.20	^{+	0.15	}_{-	0.08	}$$^{+	0.05	}_{-	0.04	}$$^{+	0.16	}_{-	0.09	} ~ \times ~ 10^{	-2	}$	\\	\hline
	[300	 ; 	500]	&	$	9.96	^{+	0.79	}_{-	0.67	} ~ \times ~ 10^{	-3	}$	&	$	6.29	^{+	0.47	}_{-	1.46	}$$^{+	0.06	}_{-	0.16	}$$^{+	0.48	}_{-	1.47	} ~ \times ~ 10^{	-3	}$	&	$	5.81	^{+	0.16	}_{-	0.06	}$$^{+	0.09	}_{-	0.08	}$$^{+	0.18	}_{-	0.10	} ~ \times ~ 10^{	-3	}$	\\	\hline
	[500	 ; 	$\infty$ ]	&	$	8.38	^{+	1.49	}_{-	1.15	} ~ \times ~ 10^{	-5	}$	&	$	3.30	^{+	0.66	}_{-	2.32	} ~ \times ~ 10^{	-5	}$	&	$	3.21	^{+	0.62	}_{-	0.03	} ~ \times ~ 10^{	-5	}$	\\	\hline 
\end{tabular}
\caption{\label{tab:PTt} The $\PTt$ differential distribution in LO, NLO and NNLO QCD. The format is $central\pm scales\pm pdf\pm total$. At LO, as well as for the last bin, only the scale error is shown. The normalisation is {\it per bin} and thus differs from the one shown in fig.~\ref{fig:PTt} (to convert between the two one needs to divide by the bin width). The MC error (not shown) is estimated in sec.~\ref{sec:PTt}.}
\end{center}
\end{table}
%

\begin{table}[h]
\begin{center}
\begin{tabular}{| c | c | c | c |}
\hline
$\modyt$ & \multicolumn{3}{|c|}{$d\sigma/d\modyt$ [pb/bin]} \\
\hline 
 & {\rm LO} & {\rm NLO} & {\rm NNLO} \\ \hline
	[	0	 ; 	0.25	]	&	$	1.94	^{+	0.84	}_{-	0.54	} ~ \times ~ 10^{	0	}$	&	$	1.96	^{+	0.09	}_{-	0.21	}$$^{+	0.06	}_{-	0.03	}$$^{+	0.11	}_{-	0.21	} ~ \times ~ 10^{	0	}$	&	$	2.04	^{+	0.06	}_{-	0.10	}$$^{+	0.05	}_{-	0.04	}$$^{+	0.08	}_{-	0.10	} ~ \times ~ 10^{	0	}$	\\	\hline
	[0.25	 ; 	0.5]	&	$	1.72	^{+	0.75	}_{-	0.48	} ~ \times ~ 10^{	0	}$	&	$	1.74	^{+	0.08	}_{-	0.19	}$$^{+	0.04	}_{-	0.04	}$$^{+	0.09	}_{-	0.19	} ~ \times ~ 10^{	0	}$	&	$	1.82	^{+	0.06	}_{-	0.09	}$$^{+	0.05	}_{-	0.03	}$$^{+	0.08	}_{-	0.10	} ~ \times ~ 10^{	0	}$	\\	\hline
	[0.5	 ; 	0.75]	&	$	1.33	^{+	0.59	}_{-	0.37	} ~ \times ~ 10^{	0	}$	&	$	1.35	^{+	0.07	}_{-	0.15	}$$^{+	0.04	}_{-	0.03	}$$^{+	0.08	}_{-	0.15	} ~ \times ~ 10^{	0	}$	&	$	1.41	^{+	0.05	}_{-	0.07	}$$^{+	0.04	}_{-	0.02	}$$^{+	0.06	}_{-	0.08	} ~ \times ~ 10^{	0	}$	\\	\hline
	[	0.75	 ; 	1	]	&	$	8.83	^{+	4.01	}_{-	2.53	} ~ \times ~ 10^{	-1	}$	&	$	8.93	^{+	0.55	}_{-	1.06	}$$^{+	0.23	}_{-	0.18	}$$^{+	0.59	}_{-	1.07	} ~ \times ~ 10^{	-1	}$	&	$	9.35	^{+	0.37	}_{-	0.54	}$$^{+	0.23	}_{-	0.16	}$$^{+	0.44	}_{-	0.56	} ~ \times ~ 10^{	-1	}$	\\	\hline
	[	1	 ; 	1.25	]	&	$	4.79	^{+	2.25	}_{-	1.40	} ~ \times ~ 10^{	-1	}$	&	$	4.84	^{+	0.35	}_{-	0.61	}$$^{+	0.16	}_{-	0.07	}$$^{+	0.39	}_{-	0.62	} ~ \times ~ 10^{	-1	}$	&	$	5.17	^{+	0.25	}_{-	0.34	}$$^{+	0.12	}_{-	0.09	}$$^{+	0.28	}_{-	0.35	} ~ \times ~ 10^{	-1	}$	\\	\hline
	[1.25	 ; 	1.5]	&	$	1.97	^{+	0.97	}_{-	0.60	} ~ \times ~ 10^{	-1	}$	&	$	1.99	^{+	0.18	}_{-	0.28	}$$^{+	0.05	}_{-	0.05	}$$^{+	0.18	}_{-	0.28	} ~ \times ~ 10^{	-1	}$	&	$	2.15	^{+	0.15	}_{-	0.16	}$$^{+	0.05	}_{-	0.05	}$$^{+	0.16	}_{-	0.16	} ~ \times ~ 10^{	-1	}$	\\	\hline
	[	1.5	 ; 	$\infty$	]	&	$	6.18	^{+	3.27	}_{-	1.96	} ~ \times ~ 10^{	-2	}$	&	$	6.22	^{+	0.69	}_{-	0.96	}$$^{+	0.23	}_{-	0.16	}$$^{+	0.73	}_{-	0.98	} ~ \times ~ 10^{	-2	}$	&	$	6.67	^{+	0.58	}_{-	0.60	}$$^{+	0.22	}_{-	0.19	}$$^{+	0.62	}_{-	0.63	} ~ \times ~ 10^{	-2	}$	\\	\hline
%
%
\hline
$\modyt$ & \multicolumn{3}{|c|}{$(1/\sigma)d\sigma/d\modyt$ [1/bin]} \\
\hline 
 & {\rm LO} & {\rm NLO} & {\rm NNLO} \\ \hline
	[	0	 ; 	0.25	]	&	$	2.94	^{+	0.02	}_{-	0.02	} ~ \times ~ 10^{	-1	}$	&	$	2.93	^{+	0.02	}_{-	0.02	}$$^{+	0.02	}_{-	0.00	}$$^{+	0.03	}_{-	0.02	} ~ \times ~ 10^{	-1	}$	&	$	2.91	^{+	0.02	}_{-	0.02	}$$^{+	0.01	}_{-	0.01	}$$^{+	0.02	}_{-	0.02	} ~ \times ~ 10^{	-1	}$	\\	\hline
	[0.25	 ; 	0.5]	&	$	2.60	^{+	0.01	}_{-	0.01	} ~ \times ~ 10^{	-1	}$	&	$	2.60	^{+	0.01	}_{-	0.01	}$$^{+	0.00	}_{-	0.02	}$$^{+	0.01	}_{-	0.02	} ~ \times ~ 10^{	-1	}$	&	$	2.61	^{+	0.01	}_{-	0.01	}$$^{+	0.01	}_{-	0.00	}$$^{+	0.01	}_{-	0.01	} ~ \times ~ 10^{	-1	}$	\\	\hline
	[0.5	 ; 	0.75]	&	$	2.01	^{+	0.00	}_{-	0.00	} ~ \times ~ 10^{	-1	}$	&	$	2.02	^{+	0.00	}_{-	0.00	}$$^{+	0.00	}_{-	0.00	}$$^{+	0.00	}_{-	0.00	} ~ \times ~ 10^{	-1	}$	&	$	2.01	^{+	0.00	}_{-	0.00	}$$^{+	0.00	}_{-	0.00	}$$^{+	0.00	}_{-	0.00	} ~ \times ~ 10^{	-1	}$	\\	\hline
	[	0.75	 ; 	1	]	&	$	1.33	^{+	0.01	}_{-	0.01	} ~ \times ~ 10^{	-1	}$	&	$	1.34	^{+	0.01	}_{-	0.01	}$$^{+	0.00	}_{-	0.01	}$$^{+	0.01	}_{-	0.01	} ~ \times ~ 10^{	-1	}$	&	$	1.34	^{+	0.00	}_{-	0.01	}$$^{+	0.00	}_{-	0.00	}$$^{+	0.00	}_{-	0.01	} ~ \times ~ 10^{	-1	}$	\\	\hline
	[	1	 ; 	1.25	]	&	$	7.24	^{+	0.14	}_{-	0.12	} ~ \times ~ 10^{	-2	}$	&	$	7.25	^{+	0.13	}_{-	0.11	}$$^{+	0.09	}_{-	0.02	}$$^{+	0.16	}_{-	0.11	} ~ \times ~ 10^{	-2	}$	&	$	7.38	^{+	0.08	}_{-	0.10	}$$^{+	0.04	}_{-	0.06	}$$^{+	0.09	}_{-	0.11	} ~ \times ~ 10^{	-2	}$	\\	\hline
	[1.25	 ; 	1.5]	&	$	2.98	^{+	0.10	}_{-	0.09	} ~ \times ~ 10^{	-2	}$	&	$	2.98	^{+	0.10	}_{-	0.09	}$$^{+	0.02	}_{-	0.06	}$$^{+	0.10	}_{-	0.10	} ~ \times ~ 10^{	-2	}$	&	$	3.07	^{+	0.09	}_{-	0.07	}$$^{+	0.04	}_{-	0.05	}$$^{+	0.10	}_{-	0.08	} ~ \times ~ 10^{	-2	}$	\\	\hline
	[	1.5	 ; 	$\infty$	]	&	$	9.33	^{+	0.56	}_{-	0.47	} ~ \times ~ 10^{	-3	}$	&	$	9.31	^{+	0.51	}_{-	0.44	}$$^{+	0.20	}_{-	0.23	}$$^{+	0.55	}_{-	0.50	} ~ \times ~ 10^{	-3	}$	&	$	9.52	^{+	0.46	}_{-	0.37	}$$^{+	0.23	}_{-	0.27	}$$^{+	0.51	}_{-	0.45	} ~ \times ~ 10^{	-3	}$	\\	\hline
\end{tabular}
\caption{\label{tab:mody} The $\modyt$ differential distribution in LO, NLO and NNLO QCD. The format is $central\pm scales\pm pdf\pm total$. At LO, as well as for the last bin, only the scale error is given. The normalisation is {\it per bin} and thus differs from the one shown in fig.~\ref{fig:mody} (to convert between the two one needs to divide by the bin width). The MC error (not shown) is estimated in sec.~\ref{sec:mody}.}
\end{center}
\end{table}
%

\begin{table}[h]
\begin{center}
\begin{tabular}{| c | c | c | c |}
\hline
$\cos\theta$ & \multicolumn{3}{|c|}{$(1/\sigma)d\sigma/d\cos\theta$ [1/bin]} \\
\hline 
 & {\rm LO} & {\rm NLO} & {\rm NNLO} \\ \hline
	[	-1	 ; 	-0.8	]	&	$	1.13	^{+	0.01	}_{-	0.00	} ~ \times ~ 10^{	-1	}$	&	$	1.08	^{+	0.01	}_{-	0.02	} ~ \times ~ 10^{	-1	}$	&	$	1.08	^{+	0.01	}_{-	0.00	} ~ \times ~ 10^{	-1	}$	\\	\hline
	[	-0.8	 ; 	-0.6	]	&	$	1.04	^{+	0.00	}_{-	0.00	} ~ \times ~ 10^{	-1	}$	&	$	9.69	^{+	0.16	}_{-	0.31	} ~ \times ~ 10^{	-2	}$	&	$	9.57	^{+	0.08	}_{-	0.03	} ~ \times ~ 10^{	-2	}$	\\	\hline
	[	-0.6	 ; 	-0.4	]	&	$	9.77	^{+	0.01	}_{-	0.01	} ~ \times ~ 10^{	-2	}$	&	$	9.09	^{+	0.17	}_{-	0.33	} ~ \times ~ 10^{	-2	}$	&	$	8.84	^{+	0.12	}_{-	0.11	} ~ \times ~ 10^{	-2	}$	\\	\hline
	[	-0.4	 ; 	-0.2	]	&	$	9.38	^{+	0.02	}_{-	0.03	} ~ \times ~ 10^{	-2	}$	&	$	8.78	^{+	0.16	}_{-	0.30	} ~ \times ~ 10^{	-2	}$	&	$	8.55	^{+	0.13	}_{-	0.08	} ~ \times ~ 10^{	-2	}$	\\	\hline
	[	-0.2	 ; 	0	]	&	$	9.18	^{+	0.03	}_{-	0.04	} ~ \times ~ 10^{	-2	}$	&	$	8.72	^{+	0.13	}_{-	0.24	} ~ \times ~ 10^{	-2	}$	&	$	8.49	^{+	0.13	}_{-	0.13	} ~ \times ~ 10^{	-2	}$	\\	\hline
	[	0	 ; 	0.2	]	&	$	9.18	^{+	0.03	}_{-	0.04	} ~ \times ~ 10^{	-2	}$	&	$	8.90	^{+	0.09	}_{-	0.15	} ~ \times ~ 10^{	-2	}$	&	$	8.75	^{+	0.09	}_{-	0.10	} ~ \times ~ 10^{	-2	}$	\\	\hline
	[	0.2	 ; 	0.4	]	&	$	9.38	^{+	0.02	}_{-	0.03	} ~ \times ~ 10^{	-2	}$	&	$	9.35	^{+	0.03	}_{-	0.03	} ~ \times ~ 10^{	-2	}$	&	$	9.24	^{+	0.05	}_{-	0.07	} ~ \times ~ 10^{	-2	}$	\\	\hline
	[	0.4	 ; 	0.6	]	&	$	9.77	^{+	0.01	}_{-	0.01	} ~ \times ~ 10^{	-2	}$	&	$	1.01	^{+	0.01	}_{-	0.01	} ~ \times ~ 10^{	-1	}$	&	$	1.02	^{+	0.00	}_{-	0.01	} ~ \times ~ 10^{	-1	}$	\\	\hline
	[	0.6	 ; 	0.8	]	&	$	1.04	^{+	0.00	}_{-	0.00	} ~ \times ~ 10^{	-1	}$	&	$	1.13	^{+	0.04	}_{-	0.02	} ~ \times ~ 10^{	-1	}$	&	$	1.16	^{+	0.01	}_{-	0.01	} ~ \times ~ 10^{	-1	}$	\\	\hline
	[	0.8	 ; 	1	]	&	$	1.13	^{+	0.01	}_{-	0.00	} ~ \times ~ 10^{	-1	}$	&	$	1.33	^{+	0.10	}_{-	0.05	} ~ \times ~ 10^{	-1	}$	&	$	1.40	^{+	0.04	}_{-	0.04	} ~ \times ~ 10^{	-1	}$	\\	\hline
\end{tabular}
\caption{\label{tab:costheta} Normalised $\cos\theta$ differential distribution in LO, NLO and NNLO QCD shown in fig.~\ref{fig:costheta}(left) (same normalisation). The format is $central\pm scales$.}
\end{center}
\end{table}
%
\begin{table}[h]
\begin{center}
\begin{tabular}{| c | c | c | c |}
\hline
{\rm Moment} & \multicolumn{3}{|c|}{Legendre Moment} \\
\hline 
 & {\rm LO} & {\rm NLO} & {\rm NNLO} \\ \hline
	1	&	$	-4.82	^{+	0.12	}_{-	0.14	} ~ \times ~ 10^{	-5	}$	&	$	1.24	^{+	0.57	}_{-	0.29	} ~ \times ~ 10^{	-1	}$	&	$	1.59	^{+	0.11	}_{-	0.19	} ~ \times ~ 10^{	-1	}$	\\	\hline
	2	&	$	1.75	^{+	0.08	}_{-	0.06	} ~ \times ~ 10^{	-1	}$	&	$	2.68	^{+	0.49	}_{-	0.28	} ~ \times ~ 10^{	-1	}$	&	$	3.17	^{+	0.29	}_{-	0.28	} ~ \times ~ 10^{	-1	}$	\\	\hline
	3	&	$	2.10	^{+	0.00	}_{-	0.04	} ~ \times ~ 10^{	-5	}$	&	$	2.45	^{+	1.17	}_{-	0.60	} ~ \times ~ 10^{	-2	}$	&	$	3.96	^{+	0.74	}_{-	0.67	} ~ \times ~ 10^{	-2	}$	\\	\hline
	4	&	$	7.47	^{+	1.68	}_{-	1.22	} ~ \times ~ 10^{	-3	}$	&	$	3.34	^{+	1.45	}_{-	0.78	} ~ \times ~ 10^{	-2	}$	&	$	5.12	^{+	1.11	}_{-	0.93	} ~ \times ~ 10^{	-2	}$	\\	\hline
	5	&	$	-1.63	^{+	0.39	}_{-	0.35	} ~ \times ~ 10^{	-5	}$	&	$	3.25	^{+	1.58	}_{-	0.76	} ~ \times ~ 10^{	-3	}$	&	$	6.88	^{+	2.14	}_{-	1.48	} ~ \times ~ 10^{	-3	}$	\\	\hline
	6	&	$	1.27	^{+	0.33	}_{-	0.23	} ~ \times ~ 10^{	-3	}$	&	$	6.35	^{+	3.06	}_{-	1.67	} ~ \times ~ 10^{	-3	}$	&	$	1.08	^{+	0.33	}_{-	0.26	} ~ \times ~ 10^{	-2	}$	\\	\hline
	7	&	$	3.70	^{+	2.00	}_{-	2.37	} ~ \times ~ 10^{	-6	}$	&	$	9.69	^{+	4.17	}_{-	2.39	} ~ \times ~ 10^{	-4	}$	&	$	1.81	^{+	0.80	}_{-	0.48	} ~ \times ~ 10^{	-3	}$	\\	\hline
	8	&	$	5.15	^{+	8.31	}_{-	5.66	} ~ \times ~ 10^{	-5	}$	&	$	1.47	^{+	0.79	}_{-	0.39	} ~ \times ~ 10^{	-3	}$	&	$	3.87	^{+	0.86	}_{-	0.76	} ~ \times ~ 10^{	-3	}$	\\	\hline
\end{tabular}
\caption{\label{tab:legmom} First eight Legendre moments $a_i$ in LO, NLO and NNLO QCD shown in fig.~\ref{fig:costheta}(right) (same normalisation). The format is $central\pm scales$.}
\end{center}
\end{table}
%


\begin{table}[h]
\begin{center}
\begin{tabular}{| c | c | c | c |}
\hline
$\dy$ & \multicolumn{3}{|c|}{$d\sigma/d\dy$ [pb/bin]} \\
\hline 
 & {\rm LO} & {\rm NLO} & {\rm NNLO} \\ \hline
	[	-2	 ; 	-1.5	]	&	$	2.26	^{+	1.13	}_{-	0.69	} ~ \times ~ 10^{	-1	}$	&	$	1.90	^{+	0.06	}_{-	0.18	}$$^{+	0.06	}_{-	0.05	}$$^{+	0.08	}_{-	0.18	} ~ \times ~ 10^{	-1	}$	&	$	1.98	^{+	0.13	}_{-	0.11	}$$^{+	0.05	}_{-	0.05	}$$^{+	0.14	}_{-	0.12	} ~ \times ~ 10^{	-1	}$	\\	\hline
	[	-1.5	 ; 	-1	]	&	$	5.21	^{+	2.40	}_{-	1.51	} ~ \times ~ 10^{	-1	}$	&	$	4.74	^{+	0.14	}_{-	0.46	}$$^{+	0.13	}_{-	0.10	}$$^{+	0.19	}_{-	0.47	} ~ \times ~ 10^{	-1	}$	&	$	4.87	^{+	0.20	}_{-	0.22	}$$^{+	0.12	}_{-	0.08	}$$^{+	0.23	}_{-	0.24	} ~ \times ~ 10^{	-1	}$	\\	\hline
	[	-1	 ; 	-0.5	]	&	$	1.06	^{+	0.47	}_{-	0.30	} ~ \times ~ 10^{	0	}$	&	$	1.00	^{+	0.03	}_{-	0.10	}$$^{+	0.03	}_{-	0.02	}$$^{+	0.04	}_{-	0.10	} ~ \times ~ 10^{	0	}$	&	$	1.03	^{+	0.03	}_{-	0.04	}$$^{+	0.03	}_{-	0.02	}$$^{+	0.04	}_{-	0.05	} ~ \times ~ 10^{	0	}$	\\	\hline
	[	-0.5	 ; 	0	]	&	$	1.51	^{+	0.65	}_{-	0.41	} ~ \times ~ 10^{	0	}$	&	$	1.48	^{+	0.06	}_{-	0.15	}$$^{+	0.04	}_{-	0.03	}$$^{+	0.07	}_{-	0.15	} ~ \times ~ 10^{	0	}$	&	$	1.52	^{+	0.04	}_{-	0.07	}$$^{+	0.04	}_{-	0.03	}$$^{+	0.06	}_{-	0.07	} ~ \times ~ 10^{	0	}$	\\	\hline
	[	0	 ; 	0.5	]	&	$	1.51	^{+	0.65	}_{-	0.41	} ~ \times ~ 10^{	0	}$	&	$	1.54	^{+	0.08	}_{-	0.17	}$$^{+	0.04	}_{-	0.03	}$$^{+	0.09	}_{-	0.17	} ~ \times ~ 10^{	0	}$	&	$	1.61	^{+	0.05	}_{-	0.08	}$$^{+	0.04	}_{-	0.03	}$$^{+	0.06	}_{-	0.08	} ~ \times ~ 10^{	0	}$	\\	\hline
	[	0.5	 ; 	1	]	&	$	1.06	^{+	0.47	}_{-	0.30	} ~ \times ~ 10^{	0	}$	&	$	1.14	^{+	0.09	}_{-	0.14	}$$^{+	0.03	}_{-	0.02	}$$^{+	0.10	}_{-	0.14	} ~ \times ~ 10^{	0	}$	&	$	1.21	^{+	0.05	}_{-	0.07	}$$^{+	0.03	}_{-	0.02	}$$^{+	0.06	}_{-	0.08	} ~ \times ~ 10^{	0	}$	\\	\hline
	[	1	 ; 	1.5	]	&	$	5.20	^{+	2.40	}_{-	1.51	} ~ \times ~ 10^{	-1	}$	&	$	5.89	^{+	0.65	}_{-	0.83	}$$^{+	0.15	}_{-	0.13	}$$^{+	0.67	}_{-	0.84	} ~ \times ~ 10^{	-1	}$	&	$	6.42	^{+	0.35	}_{-	0.48	}$$^{+	0.16	}_{-	0.12	}$$^{+	0.39	}_{-	0.49	} ~ \times ~ 10^{	-1	}$	\\	\hline
	[	1.5	 ; 	2	]	&	$	2.26	^{+	1.13	}_{-	0.69	} ~ \times ~ 10^{	-1	}$	&	$	2.67	^{+	0.40	}_{-	0.44	}$$^{+	0.08	}_{-	0.05	}$$^{+	0.41	}_{-	0.44	} ~ \times ~ 10^{	-1	}$	&	$	3.01	^{+	0.24	}_{-	0.29	}$$^{+	0.08	}_{-	0.06	}$$^{+	0.26	}_{-	0.30	} ~ \times ~ 10^{	-1	}$	\\	\hline
\end{tabular}
\caption{\label{tab:diff-y} Differential distribution in the rapidity difference $\dy$ between $t$ and $\bar t$ in LO, NLO and NNLO QCD. The format is $central\pm scales\pm pdf\pm total$. At LO only the scale error is given. The end-bins contain overflow events.}
\end{center}
\end{table}
%
\begin{table}[h]
\begin{center}
\begin{tabular}{| c | c | c |}
\hline
$\dmody$ & \multicolumn{2}{|c|}{$\AFB(\dmody)$} \\
\hline 
 & {\rm NLO} & {\rm NNLO} \\ \hline
	[	0	 ; 	0.5	]	&	$	2.14	^{+	0.96	}_{-	0.50	}$$^{+	0.06	}_{-	0.02	}$$^{+	0.97	}_{-	0.50	} ~ \times ~ 10^{	-2	}$	&	$	2.76	^{+	0.19	}_{-	0.31	}$$^{+	0.09	}_{-	0.01	}$$^{+	0.21	}_{-	0.31	} ~ \times ~ 10^{	-2	}$	\\	\hline
	[	0.5	 ; 	1	]	&	$	6.41	^{+	2.84	}_{-	1.47	}$$^{+	0.07	}_{-	0.24	}$$^{+	2.84	}_{-	1.49	} ~ \times ~ 10^{	-2	}$	&	$	8.06	^{+	0.59	}_{-	0.92	}$$^{+	0.10	}_{-	0.16	}$$^{+	0.59	}_{-	0.93	} ~ \times ~ 10^{	-2	}$	\\	\hline
	[	1	 ; 	1.5	]	&	$	1.08	^{+	0.48	}_{-	0.25	}$$^{+	0.03	}_{-	0.03	}$$^{+	0.48	}_{-	0.25	} ~ \times ~ 10^{	-1	}$	&	$	1.37	^{+	0.08	}_{-	0.15	}$$^{+	0.02	}_{-	0.03	}$$^{+	0.08	}_{-	0.15	} ~ \times ~ 10^{	-1	}$	\\	\hline
	[	1.5	 ; 	2	]	&	$	1.69	^{+	0.79	}_{-	0.40	}$$^{+	0.07	}_{-	0.04	}$$^{+	0.79	}_{-	0.40	} ~ \times ~ 10^{	-1	}$	&	$	2.07	^{+	0.08	}_{-	0.21	}$$^{+	0.06	}_{-	0.04	}$$^{+	0.10	}_{-	0.21	} ~ \times ~ 10^{	-1	}$	\\	\hline
\end{tabular}
\caption{\label{tab:afb-y} $\dmody$-dependent $\AFB$ in LO, NLO and NNLO QCD. The format is $central\pm scales\pm pdf\pm total$. At LO only the scale error is given. The highest bin contains overflow events.}
\end{center}
\end{table}
%

\begin{table}[h]
\begin{center}
\begin{tabular}{| c | c | c | c |}
\hline
$\Mtt{\dy\over \dmody}$[{\rm GeV}] & \multicolumn{3}{|c|}{$d^2\sigma/d\dy d\Mtt$ [pb/bin]} \\
\hline 
 & {\rm LO} & {\rm NLO} & {\rm NNLO} \\ \hline
	[	-$\infty$	 ; 	-650	]	&	$	1.42	^{+	0.76	}_{-	0.46	} ~ \times ~ 10^{	-1	}$	&	$	9.04	^{+	0.03	}_{-	1.85	}$$^{+	0.30	}_{-	0.20	}$$^{+	0.30	}_{-	1.86	} ~ \times ~ 10^{	-2	}$	&	$	9.01	^{+	0.71	}_{-	0.24	}$$^{+	0.24	}_{-	0.22	}$$^{+	0.75	}_{-	0.32	} ~ \times ~ 10^{	-2	}$	\\	\hline
	[	-650	 ; 	-550	]	&	$	2.88	^{+	1.41	}_{-	0.87	} ~ \times ~ 10^{	-1	}$	&	$	2.27	^{+	0.03	}_{-	0.17	}$$^{+	0.06	}_{-	0.06	}$$^{+	0.07	}_{-	0.18	} ~ \times ~ 10^{	-1	}$	&	$	2.27	^{+	0.08	}_{-	0.07	}$$^{+	0.06	}_{-	0.05	}$$^{+	0.10	}_{-	0.09	} ~ \times ~ 10^{	-1	}$	\\	\hline
	[	-550	 ; 	-450	]	&	$	8.60	^{+	3.93	}_{-	2.47	} ~ \times ~ 10^{	-1	}$	&	$	7.78	^{+	0.22	}_{-	0.74	}$$^{+	0.22	}_{-	0.15	}$$^{+	0.31	}_{-	0.75	} ~ \times ~ 10^{	-1	}$	&	$	7.88	^{+	0.21	}_{-	0.31	}$$^{+	0.21	}_{-	0.14	}$$^{+	0.30	}_{-	0.34	} ~ \times ~ 10^{	-1	}$	\\	\hline
	[	-450	 ; 	-350	]	&	$	1.99	^{+	0.84	}_{-	0.54	} ~ \times ~ 10^{	0	}$	&	$	2.02	^{+	0.10	}_{-	0.22	}$$^{+	0.05	}_{-	0.03	}$$^{+	0.12	}_{-	0.22	} ~ \times ~ 10^{	0	}$	&	$	2.10	^{+	0.06	}_{-	0.10	}$$^{+	0.05	}_{-	0.04	}$$^{+	0.08	}_{-	0.11	} ~ \times ~ 10^{	0	}$	\\	\hline
	[	-350	 ; 	-250	]	&	$	2.93	^{+	1.18	}_{-	0.77	} ~ \times ~ 10^{	-2	}$	&	$	2.77	^{+	0.08	}_{-	0.24	}$$^{+	0.08	}_{-	0.04	}$$^{+	0.11	}_{-	0.24	} ~ \times ~ 10^{	-2	}$	&	$	2.98	^{+	0.18	}_{-	0.18	}$$^{+	0.07	}_{-	0.04	}$$^{+	0.19	}_{-	0.18	} ~ \times ~ 10^{	-2	}$	\\	\hline
	[	250	 ; 	350	]	&	$	2.93	^{+	1.18	}_{-	0.77	} ~ \times ~ 10^{	-2	}$	&	$	2.83	^{+	0.08	}_{-	0.26	}$$^{+	0.05	}_{-	0.07	}$$^{+	0.10	}_{-	0.26	} ~ \times ~ 10^{	-2	}$	&	$	2.97	^{+	0.18	}_{-	0.16	}$$^{+	0.07	}_{-	0.04	}$$^{+	0.20	}_{-	0.17	} ~ \times ~ 10^{	-2	}$	\\	\hline
	[	350	 ; 	450	]	&	$	1.99	^{+	0.84	}_{-	0.54	} ~ \times ~ 10^{	0	}$	&	$	2.20	^{+	0.19	}_{-	0.27	}$$^{+	0.06	}_{-	0.04	}$$^{+	0.20	}_{-	0.27	} ~ \times ~ 10^{	0	}$	&	$	2.35	^{+	0.10	}_{-	0.14	}$$^{+	0.06	}_{-	0.04	}$$^{+	0.11	}_{-	0.15	} ~ \times ~ 10^{	0	}$	\\	\hline
	[	450	 ; 	550	]	&	$	8.59	^{+	3.93	}_{-	2.47	} ~ \times ~ 10^{	-1	}$	&	$	9.08	^{+	0.72	}_{-	1.16	}$$^{+	0.23	}_{-	0.22	}$$^{+	0.76	}_{-	1.18	} ~ \times ~ 10^{	-1	}$	&	$	9.62	^{+	0.40	}_{-	0.60	}$$^{+	0.25	}_{-	0.18	}$$^{+	0.47	}_{-	0.63	} ~ \times ~ 10^{	-1	}$	\\	\hline
	[	550	 ; 	650	]	&	$	2.88	^{+	1.41	}_{-	0.87	} ~ \times ~ 10^{	-1	}$	&	$	2.82	^{+	0.17	}_{-	0.36	}$$^{+	0.10	}_{-	0.05	}$$^{+	0.20	}_{-	0.36	} ~ \times ~ 10^{	-1	}$	&	$	2.98	^{+	0.13	}_{-	0.18	}$$^{+	0.08	}_{-	0.06	}$$^{+	0.15	}_{-	0.19	} ~ \times ~ 10^{	-1	}$	\\	\hline
	[	650	 ; 	$\infty$	]	&	$	1.42	^{+	0.76	}_{-	0.46	} ~ \times ~ 10^{	-1	}$	&	$	1.24	^{+	0.05	}_{-	0.15	}$$^{+	0.04	}_{-	0.03	}$$^{+	0.06	}_{-	0.15	} ~ \times ~ 10^{	-1	}$	&	$	1.29	^{+	0.07	}_{-	0.08	}$$^{+	0.04	}_{-	0.03	}$$^{+	0.08	}_{-	0.09	} ~ \times ~ 10^{	-1	}$	\\	\hline
\end{tabular}
\caption{\label{tab:diff-mtt} $\Mtt\times{\rm sign}(\dy)$ differential distribution in LO, NLO and NNLO QCD. The format is $central\pm scales\pm pdf\pm total$. At LO only the scale error is given.}
\end{center}
\end{table}
%
\begin{table}[h]
\begin{center}
\begin{tabular}{| c | c | c |}
\hline
$\Mtt$[{\rm GeV}] & \multicolumn{2}{|c|}{$\AFB(\Mtt)$} \\
\hline 
 & {\rm NLO} & {\rm NNLO} \\ \hline
	[	350	 ; 	450	]	&	$	4.10	^{+	1.66	}_{-	0.90	}$$^{+	0.07	}_{-	0.10	}$$^{+	1.66	}_{-	0.91	} ~ \times ~ 10^{	-2	}$	&	$	5.36	^{+	0.49	}_{-	0.61	}$$^{+	0.09	}_{-	0.07	}$$^{+	0.50	}_{-	0.62	} ~ \times ~ 10^{	-2	}$	\\	\hline
	[	450	 ; 	550	]	&	$	7.71	^{+	3.69	}_{-	1.85	}$$^{+	0.09	}_{-	0.34	}$$^{+	3.69	}_{-	1.88	} ~ \times ~ 10^{	-2	}$	&	$	9.98	^{+	0.68	}_{-	1.23	}$$^{+	0.15	}_{-	0.17	}$$^{+	0.70	}_{-	1.24	} ~ \times ~ 10^{	-2	}$	\\	\hline
	[	550	 ; 	650	]	&	$	1.08	^{+	0.61	}_{-	0.28	}$$^{+	0.09	}_{-	0.00	}$$^{+	0.62	}_{-	0.28	} ~ \times ~ 10^{	-1	}$	&	$	1.34	^{+	0.06	}_{-	0.14	}$$^{+	0.03	}_{-	0.02	}$$^{+	0.06	}_{-	0.14	} ~ \times ~ 10^{	-1	}$	\\	\hline
	[	650	 ; 	750	]	&	$	1.56	^{+	1.19	}_{-	0.46	}$$^{+	0.03	}_{-	0.03	}$$^{+	1.19	}_{-	0.47	} ~ \times ~ 10^{	-1	}$	&	$	1.77	^{+	0.04	}_{-	0.19	}$$^{+	0.03	}_{-	0.02	}$$^{+	0.06	}_{-	0.19	} ~ \times ~ 10^{	-1	}$	\\	\hline
\end{tabular}
\caption{\label{tab:afb-mtt} $\Mtt$ dependent $\AFB$ in NLO and NNLO QCD. The format is $central\pm scales\pm pdf\pm total$. The lowest and highest bins contain spillover events.}
\end{center}
\end{table}
%

\begin{table}[h]
\begin{center}
\begin{tabular}{| c | c | c | c |}
\hline
$\PTtt$ [{\rm GeV}] & \multicolumn{3}{|c|}{$d^2\sigma/d\dy d\PTtt$ [pb/bin]} \\
\hline 
 & {\rm LO} & {\rm NLO} & {\rm NNLO} \\ \hline
	[	-80	 ; 	-70	]	&	$	0$	&	$	1.13	^{+	0.74	}_{-	0.41	}$$^{+	0.04	}_{-	0.03	}$$^{+	0.74	}_{-	0.41	} ~ \times ~ 10^{	-1	}$	&	$	1.39	^{+	0.10	}_{-	0.24	}$$^{+	0.04	}_{-	0.03	}$$^{+	0.11	}_{-	0.24	} ~ \times ~ 10^{	-1	}$	\\	\hline
	[	-70	 ; 	-60	]	&	$	0$	&	$	3.98	^{+	2.49	}_{-	1.41	}$$^{+	0.13	}_{-	0.07	}$$^{+	2.49	}_{-	1.41	} ~ \times ~ 10^{	-2	}$	&	$	5.12	^{+	0.52	}_{-	0.92	}$$^{+	0.13	}_{-	0.11	}$$^{+	0.53	}_{-	0.92	} ~ \times ~ 10^{	-2	}$	\\	\hline
	[	-60	 ; 	-50	]	&	$	0$	&	$	5.81	^{+	3.61	}_{-	2.05	}$$^{+	0.17	}_{-	0.13	}$$^{+	3.61	}_{-	2.06	} ~ \times ~ 10^{	-2	}$	&	$	7.50	^{+	0.77	}_{-	1.33	}$$^{+	0.18	}_{-	0.15	}$$^{+	0.79	}_{-	1.34	} ~ \times ~ 10^{	-2	}$	\\	\hline
	[	-50	 ; 	-40	]	&	$	0$	&	$	8.89	^{+	5.47	}_{-	3.12	}$$^{+	0.28	}_{-	0.17	}$$^{+	5.47	}_{-	3.12	} ~ \times ~ 10^{	-2	}$	&	$	1.15	^{+	0.12	}_{-	0.20	}$$^{+	0.03	}_{-	0.02	}$$^{+	0.12	}_{-	0.21	} ~ \times ~ 10^{	-1	}$	\\	\hline
	[	-40	 ; 	-30	]	&	$	0$	&	$	1.46	^{+	0.89	}_{-	0.51	}$$^{+	0.05	}_{-	0.03	}$$^{+	0.89	}_{-	0.51	} ~ \times ~ 10^{	-1	}$	&	$	1.88	^{+	0.17	}_{-	0.32	}$$^{+	0.04	}_{-	0.04	}$$^{+	0.18	}_{-	0.33	} ~ \times ~ 10^{	-1	}$	\\	\hline
	[	-30	 ; 	-20	]	&	$	0$	&	$	2.71	^{+	1.64	}_{-	0.94	}$$^{+	0.08	}_{-	0.05	}$$^{+	1.64	}_{-	0.94	} ~ \times ~ 10^{	-1	}$	&	$	3.39	^{+	0.27	}_{-	0.55	}$$^{+	0.08	}_{-	0.06	}$$^{+	0.28	}_{-	0.56	} ~ \times ~ 10^{	-1	}$	\\	\hline
	[	-20	 ; 	-10	]	&	$	0$	&	$	6.54	^{+	3.91	}_{-	2.25	}$$^{+	0.18	}_{-	0.13	}$$^{+	3.91	}_{-	2.26	} ~ \times ~ 10^{	-1	}$	&	$	7.48	^{+	0.30	}_{-	1.03	}$$^{+	0.18	}_{-	0.13	}$$^{+	0.35	}_{-	1.04	} ~ \times ~ 10^{	-1	}$	\\	\hline
	[	-10	 ; 	0	]	&	$	3.31	^{+	1.47	}_{-	0.93	} $	&	$	1.77	^{+	0.16	}_{-	0.76	}$$^{+	0.05	}_{-	0.03	}$$^{+	0.17	}_{-	0.76	} ~ \times ~ 10^{	0	}$	&	$	1.59	^{+	0.11	}_{-	0.03	}$$^{+	0.04	}_{-	0.03	}$$^{+	0.12	}_{-	0.04	} ~ \times ~ 10^{	0	}$	\\	\hline
	[	0	 ; 	10	]	&	$	3.31	^{+	1.47	}_{-	0.93	} $	&	$	2.39	^{+	0.04	}_{-	0.42	}$$^{+	0.07	}_{-	0.05	}$$^{+	0.08	}_{-	0.43	} ~ \times ~ 10^{	0	}$	&	$	2.21	^{+	0.06	}_{-	0.05	}$$^{+	0.06	}_{-	0.05	}$$^{+	0.09	}_{-	0.06	} ~ \times ~ 10^{	0	}$	\\	\hline
	[	10	 ; 	20	]	&	$	0$	&	$	5.60	^{+	3.39	}_{-	1.94	}$$^{+	0.18	}_{-	0.11	}$$^{+	3.39	}_{-	1.95	} ~ \times ~ 10^{	-1	}$	&	$	7.35	^{+	0.72	}_{-	1.27	}$$^{+	0.18	}_{-	0.14	}$$^{+	0.75	}_{-	1.27	} ~ \times ~ 10^{	-1	}$	\\	\hline
	[	20	 ; 	30	]	&	$	0$	&	$	2.26	^{+	1.39	}_{-	0.79	}$$^{+	0.07	}_{-	0.05	}$$^{+	1.39	}_{-	0.79	} ~ \times ~ 10^{	-1	}$	&	$	3.17	^{+	0.46	}_{-	0.61	}$$^{+	0.08	}_{-	0.06	}$$^{+	0.46	}_{-	0.61	} ~ \times ~ 10^{	-1	}$	\\	\hline
	[	30	 ; 	40	]	&	$	0$	&	$	1.20	^{+	0.74	}_{-	0.42	}$$^{+	0.04	}_{-	0.03	}$$^{+	0.74	}_{-	0.42	} ~ \times ~ 10^{	-1	}$	&	$	1.70	^{+	0.27	}_{-	0.34	}$$^{+	0.05	}_{-	0.03	}$$^{+	0.27	}_{-	0.34	} ~ \times ~ 10^{	-1	}$	\\	\hline
	[	40	 ; 	50	]	&	$	0$	&	$	7.21	^{+	4.51	}_{-	2.56	}$$^{+	0.24	}_{-	0.15	}$$^{+	4.52	}_{-	2.56	} ~ \times ~ 10^{	-2	}$	&	$	1.03	^{+	0.17	}_{-	0.20	}$$^{+	0.03	}_{-	0.02	}$$^{+	0.17	}_{-	0.21	} ~ \times ~ 10^{	-1	}$	\\	\hline
	[	50	 ; 	60	]	&	$	0$	&	$	4.68	^{+	2.95	}_{-	1.67	}$$^{+	0.13	}_{-	0.15	}$$^{+	2.96	}_{-	1.67	} ~ \times ~ 10^{	-2	}$	&	$	6.62	^{+	1.08	}_{-	1.33	}$$^{+	0.18	}_{-	0.13	}$$^{+	1.09	}_{-	1.33	} ~ \times ~ 10^{	-2	}$	\\	\hline
	[	60	 ; 	70	]	&	$	0$	&	$	3.18	^{+	2.03	}_{-	1.14	}$$^{+	0.11	}_{-	0.07	}$$^{+	2.03	}_{-	1.14	} ~ \times ~ 10^{	-2	}$	&	$	4.50	^{+	0.74	}_{-	0.91	}$$^{+	0.11	}_{-	0.11	}$$^{+	0.74	}_{-	0.92	} ~ \times ~ 10^{	-2	}$	\\	\hline
	[	70	 ; 	80	]	&	$	0$	&	$	8.96	^{+	5.95	}_{-	3.30	}$$^{+	0.34	}_{-	0.25	}$$^{+	5.96	}_{-	3.31	} ~ \times ~ 10^{	-2	}$	&	$	1.21	^{+	0.17	}_{-	0.24	}$$^{+	0.04	}_{-	0.03	}$$^{+	0.17	}_{-	0.24	} ~ \times ~ 10^{	-1	}$	\\	\hline
\end{tabular}
\caption{\label{tab:diff-ptt} $\PTtt\times{\rm sign}(\dy)$ differential distribution in LO, NLO and NNLO QCD. The format is $central\pm scales\pm pdf\pm total$. At LO only the scale error is given.}
\end{center}
\end{table}
%
\begin{table}[h]
\begin{center}
\begin{tabular}{| c | c | c |}
\hline
$\PTtt$[{\rm GeV}] & \multicolumn{2}{|c|}{$\AFB(\PTtt)$} \\
\hline 
 & {\rm NLO} & {\rm NNLO} \\ \hline
	[	0	 ; 	10	]	&	$	+1.48	^{+	1.72	}_{-	0.51	}$$^{+	0.02	}_{-	0.02	}$$^{+	1.72	}_{-	0.52	} ~ \times ~ 10^{	-1	}$	&	$	+1.64	^{+	0.03	}_{-	0.21	}$$^{+	0.02	}_{-	0.02	}$$^{+	0.03	}_{-	0.21	} ~ \times ~ 10^{	-1	}$	\\	\hline
	[	10	 ; 	20	]	&	$	-7.71	^{+	0.22	}_{-	0.19	}$$^{+	0.35	}_{-	0.18	}$$^{+	0.41	}_{-	0.26	} ~ \times ~ 10^{	-2	}$	&	$	-8.52	^{+	44.2	}_{-	20.1	}$$^{+	0.19	}_{-	0.22	}$$^{+	44.2	}_{-	20.1	} ~ \times ~ 10^{	-3	}$	\\	\hline
	[	20	 ; 	30	]	&	$	-9.00	^{+	0.27	}_{-	0.23	}$$^{+	0.18	}_{-	0.41	}$$^{+	0.33	}_{-	0.47	} ~ \times ~ 10^{	-2	}$	&	$	-3.37	^{+	3.19	}_{-	1.61	}$$^{+	0.11	}_{-	0.07	}$$^{+	3.19	}_{-	1.62	} ~ \times ~ 10^{	-2	}$	\\	\hline
	[	30	 ; 	40	]	&	$	-9.85	^{+	0.30	}_{-	0.25	}$$^{+	0.18	}_{-	0.51	}$$^{+	0.35	}_{-	0.57	} ~ \times ~ 10^{	-2	}$	&	$	-4.79	^{+	2.95	}_{-	1.53	}$$^{+	0.18	}_{-	0.09	}$$^{+	2.95	}_{-	1.53	} ~ \times ~ 10^{	-2	}$	\\	\hline
	[	40	 ; 	50	]	&	$	-1.04	^{+	0.03	}_{-	0.03	}$$^{+	0.03	}_{-	0.04	}$$^{+	0.04	}_{-	0.05	} ~ \times ~ 10^{	-1	}$	&	$	-5.72	^{+	2.65	}_{-	1.34	}$$^{+	0.14	}_{-	0.16	}$$^{+	2.66	}_{-	1.35	} ~ \times ~ 10^{	-2	}$	\\	\hline
	[	50	 ; 	60	]	&	$	-1.08	^{+	0.03	}_{-	0.03	}$$^{+	0.01	}_{-	0.07	}$$^{+	0.04	}_{-	0.08	} ~ \times ~ 10^{	-1	}$	&	$	-6.12	^{+	2.72	}_{-	1.48	}$$^{+	0.21	}_{-	0.13	}$$^{+	2.72	}_{-	1.49	} ~ \times ~ 10^{	-2	}$	\\	\hline
	[	60	 ; 	70	]	&	$	-1.12	^{+	0.04	}_{-	0.03	}$$^{+	0.03	}_{-	0.05	}$$^{+	0.04	}_{-	0.06	} ~ \times ~ 10^{	-1	}$	&	$	-6.41	^{+	2.74	}_{-	1.43	}$$^{+	0.15	}_{-	0.21	}$$^{+	2.74	}_{-	1.45	} ~ \times ~ 10^{	-2	}$	\\	\hline
	[	70	 ; 	80	]	&	$	-1.15	^{+	0.04	}_{-	0.03	}$$^{+	0.04	}_{-	0.05	}$$^{+	0.06	}_{-	0.06	} ~ \times ~ 10^{	-1	}$	&	$	-7.03	^{+	2.95	}_{-	1.46	}$$^{+	0.24	}_{-	0.21	}$$^{+	2.96	}_{-	1.47	} ~ \times ~ 10^{	-2	}$	\\	\hline\end{tabular}
\caption{\label{tab:afb-ptt} $\PTtt$ dependent $\AFB$ in NLO and NNLO QCD. The format is $central\pm scales\pm pdf\pm total$.}
\end{center}
\end{table}


\begin{thebibliography}{99}

\bibitem{Aaltonen:2009iz} 
  T.~Aaltonen {\it et al.} [CDF Collaboration],
  Phys.\ Rev.\ Lett.\  {\bf 102}, 222003 (2009)
  doi:10.1103/PhysRevLett.102.222003
  [arXiv:0903.2850 [hep-ex]].

\bibitem{CDF:2013gna} 
  T.~Aaltonen {\it et al.} [CDF Collaboration],
  Phys.\ Rev.\ Lett.\  {\bf 111}, no. 18, 182002 (2013)
  [arXiv:1306.2357 [hep-ex]].

\bibitem{CDF:public} 
Supplemental information from the CDF Collaboration:\\ 
\url{http://www-cdf.fnal.gov/physics/new/top/2013/TopAngularXS/}

\bibitem{Abazov:2010js} 
  V.~M.~Abazov {\it et al.} [D0 Collaboration],
  Phys.\ Lett.\ B {\bf 693}, 515 (2010)
  doi:10.1016/j.physletb.2010.09.011
  [arXiv:1001.1900 [hep-ex]].
  
\bibitem{Abazov:2014vga} 
  V.~M.~Abazov {\it et al.} [D0 Collaboration],
  Phys.\ Rev.\ D {\bf 90}, no. 9, 092006 (2014)
  [arXiv:1401.5785 [hep-ex]].

\bibitem{D0:public} 
Supplemental information from the D\O\ Collaboration:\\ 
\url{http://www-d0.fnal.gov/Run2Physics/WWW/results/final/TOP/T14D/}

\bibitem{Aaltonen:2012it} 
  T.~Aaltonen {\it et al.}  [CDF Collaboration],
  Phys.\ Rev.\ D {\bf 87}, 092002 (2013)
  [arXiv:1211.1003 [hep-ex]].

\bibitem{CDF:NoteAFBslope} 
  T.~Aaltonen {\it et al.}  [CDF Collaboration],
 CDF Public Note 11161.

\bibitem{Abazov:2014cca} 
  V.~M.~Abazov {\it et al.} [D0 Collaboration],
  Phys.\ Rev.\ D {\bf 90}, no. 7, 072011 (2014)
  [arXiv:1405.0421 [hep-ex]].

\bibitem{Czakon:2014xsa} 
  M.~Czakon, P.~Fiedler and A.~Mitov,
  Phys.\ Rev.\ Lett.\  {\bf 115}, no. 5, 052001 (2015)
  [arXiv:1411.3007 [hep-ph]].

\bibitem{Wang:2015lna} 
  S.~Q.~Wang, X.~G.~Wu, Z.~G.~Si and S.~J.~Brodsky,
  Phys.\ Rev.\ D {\bf 93}, 014004 (2016)
  [arXiv:1508.03739 [hep-ph]].

\bibitem{Brodsky:2011ta} 
  S.~J.~Brodsky and X.~G.~Wu,
  Phys.\ Rev.\ D {\bf 85}, 034038 (2012)
  [Phys.\ Rev.\ D {\bf 86}, 079903 (2012)]
  doi:10.1103/PhysRevD.85.034038, 10.1103/PhysRevD.86.079903
  [arXiv:1111.6175 [hep-ph]].

\bibitem{Gao:2012ja} 
  J.~Gao, C.~S.~Li and H.~X.~Zhu,
  Phys.\ Rev.\ Lett.\  {\bf 110}, no. 4, 042001 (2013)
  [arXiv:1210.2808 [hep-ph]].

\bibitem{Brucherseifer:2013iv} 
  M.~Brucherseifer, F.~Caola and K.~Melnikov,
  JHEP {\bf 1304}, 059 (2013)
  [arXiv:1301.7133 [hep-ph]].

\bibitem{Campbell:2010ff}
  J.~M.~Campbell and R.~K.~Ellis,
  Nucl.\ Phys.\ Proc.\ Suppl.\  {\bf 205-206} (2010) 10
  [arXiv:1007.3492 [hep-ph]].

\bibitem{Alioli:2010xd}
  S.~Alioli, P.~Nason, C.~Oleari and E.~Re,
  JHEP {\bf 1006} (2010) 043
    [arXiv:1002.2581 [hep-ph]].

\bibitem{Alwall:2014hca}
  J.~Alwall {\it et al.},
  JHEP {\bf 1407} (2014) 079
    [arXiv:1405.0301 [hep-ph]].

\bibitem{Gleisberg:2008ta}
  T.~Gleisberg, S.~Hoeche, F.~Krauss, M.~Schonherr, S.~Schumann, F.~Siegert and J.~Winter,
  JHEP {\bf 0902} (2009) 007
  [arXiv:0811.4622 [hep-ph]].

\bibitem{Bevilacqua:2011xh}
  G.~Bevilacqua, M.~Czakon, M.~V.~Garzelli, A.~van Hameren, A.~Kardos, C.~G.~Papadopoulos, R.~Pittau and M.~Worek,
  Comput.\ Phys.\ Commun.\  {\bf 184} (2013) 986
  [arXiv:1110.1499 [hep-ph]].

\bibitem{Denner:2010jp}
  A.~Denner, S.~Dittmaier, S.~Kallweit and S.~Pozzorini,
  Phys.\ Rev.\ Lett.\  {\bf 106} (2011) 052001
    [arXiv:1012.3975 [hep-ph]].

\bibitem{Bevilacqua:2010qb}
  G.~Bevilacqua, M.~Czakon, A.~van Hameren, C.~G.~Papadopoulos and M.~Worek,
  JHEP {\bf 1102} (2011) 083
    [arXiv:1012.4230 [hep-ph]].

\bibitem{Denner:2012yc}
  A.~Denner, S.~Dittmaier, S.~Kallweit and S.~Pozzorini,
  JHEP {\bf 1210} (2012) 110
    [arXiv:1207.5018 [hep-ph]].

\bibitem{Frederix:2013gra}
  R.~Frederix,
  Phys.\ Rev.\ Lett.\  {\bf 112} (2014) 8,  082002
    [arXiv:1311.4893 [hep-ph]].

\bibitem{Cascioli:2013wga}
  F.~Cascioli, S.~Kallweit, P.~Maierhoefer and S.~Pozzorini,
  Eur.\ Phys.\ J.\ C {\bf 74} (2014) 3,  2783
   [arXiv:1312.0546 [hep-ph]].

\bibitem{Heinrich:2013qaa}
  G.~Heinrich, A.~Maier, R.~Nisius, J.~Schlenk and J.~Winter,
  JHEP {\bf 1406} (2014) 158
    [arXiv:1312.6659 [hep-ph]].

\bibitem{Denner:2015yca}
  A.~Denner and R.~Feger,
  JHEP {\bf 1511} (2015) 209
    [arXiv:1506.07448 [hep-ph]].

\bibitem{Bevilacqua:2015qha}
  G.~Bevilacqua, H.~B.~Hartanto, M.~Kraus and M.~Worek,
  arXiv:1509.09242 [hep-ph].

\bibitem{Bernreuther:2010ny}
  W.~Bernreuther and Z.~G.~Si,
  Nucl.\ Phys.\ B {\bf 837} (2010) 90
    [arXiv:1003.3926 [hep-ph]].

\bibitem{Melnikov:2009dn}
  K.~Melnikov and M.~Schulze,
  JHEP {\bf 0908} (2009) 049
    [arXiv:0907.3090 [hep-ph]].

\bibitem{Campbell:2012uf}
  J.~M.~Campbell and R.~K.~Ellis,
  J.\ Phys.\ G {\bf 42} (2015) 1,  015005
    [arXiv:1204.1513 [hep-ph]].

\bibitem{Campbell:2014kua}
  J.~M.~Campbell, R.~K.~Ellis, P.~Nason and E.~Re,
  JHEP {\bf 1504} (2015) 114
    [arXiv:1412.1828 [hep-ph]].

\bibitem{GehrmannDeRidder:2008ug} 
  A.~Gehrmann-De Ridder, T.~Gehrmann, E.~W.~N.~Glover and G.~Heinrich,
  Phys.\ Rev.\ Lett.\  {\bf 100}, 172001 (2008)
  [arXiv:0802.0813 [hep-ph]].
  
\bibitem{GehrmannDeRidder:2007hr} 
  A.~Gehrmann-De Ridder, T.~Gehrmann, E.~W.~N.~Glover and G.~Heinrich,
  JHEP {\bf 0712}, 094 (2007)
  [arXiv:0711.4711 [hep-ph]].

\bibitem{Weinzierl:2008iv} 
  S.~Weinzierl,
  Phys.\ Rev.\ Lett.\  {\bf 101}, 162001 (2008)
  [arXiv:0807.3241 [hep-ph]].

\bibitem{Weinzierl:2009ms} 
  S.~Weinzierl,
  JHEP {\bf 0906}, 041 (2009)
  [arXiv:0904.1077 [hep-ph]].

\bibitem{DelDuca:2015zqa} 
  V.~Del Duca, C.~Duhr, G.~Somogyi, F.~Tramontano and Z.~Trocsanyi,
  JHEP {\bf 1504}, 036 (2015)
  [arXiv:1501.07226 [hep-ph]].

\bibitem{Grazzini:2008tf} 
  M.~Grazzini,
  JHEP {\bf 0802}, 043 (2008)
  [arXiv:0801.3232 [hep-ph]].

\bibitem{Catani:2009sm} 
  S.~Catani, L.~Cieri, G.~Ferrera, D.~de Florian and M.~Grazzini,
  Phys.\ Rev.\ Lett.\  {\bf 103}, 082001 (2009)
   [arXiv:0903.2120 [hep-ph]].
  
\bibitem{Ferrera:2011bk} 
  G.~Ferrera, M.~Grazzini and F.~Tramontano,
  Phys.\ Rev.\ Lett.\  {\bf 107}, 152003 (2011)
  [arXiv:1107.1164 [hep-ph]].

\bibitem{Catani:2011qz} 
  S.~Catani, L.~Cieri, D.~de Florian, G.~Ferrera and M.~Grazzini,
  Phys.\ Rev.\ Lett.\  {\bf 108}, 072001 (2012)
  [arXiv:1110.2375 [hep-ph]].

\bibitem{Grazzini:2013bna} 
  M.~Grazzini, S.~Kallweit, D.~Rathlev and A.~Torre,
  Phys.\ Lett.\ B {\bf 731}, 204 (2014)
  [arXiv:1309.7000 [hep-ph]].
  
\bibitem{Grazzini:2015nwa} 
  M.~Grazzini, S.~Kallweit and D.~Rathlev,
  JHEP {\bf 1507}, 085 (2015)
  [arXiv:1504.01330 [hep-ph]].

\bibitem{Cascioli:2014yka} 
  F.~Cascioli {\it et al.},
  Phys.\ Lett.\ B {\bf 735}, 311 (2014)
  [arXiv:1405.2219 [hep-ph]].

\bibitem{Gehrmann:2014fva} 
  T.~Gehrmann, M.~Grazzini, S.~Kallweit, P.~Maierhoefer, A.~von Manteuffel, S.~Pozzorini, D.~Rathlev and L.~Tancredi,
  Phys.\ Rev.\ Lett.\  {\bf 113}, no. 21, 212001 (2014)
  [arXiv:1408.5243 [hep-ph]].

\bibitem{Grazzini:2015hta} 
  M.~Grazzini, S.~Kallweit and D.~Rathlev,
  Phys.\ Lett.\ B {\bf 750}, 407 (2015)
  [arXiv:1507.06257 [hep-ph]].

\bibitem{Boughezal:2015aha} 
  R.~Boughezal, C.~Focke, W.~Giele, X.~Liu and F.~Petriello,
  Phys.\ Lett.\ B {\bf 748}, 5 (2015)
  [arXiv:1505.03893 [hep-ph]].
  
\bibitem{Boughezal:2015ded} 
  R.~Boughezal, J.~M.~Campbell, R.~K.~Ellis, C.~Focke, W.~T.~Giele, X.~Liu and F.~Petriello,
  arXiv:1512.01291 [hep-ph].

\bibitem{Campbell:2016jau} 
  J.~M.~Campbell, R.~K.~Ellis and C.~Williams,
  arXiv:1601.00658 [hep-ph].

\bibitem{Ridder:2013mf} 
  A.~Gehrmann-De Ridder, T.~Gehrmann, E.~W.~N.~Glover and J.~Pires,
  Phys.\ Rev.\ Lett.\  {\bf 110}, no. 16, 162003 (2013)
  [arXiv:1301.7310 [hep-ph]].

\bibitem{Currie:2013dwa} 
  J.~Currie, A.~Gehrmann-De Ridder, E.~W.~N.~Glover and J.~Pires,
  JHEP {\bf 1401}, 110 (2014)
  [arXiv:1310.3993 [hep-ph]].

\bibitem{Chen:2014gva} 
  X.~Chen, T.~Gehrmann, E.~W.~N.~Glover and M.~Jaquier,
  Phys.\ Lett.\ B {\bf 740}, 147 (2015)
  [arXiv:1408.5325 [hep-ph]].

\bibitem{Ridder:2015dxa} 
  A.~Gehrmann-De Ridder, T.~Gehrmann, E.~W.~N.~Glover, A.~Huss and T.~A.~Morgan,
  arXiv:1507.02850 [hep-ph].
  
\bibitem{Boughezal:2011jf} 
  R.~Boughezal, K.~Melnikov and F.~Petriello,
  Phys.\ Rev.\ D {\bf 85}, 034025 (2012)
  [arXiv:1111.7041 [hep-ph]].
  
\bibitem{Brucherseifer:2013cu} 
  M.~Brucherseifer, F.~Caola and K.~Melnikov,
  Phys.\ Lett.\ B {\bf 721}, 107 (2013)
  [arXiv:1302.0444 [hep-ph]].
  
\bibitem{Boughezal:2013uia} 
  R.~Boughezal, F.~Caola, K.~Melnikov, F.~Petriello and M.~Schulze,
  JHEP {\bf 1306}, 072 (2013)
  [arXiv:1302.6216 [hep-ph]].

\bibitem{Boughezal:2015dra} 
  R.~Boughezal, F.~Caola, K.~Melnikov, F.~Petriello and M.~Schulze,
  Phys.\ Rev.\ Lett.\  {\bf 115}, no. 8, 082003 (2015)
  [arXiv:1504.07922 [hep-ph]].
  
\bibitem{Caola:2014daa} 
  F.~Caola, A.~Czarnecki, Y.~Liang, K.~Melnikov and R.~Szafron,
  Phys.\ Rev.\ D {\bf 90}, no. 5, 053004 (2014)
  [arXiv:1403.3386 [hep-ph]].

\bibitem{Brucherseifer:2014ama} 
  M.~Brucherseifer, F.~Caola and K.~Melnikov,
  Phys.\ Lett.\ B {\bf 736}, 58 (2014)
  [arXiv:1404.7116 [hep-ph]].

\bibitem{Caola:2015wna} 
  F.~Caola, K.~Melnikov and M.~Schulze,
  Phys.\ Rev.\ D {\bf 92}, no. 7, 074032 (2015)
  [arXiv:1508.02684 [hep-ph]].

\bibitem{Catani:2007vq} 
  S.~Catani and M.~Grazzini,
  Phys.\ Rev.\ Lett.\  {\bf 98}, 222002 (2007)
  [hep-ph/0703012].

\bibitem{Boughezal:2015dva} 
  R.~Boughezal, C.~Focke, X.~Liu and F.~Petriello,
  Phys.\ Rev.\ Lett.\  {\bf 115}, no. 6, 062002 (2015)
  [arXiv:1504.02131 [hep-ph]].

\bibitem{Czakon:2014oma} 
  M.~Czakon and D.~Heymes,
  Nucl.\ Phys.\ B {\bf 890}, 152 (2014)
  [arXiv:1408.2500 [hep-ph]].

\bibitem{GehrmannDeRidder:2005cm} 
  A.~Gehrmann-De Ridder, T.~Gehrmann and E.~W.~N.~Glover,
  JHEP {\bf 0509}, 056 (2005)
  [hep-ph/0505111].

\bibitem{Somogyi:2006da} 
  G.~Somogyi, Z.~Trocsanyi and V.~Del Duca,
  JHEP {\bf 0701}, 070 (2007)
  [hep-ph/0609042].

\bibitem{Somogyi:2006db} 
  G.~Somogyi and Z.~Trocsanyi,
  JHEP {\bf 0701}, 052 (2007)
  [hep-ph/0609043].

\bibitem{Somogyi:2008fc} 
  G.~Somogyi and Z.~Trocsanyi,
  JHEP {\bf 0808}, 042 (2008)
  [arXiv:0807.0509 [hep-ph]].
  
\bibitem{Bolzoni:2010bt} 
  P.~Bolzoni, G.~Somogyi and Z.~Trocsanyi,
  JHEP {\bf 1101}, 059 (2011)
  [arXiv:1011.1909 [hep-ph]].

\bibitem{Somogyi:2013yk} 
  G.~Somogyi,
  JHEP {\bf 1304}, 010 (2013)
  [arXiv:1301.3919 [hep-ph]].

\bibitem{Czakon:2010td} 
  M.~Czakon,
  Phys.\ Lett.\ B {\bf 693}, 259 (2010)
  [arXiv:1005.0274 [hep-ph]].

\bibitem{Czakon:2011ve} 
  M.~Czakon,
  Nucl.\ Phys.\ B {\bf 849}, 250 (2011)
  [arXiv:1101.0642 [hep-ph]].

\bibitem{Gaunt:2015pea} 
  J.~Gaunt, M.~Stahlhofen, F.~J.~Tackmann and J.~R.~Walsh,
  JHEP {\bf 1509}, 058 (2015)
  [arXiv:1505.04794 [hep-ph]].
  
\bibitem{Anastasiou:2000kg} 
  C.~Anastasiou, E.~W.~N.~Glover, C.~Oleari and M.~E.~Tejeda-Yeomans,
  Nucl.\ Phys.\ B {\bf 601}, 318 (2001)
  [hep-ph/0010212].

\bibitem{Anastasiou:2000ue} 
  C.~Anastasiou, E.~W.~N.~Glover, C.~Oleari and M.~E.~Tejeda-Yeomans,
  Nucl.\ Phys.\ B {\bf 601}, 341 (2001)
  [hep-ph/0011094].

\bibitem{Anastasiou:2001sv} 
  C.~Anastasiou, E.~W.~N.~Glover, C.~Oleari and M.~E.~Tejeda-Yeomans,
  Nucl.\ Phys.\ B {\bf 605}, 486 (2001)
  [hep-ph/0101304].
  
\bibitem{Glover:2001af} 
  E.~W.~N.~Glover, C.~Oleari and M.~E.~Tejeda-Yeomans,
  Nucl.\ Phys.\ B {\bf 605}, 467 (2001)
  [hep-ph/0102201].
  
\bibitem{Bern:2002tk} 
  Z.~Bern, A.~De Freitas and L.~J.~Dixon,
  JHEP {\bf 0203}, 018 (2002)
  [hep-ph/0201161].

\bibitem{Bern:2003ck} 
  Z.~Bern, A.~De Freitas and L.~J.~Dixon,
  JHEP {\bf 0306}, 028 (2003)
  [JHEP {\bf 1404}, 112 (2014)]
  [hep-ph/0304168].
  
\bibitem{Czakon:2008zk} 
  M.~Czakon,
  Phys.\ Lett.\ B {\bf 664}, 307 (2008)
  [arXiv:0803.1400 [hep-ph]].

\bibitem{Baernreuther:2013caa} 
  P.~Baernreuther, M.~Czakon and P.~Fiedler,
  JHEP {\bf 1402}, 078 (2014)
  [arXiv:1312.6279 [hep-ph]].
  
\bibitem{Bonciani:2008az} 
  R.~Bonciani, A.~Ferroglia, T.~Gehrmann, D.~Maitre and C.~Studerus,
  JHEP {\bf 0807}, 129 (2008)
  [arXiv:0806.2301 [hep-ph]].
  
\bibitem{Bonciani:2009nb} 
  R.~Bonciani, A.~Ferroglia, T.~Gehrmann and C.~Studerus,
  JHEP {\bf 0908}, 067 (2009)
  [arXiv:0906.3671 [hep-ph]].

\bibitem{Bonciani:2010mn} 
  R.~Bonciani, A.~Ferroglia, T.~Gehrmann, A.~von Manteuffel and C.~Studerus,
  JHEP {\bf 1101}, 102 (2011)
  [arXiv:1011.6661 [hep-ph]].

\bibitem{Bonciani:2013ywa} 
  R.~Bonciani, A.~Ferroglia, T.~Gehrmann, A.~von Manteuffel and C.~Studerus,
  JHEP {\bf 1312}, 038 (2013)
  [arXiv:1309.4450 [hep-ph]].

\bibitem{Gehrmann:2013cxs} 
  T.~Gehrmann, L.~Tancredi and E.~Weihs,
  JHEP {\bf 1308}, 070 (2013)
  [arXiv:1306.6344, arXiv:1306.6344 [hep-ph]].

\bibitem{Gehrmann:2014bfa} 
  T.~Gehrmann, A.~von Manteuffel, L.~Tancredi and E.~Weihs,
  JHEP {\bf 1406}, 032 (2014)
  [arXiv:1404.4853 [hep-ph]].
  
\bibitem{Henn:2014lfa} 
  J.~M.~Henn, K.~Melnikov and V.~A.~Smirnov,
  JHEP {\bf 1405}, 090 (2014)
  [arXiv:1402.7078 [hep-ph]].

\bibitem{Caola:2014lpa} 
  F.~Caola, J.~M.~Henn, K.~Melnikov and V.~A.~Smirnov,
  JHEP {\bf 1409}, 043 (2014)
  [arXiv:1404.5590 [hep-ph]].

\bibitem{Papadopoulos:2014hla} 
  C.~G.~Papadopoulos, D.~Tommasini and C.~Wever,
  JHEP {\bf 1501}, 072 (2015)
  [arXiv:1409.6114 [hep-ph]].

\bibitem{Caola:2014iua} 
  F.~Caola, J.~M.~Henn, K.~Melnikov, A.~V.~Smirnov and V.~A.~Smirnov,
  JHEP {\bf 1411}, 041 (2014)
  [arXiv:1408.6409 [hep-ph]].

\bibitem{Gehrmann:2015ora} 
  T.~Gehrmann, A.~von Manteuffel and L.~Tancredi,
  JHEP {\bf 1509}, 128 (2015)
  [arXiv:1503.04812 [hep-ph]].

\bibitem{vonManteuffel:2015msa} 
  A.~von Manteuffel and L.~Tancredi,
  JHEP {\bf 1506}, 197 (2015)
  [arXiv:1503.08835 [hep-ph]].

\bibitem{Czakon:2015owf} 
  M.~Czakon, D.~Heymes and A.~Mitov,
  arXiv:1511.00549 [hep-ph].

\bibitem{Abelof:2015lna} 
  G.~Abelof, A.~Gehrmann-De Ridder and I.~Majer,
  JHEP {\bf 1512}, 074 (2015)
  [arXiv:1506.04037 [hep-ph]].
  
\bibitem{Abelof:2014fza} 
  G.~Abelof, A.~Gehrmann-De Ridder, P.~Maierhofer and S.~Pozzorini,
  JHEP {\bf 1408}, 035 (2014)
  [arXiv:1404.6493 [hep-ph]].

\bibitem{Bonciani:2015sha} 
  R.~Bonciani, S.~Catani, M.~Grazzini, H.~Sargsyan and A.~Torre,
  Eur.\ Phys.\ J.\ C {\bf 75}, no. 12, 581 (2015)
  [arXiv:1508.03585 [hep-ph]].

\bibitem{Czakon:2013goa} 
  M.~Czakon, P.~Fiedler and A.~Mitov,
  Phys.\ Rev.\ Lett.\  {\bf 110}, no. 25, 252004 (2013)
  [arXiv:1303.6254 [hep-ph]].

\bibitem{Weinzierl:2011uz} 
  S.~Weinzierl,
  Phys.\ Rev.\ D {\bf 84}, 074007 (2011)
  [arXiv:1107.5131 [hep-ph]].

\bibitem{Czakon:2012pz} 
  M.~Czakon and A.~Mitov,
  JHEP {\bf 1301}, 080 (2013)
  [arXiv:1210.6832 [hep-ph]].

\bibitem{Czakon:2012zr} 
  M.~Czakon and A.~Mitov,
  JHEP {\bf 1212}, 054 (2012)
  [arXiv:1207.0236 [hep-ph]].

\bibitem{Baernreuther:2012ws} 
  P.~B\"arnreuther, M.~Czakon and A.~Mitov,
  Phys.\ Rev.\ Lett.\  {\bf 109}, 132001 (2012)
  [arXiv:1204.5201 [hep-ph]].
  
\bibitem{Ferroglia:2009ii} 
  A.~Ferroglia, M.~Neubert, B.~D.~Pecjak and L.~L.~Yang,
  JHEP {\bf 0911}, 062 (2009)
  [arXiv:0908.3676 [hep-ph]].

\bibitem{Anastasiou:2008vd} 
  C.~Anastasiou and S.~M.~Aybat,
  Phys.\ Rev.\ D {\bf 78}, 114006 (2008)
  [arXiv:0809.1355 [hep-ph]].

\bibitem{Dittmaier:2007wz}
 S.~Dittmaier, P.~Uwer, S.~Weinzierl,
 Phys.\ Rev.\ Lett.\  {\bf 98}, 262002 (2007)
 [hep-ph/0703120 [hep-ph]];
 Eur.\ Phys.\ J.\  {\bf C59}, 625-646 (2009)
 [arXiv:0810.0452 [hep-ph]].
  
\bibitem{Czakon:2011xx} 
  M.~Czakon and A.~Mitov,
  Comput.\ Phys.\ Commun.\  {\bf 185}, 2930 (2014)
  [arXiv:1112.5675 [hep-ph]].
  
\bibitem{Cacciari:2008zb} 
  M.~Cacciari, S.~Frixione, M.~L.~Mangano, P.~Nason and G.~Ridolfi,
  JHEP {\bf 0809}, 127 (2008)
  [arXiv:0804.2800 [hep-ph]].

\bibitem{Martin:2009iq} 
  A.~D.~Martin, W.~J.~Stirling, R.~S.~Thorne and G.~Watt,
  Eur.\ Phys.\ J.\ C {\bf 63}, 189 (2009)
  [arXiv:0901.0002 [hep-ph]].

\bibitem{Gao:2013xoa} 
  J.~Gao {\it et al.},
  Phys.\ Rev.\ D {\bf 89}, no. 3, 033009 (2014)
  [arXiv:1302.6246 [hep-ph]].

\bibitem{Ball:2012cx} 
  R.~D.~Ball {\it et al.},
  Nucl.\ Phys.\ B {\bf 867}, 244 (2013)
  [arXiv:1207.1303 [hep-ph]].
  
\bibitem{CooperSarkar:2011aa} 
  A.~M.~Cooper-Sarkar [ZEUS and H1 Collaborations],
  PoS EPS {\bf -HEP2011}, 320 (2011)
  [arXiv:1112.2107 [hep-ph]].

\bibitem{Bernreuther:2012sx} 
  W.~Bernreuther and Z.~G.~Si,
  Phys.\ Rev.\ D {\bf 86}, 034026 (2012)
  [arXiv:1205.6580 [hep-ph]].

\bibitem{Ahrens:2010zv} 
  V.~Ahrens, A.~Ferroglia, M.~Neubert, B.~D.~Pecjak and L.~L.~Yang,
  JHEP {\bf 1009}, 097 (2010)
  [arXiv:1003.5827 [hep-ph]].

\bibitem{Ferroglia:2015ivv} 
  A.~Ferroglia, B.~D.~Pecjak, D.~J.~Scott and L.~L.~Yang,
  arXiv:1512.02535 [hep-ph].

\bibitem{Ferroglia:2013zwa} 
  A.~Ferroglia, B.~D.~Pecjak and L.~L.~Yang,
  JHEP {\bf 1309}, 032 (2013)
  [arXiv:1306.1537 [hep-ph]].

\bibitem{Ahrens:2011mw} 
  V.~Ahrens, A.~Ferroglia, M.~Neubert, B.~D.~Pecjak and L.~L.~Yang,
  JHEP {\bf 1109}, 070 (2011)
  [arXiv:1103.0550 [hep-ph]].

\bibitem{Kidonakis:2010dk} 
  N.~Kidonakis,
  Phys.\ Rev.\ D {\bf 82}, 114030 (2010)
  [arXiv:1009.4935 [hep-ph]].

\bibitem{Kidonakis:2011zn} 
  N.~Kidonakis,
  Phys.\ Rev.\ D {\bf 84}, 011504 (2011)
  [arXiv:1105.5167 [hep-ph]].

\bibitem{Melnikov:2010iu} 
  K.~Melnikov and M.~Schulze,
  Nucl.\ Phys.\ B {\bf 840}, 129 (2010)
  [arXiv:1004.3284 [hep-ph]].

\bibitem{Melnikov:2011qx} 
  K.~Melnikov, A.~Scharf and M.~Schulze,
  Phys.\ Rev.\ D {\bf 85}, 054002 (2012)
  [arXiv:1111.4991 [hep-ph]].
    
\bibitem{Hoeche:2013mua} 
  S.~Hoeche, J.~Huang, G.~Luisoni, M.~Schoenherr and J.~Winter,
  Phys.\ Rev.\ D {\bf 88}, no. 1, 014040 (2013)
  [arXiv:1306.2703 [hep-ph]].

\bibitem{Gripaios:2013rda} 
  B.~Gripaios, A.~Papaefstathiou and B.~Webber,
  JHEP {\bf 1311}, 105 (2013)
  [arXiv:1309.0810 [hep-ph]].

\bibitem{Almeida:2008ug} 
  L.~G.~Almeida, G.~F.~Sterman and W.~Vogelsang,
  Phys.\ Rev.\ D {\bf 78}, 014008 (2008)
  [arXiv:0805.1885 [hep-ph]].

\bibitem{Ahrens:2011uf} 
  V.~Ahrens, A.~Ferroglia, M.~Neubert, B.~D.~Pecjak and L.~L.~Yang,
  Phys.\ Rev.\ D {\bf 84}, 074004 (2011)
  [arXiv:1106.6051 [hep-ph]].

\bibitem{Skands:2012mm} 
  P.~Skands, B.~Webber and J.~Winter,
  JHEP {\bf 1207}, 151 (2012)
  [arXiv:1205.1466 [hep-ph]].

\bibitem{Kidonakis:2015ona} 
  N.~Kidonakis,
  Phys.\ Rev.\ D {\bf 91}, no. 7, 071502 (2015)
  [arXiv:1501.01581 [hep-ph]].

\bibitem{Cacciari:2011hy} 
  M.~Cacciari, M.~Czakon, M.~Mangano, A.~Mitov and P.~Nason,
  Phys.\ Lett.\ B {\bf 710}, 612 (2012)
  [arXiv:1111.5869 [hep-ph]].

\bibitem{Mrenna:1996cz} 
  S.~Mrenna and C.~P.~Yuan,
  Phys.\ Rev.\ D {\bf 55}, 120 (1997)
  [hep-ph/9606363].

\bibitem{Li:2013mia} 
  H.~T.~Li, C.~S.~Li, D.~Y.~Shao, L.~L.~Yang and H.~X.~Zhu,
  Phys.\ Rev.\ D {\bf 88}, 074004 (2013)
  [arXiv:1307.2464].

\bibitem{Czakon:2015pga} 
  M.~Czakon, A.~Mitov and J.~Rojo,
  arXiv:1501.01112 [hep-ph].

\bibitem{Czakon:2013tha} 
  M.~Czakon, M.~L.~Mangano, A.~Mitov and J.~Rojo,
  JHEP {\bf 1307}, 167 (2013)
  [arXiv:1303.7215 [hep-ph]].

\bibitem{Aad:2014zka} 
  G.~Aad {\it et al.} [ATLAS Collaboration],
  Phys.\ Rev.\ D {\bf 90}, no. 7, 072004 (2014)
  [arXiv:1407.0371 [hep-ex]].
    
\end{thebibliography}
\end{document}